\documentclass{article}
\usepackage{mystyle}
\usepackage{xspace}

\newcommand{\rmS}{\mathrm{S}}
\newcommand{\obs}{\mathrm{obs}}
\newcommand{\symmpi}{\text{\normalfont\ttfamily C-SymmPI}}
\newcommand{\pen}{\mathrm{penalty}}
\newcommand{\inte}{\mathrm{int}}
\newcommand{\TV}{\mathrm{TV}}
\newcommand{\proj}{\mathrm{proj}}
\newcommand{\inn}{\mathrm{in}}
\newcommand{\out}{\mathrm{out}}
\newcommand{\samp}{\mathrm{samp}}
\newcommand{\stab}[1]{\mathrm{Stab}(#1)}
\newcommand{\csymmpi}{\texttt{C-SymmPI}\xspace}

\begin{document}
\title{Conditional Predictive Inference for General Structured Data with Group Symmetries}
\author{
    Yichen Shen\thanks{Data Science Institute, University of Chicago. Email: \texttt{ycshen@uchicago.edu}} \qquad
    Mengxin Yu\thanks{Department of Statistics and Data Science, Washington University in St. Louis. Email: \texttt{myu@wustl.edu}}
}
\maketitle
\begin{abstract}

We study distribution-free predictive inference for structured data with group symmetries, aiming to provide near-conditional coverage beyond observation-level exchangeability. Existing methods for structured data largely provide marginal coverage, which can mask substantial heterogeneity across observations. We introduce \texttt{C-SymmPI}, a unified framework that controls function-weighted coverage errors over a user-specified function class, with the flexibility to condition on any chosen representation of the observed data. We establish theoretical guarantees under both distributional invariance and distribution shift, and we derive convergence rates for linear and RKHS function classes. Experiments on two-layer hierarchical and network-structured data demonstrate improved conditional coverage and adaptation to local heterogeneity.
\end{abstract}

\tableofcontents

\clearpage
\section{Introduction}
\label{sec:intro}

In distribution-free predictive inference, we observe a partially observed version $Z_\obs$ of an underlying complete data object $Z$. The goal is to construct a prediction set $\hat{C}(Z_\obs)$ that contains the target $Z$, without making strong distributional assumptions. A widely used approach is conformal prediction \citep{vovk2005algorithmic,angelopoulos2021gentle}, which guarantees \emph{marginal coverage} under the exchangeability assumption:
$$
\mathbb{P}\rbr{Z \in \hat{C}(Z_\obs)}\geq 1 - \alpha,
$$
where $\alpha \in (0,1)$ is a user-specified miscoverage level. This guarantee is marginal in the sense that it averages over the randomness in the training, calibration, and test data. As a result, it may exhibit local miscoverage when the data distribution is heterogeneous across different realizations of $Z_{\obs}$. In other words, for some realizations of $Z_\obs$, the prediction set $\hat{C}(Z_\obs)$ may exhibit under- or over-coverage. Therefore, one may seek a stronger guarantee that holds conditionally on each realization of $Z_\obs$, known as \emph{conditional coverage}:
$$
\mathbb{P}\rbr{Z \in \hat{C}(Z_\obs) \mid Z_\obs}\geq 1 - \alpha.
$$

Without distributional assumptions, exact conditional coverage cannot be achieved in general \citep{vovk2005algorithmic,foygel2021limits}. Several recent papers therefore pursue \emph{near-conditional coverage} instead \citep{hore2024conformalpredictionlocalweights, gibbs2025conformal, lee2025conditionalpredictiveinferencelkcoverage}. However, these methods typically rely on the global exchangeability of the data \citep{romano2019conformalized, hore2024conformalpredictionlocalweights, zhang2024posteriorconformalprediction, gibbs2025conformal, lee2025conditionalpredictiveinferencelkcoverage}. This assumption is frequently violated for structured data, such as graph-, cluster-, or tree-structured data.

Recently, \citet{dobriban2024symmpipredictiveinferencedata} proposed SymmPI, a unified framework for constructing valid prediction sets under group symmetries (see Section~\ref{sec:symmpi_framework}). Group symmetries extend classical exchangeability, so SymmPI applies to graph-structured data and to hierarchical data arising in cluster randomized trials.
While SymmPI extends validity beyond global exchangeability, it \emph{mainly
guarantees marginal coverage}. In heterogeneous settings, marginal
coverage may be insufficient, motivating the need for stronger conditional
guarantees. This need arises naturally in several real-world scenarios:

\begin{itemize}
\item[(i)] In cluster randomized trials, heterogeneity arises at multiple levels (both across and within clusters). Different clusters may serve distinct patient populations or follow different protocols, inducing cluster-specific effects. Individuals within each cluster also differ in their characteristics and responses. Marginal guarantees \citep{dobriban2024symmpipredictiveinferencedata,wang2024conformal} average over all clusters and individuals. They can therefore mask prediction sets that are too narrow for high-risk clusters and too wide for low-risk ones. For clinical decision-making, conditional coverage that provides cluster- or individual-level validity is often preferred.
\item[(ii)] Networks are often heterogeneous in their connectivity structure, with dense communities and sparse bridging regions. A marginal guarantee may fail to provide reliable coverage for structurally distinct nodes. For example, it may exhibit undercoverage for nodes in sparse regions, while exhibiting overcoverage for nodes in dense communities. This motivates guarantees conditional on a node's local topology, which allow robust inference across structurally distinct nodes.
\end{itemize}

In both scenarios, the marginal coverage provided by existing works \citep{dobriban2024symmpipredictiveinferencedata, wang2024conformal} may not adequately capture the heterogeneity inherent in the data. Motivated by this limitation, we aim to address the following question:
\begin{quote}
\textit{Can we achieve near-conditional predictive inference guarantees for broader classes of data structures that go beyond exchangeability?}
\end{quote}

We answer this question affirmatively. Our key idea is to replace the single orbit quantile of SymmPI with a threshold learned by quantile regression along the group orbit. The threshold then adapts to the observed data, while the group symmetry preserves validity. Our key contributions are summarized below.

\subsection{Our Contributions}
\paragraph{A unified framework for near-conditional inference.}
We introduce \texttt{C-SymmPI}, a unified framework for \emph{near-conditional} predictive inference for data with group symmetries, such as networks and clustered data, where observation-level exchangeability may fail. \texttt{C-SymmPI} allows (near) conditioning on an \emph{arbitrarily} chosen representation $\eta(Z_{\obs})$ of the observed data, so that the conditioning information can be tailored to the inferential goal. For example, the representation can be the full observed data ($\eta=\mathrm{id}$) or only the test covariate ($\eta(Z_{\obs})=X_{n+1}$). For large or infinite groups, a sampling-based approximation of the group average reduces computation while retaining a near-conditional guarantee.

\paragraph{Statistical guarantees.}
We adopt a relaxed multi-accuracy formulation that characterizes departures from conditional coverage through function-weighted averages of coverage errors. Within this formulation, we establish finite-sample guarantees for \texttt{C-SymmPI}, both under exact symmetry and under departures from it. We also derive explicit convergence rates for linear and RKHS function classes. These rates are governed by an effective sample size determined by the symmetry, recovering the known rates under exchangeability and improving as the symmetry becomes richer.

\paragraph{Broad applicability.} We demonstrate \texttt{C-SymmPI} in two structured settings where observation-level exchangeability fails. For \textbf{two-layer hierarchical models} with random cluster sizes, the group structure induces a principled weighting of the calibration scores. We further specialize the framework to cluster randomized trials, with near-conditional guarantees for both individual- and cluster-level treatment effects. For \textbf{network-structured data}, we adapt the framework to network-assisted regression, where \texttt{C-SymmPI} leverages network topology to provide topology-aware conditional guarantees. In the classical exchangeable setting, \texttt{C-SymmPI} recovers existing conditional conformal guarantees as a special case. Similar constructions extend to other structured data with group symmetries, such as trees and images.

\begin{figure}[htbp]
    \centering
    \begin{minipage}[t]{0.32\textwidth}
        \centering
        \includegraphics[width=\linewidth]{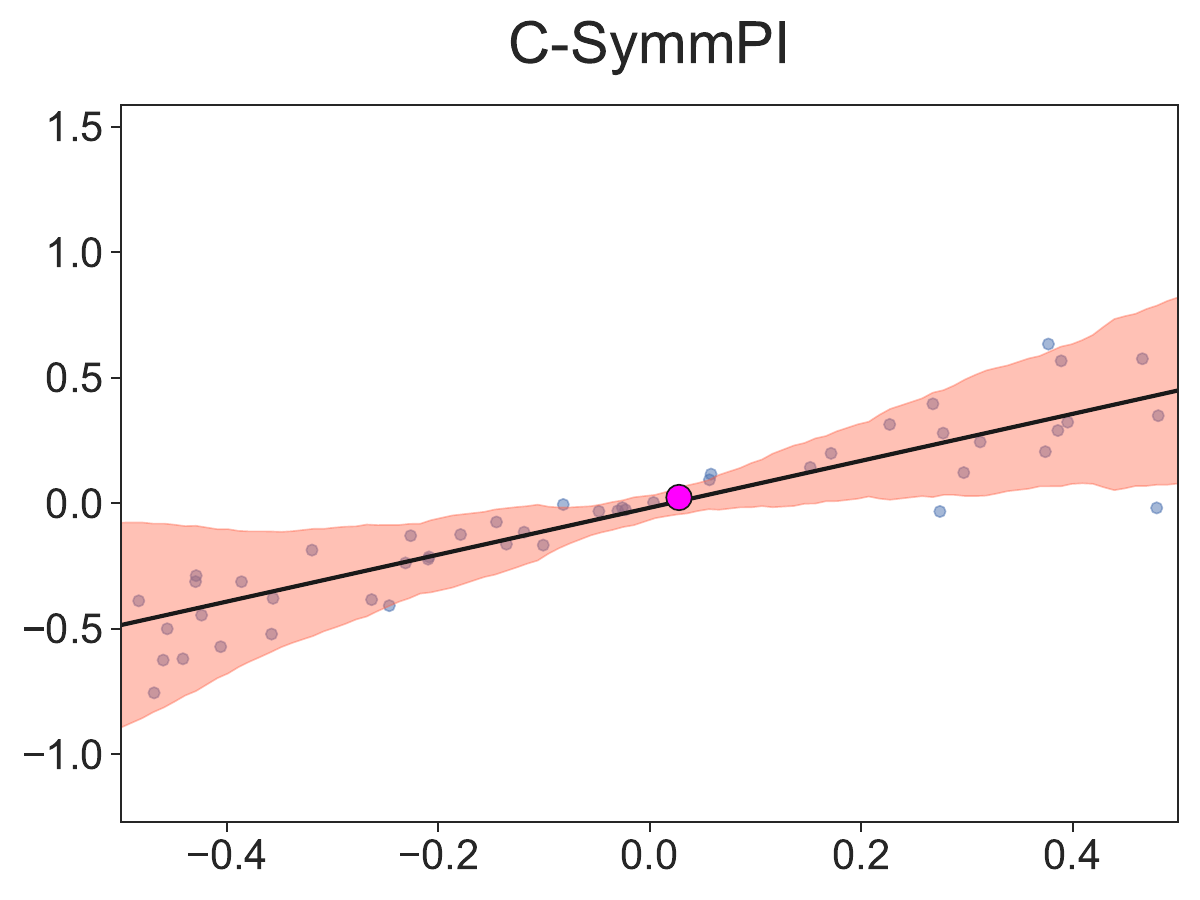}
    \end{minipage}
    \hfill
    \begin{minipage}[t]{0.32\textwidth}
        \centering
        \includegraphics[width=\linewidth]{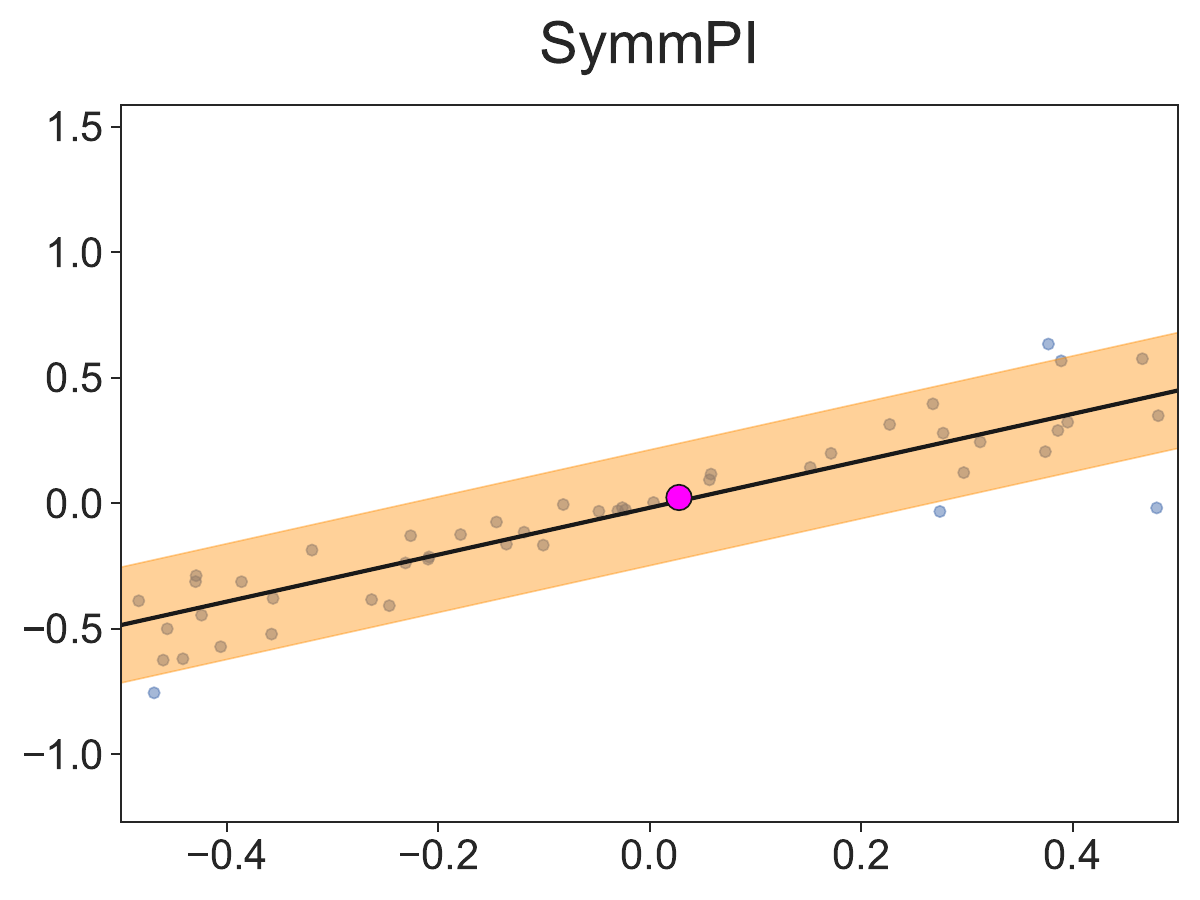}
    \end{minipage}
    \hfill
    \begin{minipage}[t]{0.32\textwidth}
        \centering
        \includegraphics[width=\linewidth]{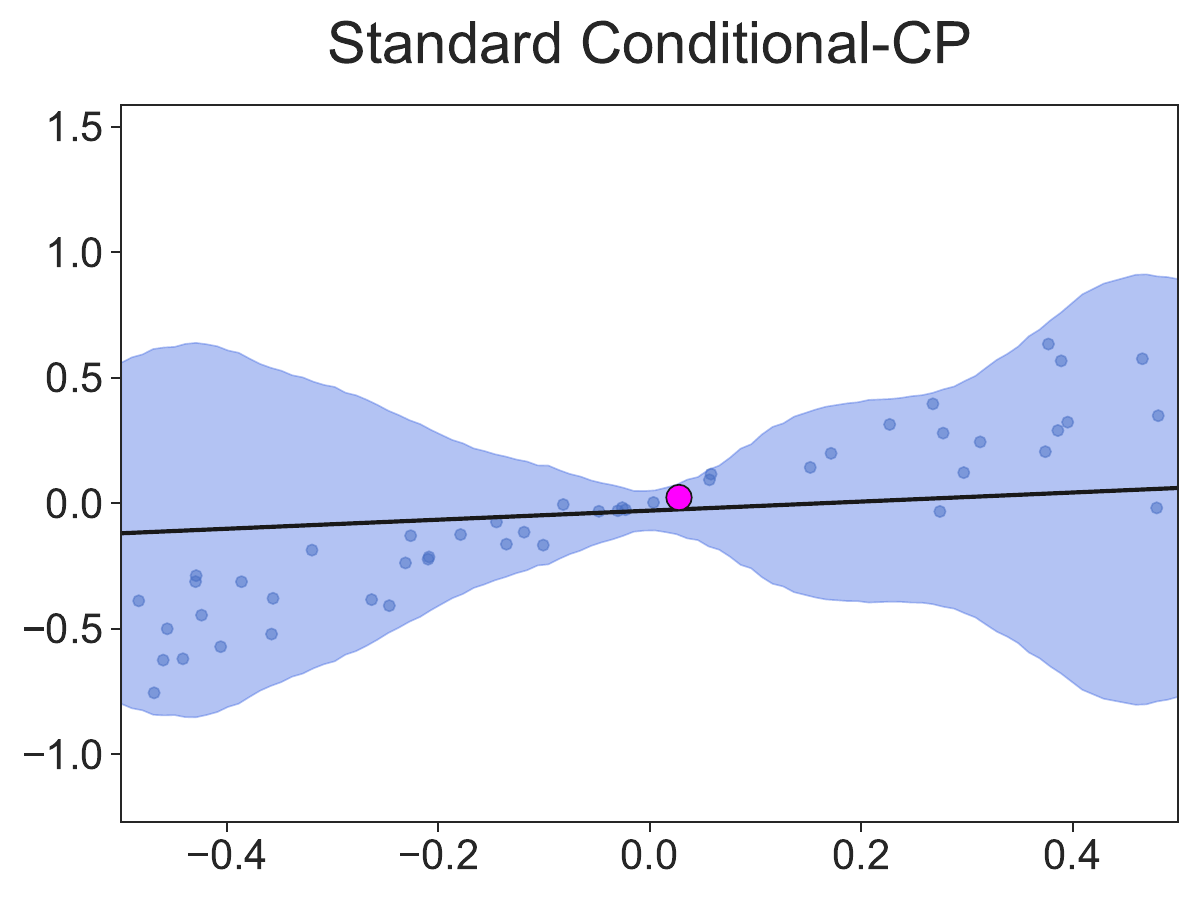}
    \end{minipage}
    \caption{Comparison of prediction sets constructed by \texttt{C-SymmPI} (our method), SymmPI \citep{dobriban2024symmpipredictiveinferencedata}, and Standard Conditional-CP \citep{gibbs2025conformal} on simulated two-layer hierarchical data with heterogeneous noise. Within each cluster, the noise variance is lower near the center and higher in the tails. \texttt{C-SymmPI} adapts to this heterogeneity, producing narrower intervals in low-variance regions and wider ones in high-variance regions. The competing methods adapt less to the varying noise levels. Details are provided in Section~\ref{sec:numerical_simulation}.}
    \label{fig:two_layer_experiments_intro}
\end{figure}

\paragraph{Empirical validation.} In numerical simulations across a range of settings, \csymmpi produces more adaptive and robust prediction intervals than SymmPI and other baselines (see Figure~\ref{fig:two_layer_experiments_intro} for an illustration in two-layer hierarchical data). In particular, \csymmpi automatically adapts to local heterogeneity in complex data structures. We further analyze the PPACT cluster randomized trial \citep{debar2022primary} and the Cora citation network \citep{mccallum2000automating}. In both applications, \csymmpi outperforms existing approaches \citep{wang2024conformal, lunde2023conformalpredictionnetworkassistedregression} that adapt to individual- or structure-level heterogeneity, while providing the first near-conditional coverage guarantees in these settings. Code is available at \url{https://github.com/ycshen0117/C-SymmPI}.

\subsection{Related Work}

\paragraph{Predictive inference beyond exchangeability.}
Conformal prediction \citep{vovk2005algorithmic, papadopoulos2002inductive} achieves marginal coverage under exchangeability; see \citet{vovk2005algorithmic, angelopoulos2021gentle, shafer2008tutorial, angelopoulos2024theoretical} for reviews. As exchangeability is often violated in practice, many works extend predictive inference to non-exchangeable data. Examples include subsampling, CDF pooling, and specialized conformal procedures for hierarchical data \citep{dunn2023distribution, lee2023distribution, wang2024conformal}, topology-aware methods for networks \citep{lunde2023conformalpredictionnetworkassistedregression, huang2023uncertainty}, and block permutation schemes for time series \citep{chernozhukov2018exact}. Under distribution shift, weighted conformal prediction maintains validity under covariate or outcome shift \citep{tibshirani2019conformal,park2021pac,qiu2023prediction, yang2024doubly,si2023pac,ai2024not}, and weighted quantiles with randomized adjustments improve robustness \citep{barber2023conformal}. \citet{dobriban2024symmpipredictiveinferencedata} unify many structured settings through group symmetries, providing marginal coverage for networks, two-layer hierarchical models, trees, and imaging data.

\paragraph{Limits of conditional coverage.}
\citet{vovk2005algorithmic} showed that exact conditional coverage is unattainable for continuous features without producing trivial prediction sets. Further hardness results hold even under relaxed inferential targets \citep{foygel2021limits} and for inference on conditional means or medians \citep{barber2020distributionfreeinferencepossiblebinary, lee2021distribution, medarametla2021distribution}.

\paragraph{Relaxed conditional guarantees.}
For test-conditional coverage, some methods improve empirical conditional performance while formally targeting marginal coverage \citep{romano2019conformalized, chernozhukov2021distributional, guan2023localized, xie2024boosted}, while others guarantee multi-accuracy or local coverage \citep{gibbs2025conformal, hore2024conformalpredictionlocalweights}. Training-conditional guarantees hold given the training set but remain marginal over the test point \citep{vovk2012conditional, bian2023training, liang2025algorithmic, gibbs2025characterizing}. Several works target both. \citet{zhang2024posteriorconformalprediction} achieves approximate training- and test-conditional coverage by modeling conformity scores with a mixture distribution. \citet{duchi2025observationssampleconditionalcoverageconformal} shows that quantile regression yields exact training-conditional and approximate test-conditional coverage for linear function classes. \citet{lee2025conditionalpredictiveinferencelkcoverage} provides training- and test-conditional $\ell_k$-coverage guarantees over various function classes.

\paragraph{Comparison with our work.}
None of these methods applies to general non-exchangeable structured data while providing near training- and test-conditional guarantees, or, more generally, guarantees conditional on arbitrary representations of the observed data. Achieving this requires both adapting to local heterogeneity given $Z_{\obs}$ and handling the non-exchangeable structure of the data. Our framework addresses both. Table~\ref{tab:related_work} compares our work with prior methods.

\begin{table}[t]
\centering
\label{tab:related_work}
\resizebox{\textwidth}{!}{
\begin{tabular}{@{}llccc@{}}
\toprule
\multirow{2}{*}{Paper} & \multirow{2}{*}{Data Structure} & \multirow{2}{*}{\makecell{Beyond \\ Exchangeability}} & \multicolumn{2}{c}{Near-Conditional Coverage} \\
\cmidrule(l){4-5}
& & & Test Level & Training Level \\
\midrule
\citet{tibshirani2019conformal} & \multirow{2}{*}{Data with distribution shift} & \multirow{2}{*}{\checkmark} & \multirow{2}{*}{} & \multirow{2}{*}{} \\
\citet{barber2023conformal} & & & & \\
\midrule
\citet{lunde2023conformalpredictionnetworkassistedregression} & Jointly exchangeable network & & & \\
\midrule
\citet{lee2023distribution} & \multirow{2}{*}{Two-layer hierarchical model} & \multirow{2}{*}{\checkmark} & \multirow{2}{*}{} & \multirow{2}{*}{} \\
\citet{wang2024conformal} & & & & \\
\midrule
\citet{dobriban2024symmpipredictiveinferencedata} & Data under group symmetries& \checkmark & & \\
\midrule
\citet{hore2024conformalpredictionlocalweights} & \multirow{2}{*}{i.i.d.\ data} & \multirow{2}{*}{} & \multirow{2}{*}{\checkmark} & \multirow{2}{*}{} \\
\citet{gibbs2025conformal} & & & & \\
\midrule
\citet{bian2023training} & \multirow{3}{*}{i.i.d.\ data} & \multirow{3}{*}{} & \multirow{3}{*}{} & \multirow{3}{*}{\checkmark} \\
\citet{liang2025algorithmic} & & & & \\
\citet{gibbs2025characterizing} & & & & \\
\midrule
\citet{duchi2025observationssampleconditionalcoverageconformal} &\multirow{2}{*}{i.i.d.\ data} & \multirow{2}{*}{} & \multirow{2}{*}{\checkmark} & \multirow{2}{*}{\checkmark} \\
\citet{lee2025conditionalpredictiveinferencelkcoverage} & & & & \\
\midrule
\textbf{Our work} & Data under group symmetries& \checkmark & \checkmark & \checkmark \\
\bottomrule
\end{tabular}
}
\caption{Comparison of related work.}
\end{table}

\subsection{Outline}
The rest of this paper is organized as follows. In Section~\ref{sec:preliminaries}, we introduce the necessary background on the SymmPI framework. Section~\ref{sec:main_results} presents our \csymmpi methodology, along with its theoretical guarantees and computational considerations. In Section~\ref{sec:examples}, we illustrate the versatility of our framework through several key examples. Section~\ref{sec:experiments} provides empirical validation of our method through simulations and real-data analyses. Finally, we conclude in Section~\ref{sec:discussion} with a summary and a discussion of future directions.
\section{Preliminaries}
\label{sec:preliminaries}

In this section, we review the group-theoretic framework for predictive inference \citep{dobriban2024symmpipredictiveinferencedata} that underlies our approach, using split conformal prediction \citep{papadopoulos2002inductive,vovk2005algorithmic,angelopoulos2021gentle} as a running example. Our methodological development relies on several concepts from group theory, including group actions, orbits, stabilizers, cosets, and the uniform (Haar) distribution on a group. A self-contained review of these concepts is provided in Appendix~\ref{sec:group_theory_review}. For a comprehensive treatment, see \citet{artin2018algebra, giri1996group, diaconis1988group, nachbin1976haar}.

\subsection{Group-Theoretic Predictive Inference}
\label{sec:symmpi_framework}
The framework describes a predictive inference procedure through three ingredients, namely an invariance assumption on the data, a transformation of the data into scores, and a threshold computed from the scores. It accommodates symmetry structures well beyond exchangeability (see Appendix~\ref{sec:additional_examples_symmpi} and \citet{dobriban2024symmpipredictiveinferencedata} for further examples). We describe each ingredient below and illustrate it with split conformal prediction.

\paragraph{Problem setup and distributional invariance.}

We consider a complete data vector $Z$ taking values in a measurable space $\cZ$. We assume that $Z$ is \emph{$\cG$-distributionally invariant}: $$Z \deq G\cdot Z, \qquad G \sim U.$$ Here $U$ is the uniform distribution on $\cG$.\footnote{Throughout the paper, we assume $\cG$ is compact. For finite $\cG$, $U$ is the discrete uniform distribution; in general, it is the unique probability measure on $\cG$ invariant under left and right multiplication. See Appendix~\ref{sec:group_theory_review}.}, and $g\cdot z$ denotes the (left) group action of $g\in\cG$ on $z\in\cZ$ (e.g., a permutation $g$ reordering the entries of $z$), which satisfies $g\cdot(h\cdot z)=(gh)\cdot z$. In words, the distribution of $Z$ is unchanged under the group action.\footnote{Here, ``$\deq$'' denotes equality in distribution.  We adopt this notation throughout this paper.}

In split conformal prediction, we observe labeled samples $(X_1, Y_1), \dots, (X_n, Y_n)$ and a test input $X_{n+1}$, and we write $Z = (Z_1,\cdots,Z_{n+1})^\top = ((X_1, Y_1), \dots, (X_{n+1}, Y_{n+1}))^\top$. When $\mathcal{G} = \rmS_{n+1}$ acts on $Z$ by the permutation action $g\cdot Z := (Z_{g^{-1}(1)}, \dots, Z_{g^{-1}(n+1)})^\top$, this assumption corresponds to the exchangeability of the data points $\{(X_i, Y_i)\}_{i=1}^{n+1}$.  More generally, distributional invariance covers many non-exchangeable structures, such as network data, two-layer hierarchical data, and rotationally invariant data \citep{dobriban2024symmpipredictiveinferencedata}.

Since only part of the data is observed in practice, we formalize what is observed through an \emph{observation function} $\obs(\cdot):\cZ \to \cO$, where $\cO$ is a measurable space. In the split conformal prediction example, the observation is
$
\obs(Z)=\bigl((X_1,Y_1),\ldots,(X_n,Y_n),X_{n+1}\bigr)^\top .$

\paragraph{Transformations.}

We map $Z$ via a \emph{transformation} $V: \cZ \to \tilde{\cZ}$, where $\tilde{\cZ}$ is another measurable space on which $\cG$ also acts. With a slight abuse of notation, we denote this action by ``$\cdot$'' as well. We assume that $V$ is \emph{$\cG$-deterministically equivariant}: $$V(g\cdot z) = g\cdot V(z),$$ for all $g \in \cG$ and $z \in \cZ$.\footnote{\citet{dobriban2024symmpipredictiveinferencedata} consider a weaker notion of \emph{$\cG$-distributional equivariance} for a mapping $V$, requiring that for $G \sim U$, $
V(G\cdot Z) \deq G\cdot V(Z).$
In practice, however, transformations are typically chosen to satisfy the stronger equivariance condition.} That is, $V$ commutes with the group action at every point of $\cZ$. Throughout, we write $\tilde z:=V(z)$ and $\tilde Z:=V(Z)$ for the transformed data, so that deterministic equivariance reads $V(g\cdot z)=g\cdot\tilde z$.

In split conformal prediction, given a pretrained model $\hat{\mu}: \cX \to \cY$,\footnote{Assuming that $\hat{\mu}$ is pretrained and fixed corresponds to the standard split conformal setup in \citet{papadopoulos2002inductive}, where one subset of the data is used to fit $\hat{\mu}$ and an independent calibration subset is used to compute the scores and construct the prediction set.} $V$ maps the dataset $Z$ to the vector of nonconformity scores $\tilde{Z}= (S_1,\dots,S_{n+1})^\top$, where $S_i = |Y_i - \hat{\mu}(X_i)|$. Under the permutation action on both $\cZ$ and $\tilde{\cZ}$, for any $g \in \rmS_{n+1}$ we have
$$
V(g\cdot Z)
=
\left(
|Y_{g^{-1}(1)} - \hat{\mu}(X_{g^{-1}(1)})|,
\dots,
|Y_{g^{-1}(n+1)} - \hat{\mu}(X_{g^{-1}(n+1)})|
\right)^\top
=
g\cdot\tilde{Z}.
$$
Therefore, $V$ is an $\rmS_{n+1}$-deterministically equivariant mapping.

For other data structures, such as networks or images, natural choices of $V$ include equivariant graph neural networks and translation- or rotation-equivariant convolutional neural networks \citep{lecun1989backpropagation, finzi2020generalizing}.

\paragraph{Threshold and Prediction Set.}

Next, we construct the prediction set by specifying a threshold. We consider a \emph{test function} $\psi: \tilde{\cZ} \to \RR$, where smaller values of $\psi(\tilde{Z})$ indicate more plausible realizations that should be included in the prediction set. For a target coverage level $1 - \alpha \in (0,1)$ and a realization $z$ of $Z$, the \emph{threshold} $t_{\tilde{z}}$ is the $(1-\alpha)$-quantile of $\psi(G\cdot\tilde{z})$ over $G \sim U$,
\[
t_{\tilde{z}} = Q_{1-\alpha}\bigl(\psi(G\cdot\tilde{z}),\, G \sim U\bigr).
\]
The resulting prediction set is defined as
\begin{equation}
\label{eq:standard_symmpi_set}
T^{\mathrm{SymmPI}}(z_{\obs})
=
\cbr{z \in \cZ : \psi(\tilde{z}) \le t_{\tilde{z}},\; \obs(z) = z_{\obs}}.
\end{equation}
\citet{dobriban2024symmpipredictiveinferencedata} establish a finite-sample marginal coverage guarantee for this prediction region, ensuring that
$$
\PP\!\left(Z \in T^{\mathrm{SymmPI}}(Z_{\obs})\right) \ge 1 - \alpha .
$$
To recover split conformal prediction, take the test function $\psi(\cdot) = e_{n+1}^\top \cdot$, so that $\psi(\tilde{Z}) = |Y_{n+1} - \hat{\mu}(X_{n+1})|$ is the absolute residual of the unobserved outcome $Y_{n+1}$.
When $\mathcal{G} = S_{n+1}$, the threshold becomes
$$t_{\tilde{Z}}
=
Q_{1-\alpha}\!\Big(
\big\{|Y_1-\hat\mu(X_1)|,\dots,|Y_n-\hat\mu(X_n)|,|Y_{n+1}-\hat\mu(X_{n+1})|\big\}
\Big).
$$
A candidate value $y$ is regarded as plausible if it belongs to the prediction set
\begin{align*}
\hat C(X_{n+1})
=
\bigg\{
y :
|y-\hat\mu(X_{n+1})|
\le
Q_{1-\alpha}\!\Big(
\{|Y_1-\hat\mu(X_1)|,\dots,|Y_n-\hat\mu(X_n)|,|y-\hat\mu(X_{n+1})|\}
\Big)
\bigg\},
\end{align*}
which coincides with the standard split conformal prediction set.\footnote{
This prediction set is equivalent to the one obtained by using the $\lceil (1-\alpha)(n+1) \rceil / n$ empirical quantile of the conformity scores $S_1,\dots,S_n$ as the threshold.
}

In many applications, however, the \emph{conditional coverage} $\PP\!\left(Z \in T^{\mathrm{SymmPI}}(Z_{\obs}) \mid Z_{\obs}\right) \ge 1 - \alpha$ is of greater interest than the marginal guarantee, as it helps mitigate under- or over-coverage across different realizations of the observed sample. In Section~\ref{sec:multi_accuracy}, we introduce (relaxed) multi-accuracy, a relaxation of exact conditional coverage.

\subsection{Multi-Accuracy}
\label{sec:multi_accuracy}

We now introduce the \emph{multi-accuracy} condition \citep{hebert2018multicalibration, kim2019multiaccuracy}.

\begin{proposition}
    \label{prop:conditional_coverage}
    For any prediction set $\hat{C}:\cO\rightrightarrows \cZ$\footnote{The notation ``$\rightrightarrows$'' indicates that $\hat{C}$ maps each element in $\mathcal{O}$ to a subset of $\mathcal{Z}$.}, the following two conditions are equivalent:
    \begin{itemize}
        \item[(i)] $\PP\left(Z \in \hat{C}(Z_\obs) \mid Z_\obs = z_{\obs}\right) = 1 - \alpha$, for $z_{\obs} \in \cO$ almost surely.
        \item[(ii)] $\EE\left[f(Z_{\obs})\left(\1\{Z \in \hat{C}(Z_{\obs})\} - (1 - \alpha)\right)\right] = 0$,
        for all bounded non-negative functions $f: \cO \to \RR$.\footnote{All the functions we consider throughout this paper are assumed to be measurable.}
    \end{itemize}
\end{proposition}

\begin{remark}
    In (i), the probability is taken over the randomness of the unobserved components of $Z$, conditional on $Z_{\obs} = z_{\obs}$. In (ii), the expectation is taken over the randomness of the full data $Z$.
\end{remark}

Proposition~\ref{prop:conditional_coverage}, proved in Section~\ref{sec:proof_conditional_coverage}, re-expresses conditional coverage through \emph{function-weighted averages} rather than pointwise conditional probabilities. Since distribution-free methods generally cannot achieve exact conditional coverage in finite samples \citep{vovk2005algorithmic, foygel2021limits}, we instead consider the relaxed condition
\begin{equation}
    \label{eq:relaxed_conditional_coverage}
    \left|\EE\left[f(Z_{\obs})\left(\1\{Z \in \hat{C}(Z_{\obs})\} - (1 - \alpha)\right)\right]\right| \leq \varepsilon, \text{ for all } f\in\cF,
\end{equation}
where $\cF \subseteq \cbr{\cO \to \RR}$ is a pre-specified function class, and $\varepsilon > 0$ is a small tolerance parameter. This relaxes multi-accuracy in two ways. The equality is replaced by an inequality with tolerance $\varepsilon$, and the condition is required only for $f\in\cF$ rather than for all bounded non-negative functions.

\begin{remark}
We can similarly generalize our framework to the setting of \cite{lee2025conditionalpredictiveinferencelkcoverage}, which considers $\ell_k$-coverage over $f \in \cF$. As this extension is not our primary focus, we omit the details.
\end{remark}

\section{Methodology: C-SymmPI}
\label{sec:main_results}
In this section, we introduce our method, \symmpi, which provides near-conditional coverage for general data structures with group symmetries. Section~\ref{sec:prediction_region} constructs the method by replacing the single orbit quantile of SymmPI with a threshold learned along the orbit, and Section~\ref{sec:flexible_conditioning} shows how to condition on a chosen representation of the observed data. We then establish coverage guarantees in Section~\ref{sec:theoretical_properties} and convergence rates for linear and RKHS function classes in Section~\ref{sec:choice_of_function_class}.

\subsection{Constructing Prediction Set}
\label{sec:prediction_region}
SymmPI (Section~\ref{sec:symmpi_framework}) attains marginal coverage by comparing the test statistic with a \emph{single} quantile of $\psi$ over the orbit of the transformed data. \csymmpi instead lets this quantile vary along the orbit, depending on the observed part of each orbit element. The resulting threshold depends on the observed data and can therefore adapt to local heterogeneity.

\paragraph{SymmPI as orbit quantile.}
Let $\ell_\alpha(t, s) := \alpha[t - s]_+ + (1-\alpha)[s - t]_+$ denote the pinball loss, where $[x]_+ = \max\{x,0\}$. For any real random variable $S$, a $(1-\alpha)$-quantile of $S$ minimizes $t\mapsto\EE[\ell_\alpha(t,S)]$ over $t\in\RR$ \citep{koenker1978regression}. Taking $S=\psi(G\cdot\tilde z)$ with $G\sim U$, the SymmPI threshold of Section~\ref{sec:symmpi_framework} is therefore a solution of
\begin{equation}
    \label{eq:symmpi_pinball}
    t_{\tilde z}\in\argmin_{t\in\RR}\ \EE_{G\sim U}\Bigl[\ell_\alpha\bigl(t,\ \psi(G\cdot\tilde z)\bigr)\Bigr].
\end{equation}
In other words, SymmPI fits a \emph{constant} to the orbit values $\{\psi(g\cdot\tilde z): g\in\cG\}$, so every element of the orbit is compared with the same cutoff. Because this cutoff cannot distinguish orbit elements with different observed parts, the resulting guarantee is only marginal.

\paragraph{Orbit quantile regression.}
Each orbit element $g\cdot z$ carries an observable part $\obs(g\cdot z)\in\cO$. We replace the constant $t$ in \eqref{eq:symmpi_pinball} by a function $t:\cO\to\RR$ evaluated at the observable part:
\begin{equation}
    \label{eq:orbit_regression}
    \min_{t(\cdot)\in\{\cO\to\RR\}}\ \EE_{G\sim U}\Bigl[\ell_\alpha\bigl(t(\obs(G\cdot z)),\ \psi(G\cdot\tilde z)\bigr)\Bigr].
\end{equation}
This is a quantile regression in which the orbit plays the role of the data set. The inputs are $\obs(G\cdot z)$, the responses are $\psi(G\cdot\tilde z)$, and the sampling distribution is induced by the uniform distribution $U$ on $\cG$.

\paragraph{Regularization.}
Over a rich function class, the regression \eqref{eq:orbit_regression} may interpolate the orbit values, leading to trivial prediction sets. To address this issue, we optimize over a \emph{linear space} $\cF\subseteq\{\cO\to\RR\}$ of candidate threshold functions, such as $\cF=\{\cO\to\RR\}$ or the linear and RKHS classes in Section~\ref{sec:choice_of_function_class}, and add a regularization term $\cR:\cF\to \RR$ to control overfitting. We impose the following mild conditions on the regularizer:

\begin{assumption}
    \label{assumption:regularization}
    The regularizer $\cR:\cF\to\RR$ satisfies the following:
    \begin{itemize}
        \item[(i)] $\cR$ is convex on $\cF$.
        \item[(ii)] For all $t, f\in\cF$, the directional derivative $D_{f}\cR(t):=\displaystyle\lim_{\varepsilon\to 0}\frac{\cR(t+\varepsilon f) - \cR(t)}{\varepsilon}$ exists.
    \end{itemize}
\end{assumption}
Common examples of $\cR$ include the squared $L^2$ norm $\cR(t)=\lambda\|t\|_2^2$ on $\cF\subseteq L^2$, the squared RKHS norm $\cR(t)=\lambda\|t\|_K^2$ on an RKHS $\cF=\cF_K$, and the trivial regularizer $\cR\equiv0$ for simple function classes. This leads to the adaptive threshold function of \csymmpi: for each $z\in\cZ$,
\begin{equation}
    \label{eq:hat_t}
    \hat{t}_{z}(\cdot) = \argmin_{t(\cdot)\in\cF}\EE_{G\sim U}\left[\ell_\alpha\bigl(t(\obs(G\cdot z)), \psi(G\cdot\tilde z)\bigr)\right] + \cR(t).
\end{equation}
The theory in Section~\ref{sec:theoretical_properties} builds on \eqref{eq:hat_t}. As we discuss in Section~\ref{sec:flexible_conditioning}, the choice of $\cF$ also determines the information on which coverage is conditioned.

\paragraph{C-SymmPI prediction set.}

Replacing the quantile threshold in \eqref{eq:standard_symmpi_set} with the learned threshold \eqref{eq:hat_t}, we define the \csymmpi prediction set as
\begin{equation}
    \label{eq:symmpi}
    T^{\symmpi}(z_{\obs}) = \cbr{z \in \cZ : \psi(\tilde z) \leq \hat{t}_{z}(z_\obs), \obs(z) = z_{\obs}}.
\end{equation}
The complete procedure is summarized in Algorithm~\ref{alg:symmpi}. For each completion $z$ of the observed data $z_{\obs}$,\footnote{A completion $z$ of the observed data $z_{\obs}$ is any element $z\in\cZ$ satisfying $\obs(z)=z_{\obs}$.} the algorithm solves \eqref{eq:hat_t} to obtain the threshold function $\hat{t}_z(\cdot)$ and includes $z$ in the prediction set whenever $\psi(\tilde z) \leq \hat{t}_{z}(z_{\obs})$.

\begin{algorithm}[ht]
    \caption{\symmpi: Conditional predictive inference for data with group symmetries}
    \label{alg:symmpi}
    \begin{algorithmic}[1]
    \Require Data $z$ satisfying distributional invariance under group $\mathcal{G}$. Observation function $\obs : \mathcal{Z} \to \mathcal{O}$. Observed data $z_{\obs}$. Deterministically equivariant map $V : \mathcal{Z} \to \tilde{\mathcal{Z}}$. Test function $\psi : \tilde{\mathcal{Z}} \to \mathbb{R}$. Miscoverage level $\alpha \in (0, 1)$. Threshold class $\cF$ and penalty term $\cR$ satisfying Assumption~\ref{assumption:regularization}.
    \State Initialize $T^{\symmpi}(z_{\obs}) = \emptyset$.
    \For{$z \in \mathcal{Z}$ satisfying $\obs(z) = z_{\obs}$}
        \State Solve the optimization problem \eqref{eq:hat_t} to obtain $\hat{t}_{z}(\cdot)$. \Comment{Adaptive threshold function.}
        \If {$\psi(\tilde z) \leq \hat{t}_{z}(z_\obs)$}
            \State Set $T^{\symmpi}(z_{\obs}) \gets T^{\symmpi}(z_{\obs}) \cup \{z\}$.
        \EndIf
    \EndFor
    \Ensure Prediction region $T^{\symmpi}(z_{\obs})$.
    \end{algorithmic}
\end{algorithm}

\begin{remark}
    In practice, we take $\cF$ in \eqref{eq:hat_t} to be a rich parameterized function class, which reduces the infinite-dimensional optimization over functions to a tractable problem (see Section~\ref{sec:choice_of_function_class}).
\end{remark}

\subsection{Flexible Conditioning}
\label{sec:flexible_conditioning}
In many applications, one is interested in coverage conditional on \emph{selected features or summaries of the observed data} rather than on all of $z_\obs$, for instance the covariates of the test unit or local statistics in its neighborhood. \csymmpi accommodates such goals through the choice of the threshold class. Since the guarantees in Section~\ref{sec:theoretical_properties} hold for weighting functions $f\in\cF$, the coverage is conditioned on exactly the information that the threshold is allowed to depend on.

To condition on a representation of the observed data, we take a map $\eta:\cO\to\cW$ into a representation space $\cW$ and let the threshold depend on $z_\obs$ only through $\eta(z_\obs)$. That is, we use the class $\cF=\{t\circ\eta: t\in\cF_\cW\}$ in \eqref{eq:hat_t}, where $\cF_\cW\subseteq\{\cW\to\RR\}$ is a space of functions on $\cW$. The penalty $\cR:\cF_\cW\to\RR$ satisfies Assumption~\ref{assumption:regularization} with $\cF$ replaced by $\cF_\cW$. Writing $\proj(z):=\eta(\obs(z))$ and $z_\proj:=\proj(z)$, the threshold \eqref{eq:hat_t} then takes the explicit form
\begin{equation}
    \label{eq:hat_t_eta}
    \hat{t}_z^{\,\proj}(\cdot)
    = \argmin_{t(\cdot)\in\cF_\cW}
    \EE_{G\sim U}\left[
        \ell_\alpha\bigl(t(\proj(G\cdot z)), \psi(G\cdot\tilde z)\bigr)
    \right] + \cR(t),
\end{equation}
and the prediction set \eqref{eq:symmpi} becomes
\begin{equation}
    \label{eq:symmpi_eta}
    T^{\symmpi}(z_{\obs}) = \cbr{z \in \cZ : \psi(\tilde z) \leq \hat{t}_z^{\,\proj}(z_\proj),\ \obs(z) = z_{\obs}}.
\end{equation}

The choice of $\eta$ determines the type of conditional guarantee provided. Taking $\eta$ to be the identity yields coverage conditional on the full observed data, including both the training data and the test input. In contrast, setting $\eta(Z_\obs)=X_{n+1}$ in the exchangeable conformal prediction setting yields coverage conditional only on the test covariate. The choice also affects computation, since a lower-dimensional representation simplifies threshold estimation and lets many coinciding orbit terms in \eqref{eq:hat_t_eta} be merged (see Section~\ref{sec:hierarchical} and Appendix~\ref{sec:supervised_learning_iid}). All examples in Section~\ref{sec:examples} and all experiments in Section~\ref{sec:experiments} use this form, with $\eta$ extracting covariates of the target unit.

\begin{remark}
    \label{rem:not_new_method}
    Conditioning on $\eta(z_\obs)$ does not lead to a new procedure. The threshold \eqref{eq:hat_t_eta} and the prediction set \eqref{eq:symmpi_eta} are exactly \eqref{eq:hat_t} and \eqref{eq:symmpi} with the threshold class $\{t\circ\eta: t\in\cF_\cW\}$. Consequently, all results in Sections~\ref{sec:theoretical_properties} and~\ref{sec:choice_of_function_class} apply directly, with $\obs(\cdot)$ replaced by $\proj(\cdot)$. Corollary~\ref{cor:eta_coverage} states the resulting coverage guarantee explicitly.
\end{remark}

\subsection{Theoretical Results}
\label{sec:theoretical_properties}
In this section, we first present two main theorems that establish near-conditional coverage guarantees under both the distributionally invariant setting and the distribution shift setting, and then discuss large or infinite groups. Section~\ref{sec:choice_of_function_class} specializes these results to linear and RKHS function classes.

\subsubsection{Near-Conditional Coverage under Distributional Invariance}
\label{sec:coverage_invariance}

We first establish near-conditional coverage under distributional invariance when the group is finite, deferring the proof to Section~\ref{sec:proof_coverage}. Compact groups that may be infinite are discussed in Section~\ref{sec:large_groups}.
\begin{theorem}
    \label{thm:coverage}
    Let $\mathcal{G}$ be a finite group equipped with the uniform distribution $U$. Suppose the complete data $Z \in \mathcal{Z}$ satisfies the distributional invariance property $Z \stackrel{d}{=} G\cdot Z$ for $G \sim U$, where ``$\cdot$'' denotes an action of $\mathcal{G}$ on $\mathcal{Z}$. Let the observed data be $Z_{\obs} = \obs(Z)$, where $\obs: \mathcal{Z} \to \mathcal{O}$ is an observation function for some space $\mathcal{O}$. Consider a $\mathcal{G}$-deterministically equivariant function $V: \mathcal{Z} \to \tilde{\mathcal{Z}}$ for some space $\tilde{\mathcal{Z}}$, and a test function $\psi: \tilde{\mathcal{Z}} \to \mathbb{R}$. Denote $\alpha \in (0,1)$ as the miscoverage level, $\cF\subseteq\{\cO\to\RR\}$ as the linear space of threshold functions in \eqref{eq:hat_t}, and $\cR$ as the penalty term on $\cF$ satisfying Assumption~\ref{assumption:regularization}. Then, for any $f\in\cF$ with $\EE|f(Z_\obs)|<\infty$ and $\EE|D_f\cR(\hat t_Z)|<\infty$, the \csymmpi prediction set defined in (\ref{eq:symmpi}) satisfies
    $$
        \abr{\EE_{Z}\Bigl[f(Z_{\obs})\Bigl(\1\{Z\in T^{\symmpi}(Z_{\obs})\} - (1 - \alpha)\Bigr)\Bigr]} \leq \varepsilon_\pen + \varepsilon_\inte,
    $$
    where the penalty error and interpolation error are defined as
    $$
    \varepsilon_\pen = \abr{\EE_{Z}\left[D_{f}\cR(\hat{t}_{Z})\right]}, \quad \varepsilon_\inte = \mathbb{E}_{Z}\left[|f(Z_\obs)| \1\{\psi(\tilde Z) = \hat{t}_{Z}(Z_\obs)\}\right].
    $$
\end{theorem}

    Theorem~\ref{thm:coverage} follows the relaxed multi-accuracy perspective introduced in Section~\ref{sec:multi_accuracy}. Compared with the uniform guarantee in \eqref{eq:relaxed_conditional_coverage}, the bound in Theorem~\ref{thm:coverage} holds for a fixed weighting function $f$. By restricting $f$ to suitable subsets of $\cF$, however, one can bound the penalty error $\varepsilon_\pen$ and the interpolation error $\varepsilon_\inte$ uniformly over the subset, so that the result holds simultaneously for all such $f$ (see Section~\ref{sec:choice_of_function_class} for examples).

\begin{remark}[Penalty error]
    The term $\varepsilon_\pen$ captures the bias introduced by the regularization $\cR(t)$ in the quantile regression. When the function class for $t(\cdot)$ is rich, regularization is typically required to prevent the learned threshold function $\hat{t}_Z$ from interpolating the orbit values $\psi(g_i\cdot\tilde z)$. In such cases, one may use a regularization parameter $\lambda>0$ (e.g., $\cR(t)=\lambda\|t\|_2^2$), and choosing $\lambda$ sufficiently small keeps $\varepsilon_\pen$ small. For simple function classes (e.g., the low-dimensional linear class in Section~\ref{sec:linear_function_class}), we may set the regularization to zero, which gives $\varepsilon_\pen = 0$.
\end{remark}

\begin{remark}[Interpolation error]
The term $\varepsilon_\inte$ accounts for the error arising when the test statistic $\psi(\tilde Z)$ equals the learned threshold $\hat{t}_Z(Z_\obs)$. It arises because the subgradient of the pinball loss is not unique at the origin, which corresponds to the event $\psi(\tilde Z)=\hat{t}_Z(Z_\obs)$. We analyze the magnitude of $\varepsilon_\inte$ for specific function classes in Section~\ref{sec:choice_of_function_class}.
\end{remark}

\paragraph{Conditioning on a representation.} Since the threshold \eqref{eq:hat_t_eta} is \eqref{eq:hat_t} with the class $\{t\circ\eta: t\in\cF_\cW\}$, Theorem~\ref{thm:coverage} applies directly and yields guarantees for weighting functions of $Z_\proj:=\eta(Z_\obs)$.
\begin{corollary}
    \label{cor:eta_coverage}
    Under the conditions of Theorem~\ref{thm:coverage}, for any $f\in\cF_\cW$ with $\EE|f(Z_\proj)|<\infty$ and $\EE|D_f\cR(\hat t_Z^{\,\proj})|<\infty$, the prediction set \eqref{eq:symmpi_eta} satisfies
    $$
        \abr{\EE_{Z}\Bigl[f(Z_{\proj})\Bigl(\1\{Z\in T^{\symmpi}(Z_{\obs})\} - (1 - \alpha)\Bigr)\Bigr]} \leq \varepsilon_\pen + \varepsilon_\inte,
    $$
    where
    $$
    \varepsilon_\pen = \abr{\EE_{Z}\left[D_{f}\cR(\hat{t}_Z^{\,\proj})\right]}, \quad \varepsilon_\inte = \mathbb{E}_{Z}\left[|f(Z_\proj)| \1\{\psi(\tilde Z) = \hat{t}_Z^{\,\proj}(Z_\proj)\}\right].
    $$
\end{corollary}
Taking $\eta$ to be the identity recovers Theorem~\ref{thm:coverage}, which conditions on the full observed data, including the training data. Taking $\eta(Z_\obs)=X_{n+1}$ in the exchangeable setting gives guarantees relative to functions of the test covariate (Appendix~\ref{sec:supervised_learning_iid}). We next discuss the near-conditional coverage guarantee under distribution shift.

\subsubsection{Near-Conditional Coverage under Distributional Shift}
\label{sec:coverage_shift}

The guarantee in Theorem~\ref{thm:coverage} relies on the $\mathcal{G}$-distributional invariance property $Z \deq G\cdot Z$. In practice, this exact symmetry assumption may be violated. Let $\PP$ denote the distribution of $Z$, which need not be $\cG$-invariant, and let $\QQ$ denote the distribution of $Z_G := G\cdot Z$, where $G\sim U$ is independent of $Z$. Further denote $Z_G^\obs := \obs(Z_G)$ and $\tilde{Z}_G := V(Z_G)$. For a bounded function $f:\cO\to\RR$, define the statistic
$$
\xi_f(z):=\Bigl(f(\obs(z)),\ \1\bigl\{\psi(\tilde z)>\hat t_z(\obs(z))\bigr\}\Bigr)\in\RR\times\{0,1\},\qquad z\in\cZ,
$$
and let $\PP\circ\xi_f^{-1}$ and $\QQ\circ\xi_f^{-1}$ denote the distributions of $\xi_f(Z)$ and $\xi_f(Z_G)$, respectively. The following theorem, proved in Section~\ref{sec:proof_distribution_shift}, gives the near-conditional coverage guarantee of \csymmpi without distributional invariance.

\begin{theorem}
    \label{thm:distribution_shift}
    Let $\PP$, $\QQ$, $Z_G$, $Z_G^\obs$ and $\tilde Z_G$ be defined as above. Under the conditions of Theorem~\ref{thm:coverage} except the distributional invariance property, for any bounded function $f\in\cF$ with $\EE_{G,Z}|D_f\cR(\hat t_{Z_G})|<\infty$, the following bound holds:
    $$
        \abr{\EE_{Z}\Bigl[f(Z_{\obs})\Bigl(\1\{Z\in T^{\symmpi}(Z_{\obs})\} - (1 - \alpha)\Bigr)\Bigr]} \leq \varepsilon_{\pen} + \varepsilon_{\inte} + C_{f}\|f\|_{\infty}\,\TV\bigl(\PP\circ\xi_f^{-1}, \QQ\circ\xi_f^{-1}\bigr),
    $$
    where $\TV(\mu,\nu):=\sup_{A}|\mu(A)-\nu(A)|$ is the total variation distance, the constant $C_f=1$ if $f$ does not change sign and $C_f=2\max(\alpha,1-\alpha)$ otherwise, and
    $$
    \varepsilon_{\pen} = \abr{\EE_{G,Z}\left[D_{f}\cR(\hat{t}_{Z_G})\right]}, \qquad
    \varepsilon_{\inte} = \EE_{G,Z}\left[|f(Z_G^\obs)|\1\{\psi(\tilde{Z}_G) = \hat{t}_{Z_G}(Z_G^\obs)\}\right].
    $$
\end{theorem}

Compared with Theorem~\ref{thm:coverage}, the bound has an additional term, which measures the departure from invariance through the total variation distance between the distributions of $\xi_f(Z)$ and $\xi_f(Z_G)$. If the $\cG$-distributional invariance holds, then $\QQ=\PP$, the last term vanishes, and Theorem~\ref{thm:distribution_shift} reduces to Theorem~\ref{thm:coverage}. The theorem also extends to the threshold in \eqref{eq:hat_t_eta}, yielding a near-conditional coverage guarantee with respect to $\eta(Z_{\obs})$. The argument follows that of Corollary~\ref{cor:eta_coverage}, so we omit the details.

\subsubsection{Large or Infinite Groups}
\label{sec:large_groups}

Theorems~\ref{thm:coverage} and~\ref{thm:distribution_shift} assume that $\cG$ is finite. When $\cG$ is large or infinite (e.g., the rotation group $\mathrm{SO}(p)$), the expectation over $G\sim U$ in \eqref{eq:hat_t} may be costly or infeasible to evaluate. It can be replaced by an average over finitely many sampled group elements together with the identity. The resulting procedure, presented in Appendix~\ref{sec:sampled_adaptive_symmpi}, retains a near-conditional guarantee of the same form, averaged over the sampling randomness.

\subsection{Specifying Function Classes}
\label{sec:choice_of_function_class}

The coverage error in Theorem~\ref{thm:coverage} depends on the threshold class $\cF$. In this subsection, we quantify it for two commonly used classes, namely low-dimensional linear classes and reproducing kernel Hilbert spaces (RKHSs). We first rewrite the threshold in terms of the distinct elements of the orbit (Section~\ref{sec:distinct_orbit_elements}), whose number plays the role of the sample size, and then derive convergence rates for the linear class (Section~\ref{sec:linear_function_class}) and the RKHS (Section~\ref{sec:rkhs}).

Throughout this subsection, we assume that $\cG$ is finite. All results also apply to the threshold \eqref{eq:hat_t_eta} under flexible conditioning, with $\obs(\cdot)$ replaced by $\proj(\cdot)=\eta(\obs(\cdot))$ throughout.

\subsubsection{Distinct Orbit Elements}
\label{sec:distinct_orbit_elements}

The objective in \eqref{eq:hat_t} includes one loss term for each of the $|\cG|$ group elements, although many of these terms coincide. For instance, with $n+1$ exchangeable points, $\eta(z_\obs)=x_{n+1}$, and $\psi(z)=s(y_{n+1},x_{n+1})$, i.e., the covariate and the nonconformity score of the $(n+1)$-th point (Section~\ref{sec:flexible_conditioning}), a permutation affects the loss only through the point it moves to the last position, so the $(n+1)!$ terms take only $n+1$ distinct values. We now remove this redundancy from a group-theoretic perspective.

Each group element $g$ contributes the loss term $\ell_\alpha(t(\obs(g\cdot z)),\psi(g\cdot\tilde z))$ to \eqref{eq:hat_t}, which depends on $g$ only through the input--score pair $F(g\cdot z)$, where $F(z):=(\obs(z),\psi(\tilde z))$ (under flexible conditioning, $\obs(\cdot)$ is replaced by $\proj(\cdot)=\eta(\obs(\cdot))$ throughout). Let $\cH:=\{h\in\cG: F(h\cdot z)=F(z),\ \forall z\in\cZ\}$ be the stabilizer of $F$, i.e., the group elements that leave this pair unchanged; in the example above, these are the permutations that fix the test point and only reorder the training points. Then $g$ and $hg$ give the same loss term for every $h\in\cH$, so the loss is constant on each right coset $\cH g$. By the orbit--stabilizer theorem (see Section~\ref{sec:group_theory_review}), there are $[\cG:\cH]:=|\cG|/|\cH|$ such cosets, called the index of $\cH$ in $\cG$, each containing $|\cH|$ elements.

Choose coset representatives $g_1=e,g_2,\dots,g_{[\cG:\cH]}$, and write
$$
\bigl(o_i(z),s_i(z)\bigr):=F(g_i\cdot z),\qquad i=1,\dots,[\cG:\cH],
$$
for the input and the score of the $i$-th distinct orbit element, associated with the representative $g_i$. In particular, $(o_1(z),s_1(z))=(z_\obs,\psi(\tilde z))$ is the original input--score pair of the test point.
The threshold \eqref{eq:hat_t} then becomes an unweighted quantile regression on these $[\cG:\cH]$ pairs,
\begin{equation}
    \label{eq:hat_t_distinct}
    \hat t_z(\cdot)=\argmin_{t(\cdot)\in\cF}\ \frac{1}{[\cG:\cH]}\sum_{i=1}^{[\cG:\cH]}\ell_\alpha\bigl(t(o_i(z)),s_i(z)\bigr)+\cR(t).
\end{equation}
Under flexible conditioning, $\obs(\cdot)$ is replaced by $\proj(\cdot)=\eta(\obs(\cdot))$ in the above procedure throughout with $F(z):=(\eta(\obs(z)),\psi(\tilde{z})).$
Thus $[\cG:\cH]$ mimics the role of the sample size: it equals $n+1$ for $n+1$ exchangeable points with properly chosen $\eta(\cdot)$ and $\psi(\cdot)$ (the detailed set up is presented in Appendix~\ref{sec:supervised_learning_iid}) and $Kn$ for $K$ clusters of equal size $n$ (Section~\ref{sec:hierarchical}).

\subsubsection{Linear Function Class}
\label{sec:linear_function_class}

Let $\cF_L$ be a $d$-dimensional linear function class induced by a feature map $\varphi:\cO \to \RR^d$:
$$
\cF_{L} = \left\{f:\cO \to \RR \mid f(\cdot) = \langle\theta, \varphi(\cdot)\rangle,\ \theta\in\RR^d \right\}.
$$
We impose the following boundedness condition on the feature map:
\begin{assumption}
    \label{assumption:bounded_feature_map}
    There exists a constant $b_{\varphi} > 0$ such that $\|\varphi(\cdot)\|_2 \leq b_{\varphi}$.
\end{assumption}

We take $\cF=\mathcal{F}_L$ in \eqref{eq:hat_t_distinct}, and set the regularization term $\mathcal{R}(t) \equiv 0$ as the problem is low-dimensional. Under the parametrization $\hat t^{\,L}_z(\cdot)=\langle \hat\theta_z,\varphi(\cdot)\rangle$, \eqref{eq:hat_t_distinct} reduces to an optimization over $\theta\in\RR^d$: for each $z\in\cZ$,
\begin{align}\label{linear_loss}
    \hat{\theta}_z = \argmin_{\theta\in\RR^d} \frac{1}{[\cG:\cH]} \sum_{i=1}^{[\cG:\cH]} \ell_\alpha\bigl(\langle\theta, \varphi(o_i(z))\rangle, s_i(z)\bigr).
\end{align}
This is a linear program whose objective is bounded below by zero, so the minimum is attained. If the minimizer is not unique, we take $\hat\theta_z$ to be the minimizer with the smallest Euclidean norm, which is unique and depends on $z$ only through the objective in \eqref{linear_loss}. Here $\cH:=\{h\in\cG : F(h\cdot z)=F(z)
\text{ for all } z\in\cZ\}$ is the stabilizer of the feature--score pair
$F(z):=(\varphi(\obs(z)),\psi(\tilde z))$, and
$[\cG:\cH]$ is the number of its right cosets in $\cG$. The resulting \csymmpi prediction set is:
\begin{equation}
    \label{eq:symmpi_linear}
    T^{\symmpi}_L(z_{\obs}) = \cbr{z\in\cZ:\psi(\tilde z)\leq \langle\hat{\theta}_z, \varphi(z_\obs)\rangle,\obs(z)=z_{\obs}}.
\end{equation}

To control the interpolation error, we require the score of each distinct orbit element to be continuously distributed given the features and the other scores. We stack the features of the distinct orbit elements into a matrix and their scores into a vector:
$$
\Phi := \begin{pmatrix}\varphi(o_1(Z))^\top\\ \vdots\\ \varphi(o_{[\cG:\cH]}(Z))^\top\end{pmatrix}\in\RR^{[\cG:\cH]\times d},
\qquad
\Psi := \begin{pmatrix}s_1(Z)\\ \vdots\\ s_{[\cG:\cH]}(Z)\end{pmatrix}\in\RR^{[\cG:\cH]}.
$$
The following theorem, proved in Section~\ref{sec:proofs_linear_class}, establishes the convergence rate of the near-conditional coverage error for the low-dimensional linear class.

\begin{theorem}
\label{thm:linear_class}
    Let $\cF_{L}\subseteq\{\cO\to\RR\}$ denote the class of $d$-dimensional linear functions with $d\leq [\cG:\cH]$ that satisfies Assumption~\ref{assumption:bounded_feature_map}, and suppose that, conditionally on $\Phi$ and on the other entries of $\Psi$, each entry of $\Psi$ has a continuous distribution.
 Then for all $f(\cdot) = \langle\theta, \varphi(\cdot)\rangle \in \cF_L$ with $\|\theta\|_2\leq b_\theta$, the \texttt{C-SymmPI} prediction set defined in \eqref{eq:symmpi_linear} satisfies
    $$
        \left|\EE_{Z}\left[f(Z_{\obs})\left(\1\{Z\in T^{\symmpi}_L(Z_{\obs})\}-(1-\alpha)\right)\right]\right| \leq \frac{b_{\varphi}b_\theta d}{[\cG:\cH]}.
    $$
\end{theorem}

\begin{remark}
    \label{rem:linear_continuity}
    The continuity condition holds whenever each response carries its own continuous noise, e.g., $Y_j^{(i)} = m_i(X_j^{(i)}) + \sigma_i(X_j^{(i)})\,\varepsilon_j^{(i)}$ in the hierarchical model of Section~\ref{sec:hierarchical} with clusters of equal size, where the cluster-specific functions $m_i,\sigma_i$ may be random, $\sigma_i(X_j^{(i)})\neq0$ almost surely, and the noise $\varepsilon_j^{(i)}$ is continuous and independent of everything else. 
\end{remark}
\begin{remark}
    The bound reads as dimension over the number $[\cG:\cH]$ of distinct orbit elements. It gives $\cO(d/n)$ for $n+1$ exchangeable points (Appendix~\ref{sec:supervised_learning_iid}) and $\cO(d/(Kn))$ for $K$ clusters of size $n$; Corollary~\ref{cor:hierarchical_rate} extends the latter to clusters of different sizes (Section~\ref{sec:hierarchical}). Richer symmetry yields more distinct orbit elements and hence a tighter bound.
\end{remark}

\subsubsection{RKHS Function Class}
\label{sec:rkhs}
Let $K:\cO\times\cO\to\RR$ be a positive semidefinite kernel that induces an RKHS $\cF_{K}$ with inner product $\langle \cdot, \cdot \rangle_K$ and norm $\|\cdot\|_K$. Common choices of $K$ include linear, polynomial, and Gaussian kernels. We assume that the kernel is bounded.
\begin{assumption}
    \label{assumption:bounded_kernel}
    There exists a constant $\kappa > 0$ such that $K(o, o) \leq \kappa^2$ for all $o\in\cO$.
\end{assumption}

We take $\cF=\cF_K$ in \eqref{eq:hat_t} with the penalty $\cR(t)=\lambda\|t\|_K^2$, $\lambda>0$. Since a rich RKHS can interpolate the orbit scores, the penalty is needed to prevent overfitting, at the price of a bias that vanishes as $\lambda\to0$. By \eqref{eq:hat_t_distinct}, the adaptive threshold is
\begin{equation}
    \label{eq:hat_t_RKHS}
    \hat{t}_z^{\,K}(\cdot) = \argmin_{t(\cdot) \in \cF_{K}} \frac{1}{[\cG:\cH]} \sum_{i=1}^{[\cG:\cH]} \ell_\alpha\bigl(t(o_i(z)), s_i(z)\bigr) + \lambda\|t\|_K^2,
\end{equation}
and the \csymmpi prediction region is
\begin{equation}
    \label{eq:symmpi_RKHS}
    T^{\symmpi}_K(z_{\obs}) = \cbr{z\in\cZ:\psi(\tilde z)\leq \hat{t}_{z}^{\,K}(z_\obs),\obs(z)=z_{\obs}}.
\end{equation}
By the representer theorem, \eqref{eq:hat_t_RKHS} reduces to a finite-dimensional convex program; see Appendix~\ref{sec:rkhs_details}.

We also need a technical assumption to control the interpolation error. Roughly speaking, we assume that $\psi(\tilde Z)$ has a density bounded by $b_\psi>0$, conditionally on the information used to fit the threshold without the test point. The precise statement is Assumption~\ref{assumption:conditional_density} in Appendix~\ref{sec:rkhs_details}. In the exchangeable setting of Appendix~\ref{sec:supervised_learning_iid}, this reduces to a bounded conditional density of the score given the covariate (Corollary~\ref{cor:RKHS_iid}).

The following theorem provides the near-conditional coverage error when using an RKHS. The proof is provided in Section~\ref{sec:proofs_rkhs}.
\begin{theorem}
    \label{thm:RKHS}
    Suppose Assumptions~\ref{assumption:bounded_kernel} and~\ref{assumption:conditional_density} hold. Then, for all $f \in \cF_{K}$ with $\|f\|_K\leq M$ and all $\lambda>0$, the \csymmpi prediction region defined in \eqref{eq:symmpi_RKHS} satisfies
    \begin{equation}
        \label{eq:RKHS_bound}
        \left|\EE_{Z}\left[f(Z_{\obs})\left(\1\{Z\in T^{\symmpi}_K(Z_{\obs})\}-(1-\alpha)\right)\right]\right| \leq 2\lambda\bigl|\EE_Z\langle\hat t^{\,K}_Z,f\rangle_K\bigr| + \frac{b_{\psi}\kappa^3 M}{\lambda[\cG:\cH]}.
    \end{equation}
    Moreover, if $\lambda=[\cG:\cH]^{-1/2}$ and $\EE_Z\|\hat t^{\,K}_Z\|_K\leq M_0$, then
    $$
    \left|\EE_{Z}\left[f(Z_{\obs})\left(\1\{Z\in T^{\symmpi}_K(Z_{\obs})\}-(1-\alpha)\right)\right]\right| \leq \frac{2MM_0 + b_{\psi}\kappa^3 M}{\sqrt{[\cG:\cH]}}.
    $$
\end{theorem}

\begin{remark}
The penalty parameter \(\lambda\) controls the trade-off between the two terms in \eqref{eq:RKHS_bound}. A larger \(\lambda\) increases the bias in the fitted threshold (the first term) but improves its stability, reducing the probability that the test score falls exactly on the threshold (the second term). Balancing these terms yields a rate of order \([\cG:\cH]^{-1/2}\). The slower rate relative to Theorem~\ref{thm:linear_class} reflects the cost of allowing an infinite-dimensional function class. As in that theorem, the bound tightens as the index \([\cG:\cH]\) grows. For example, with \(n+1\) exchangeable points, the rate is \(\cO(n^{-1/2})\) (Appendix~\ref{sec:supervised_learning_iid}).
\end{remark}

\section{Examples}
\label{sec:examples}

In this section, we illustrate the broader applicability of the \csymmpi framework across a range of data settings. Specifically, we consider (i) two-layer hierarchical models, including cluster randomized trials (Section~\ref{sec:hierarchical}), and (ii) network-structured data (Section~\ref{sec:network_structured_data_main}).
For each setting, we identify the underlying group symmetries, instantiate the components of the \csymmpi framework, and characterize the resulting near-conditional coverage guarantees. In the classical exchangeable setting, \csymmpi recovers the conditional conformal prediction method of \citet{gibbs2025conformal}, with a coverage error of order $d/n$ for a $d$-dimensional linear class and $1/\sqrt{n}$ for an RKHS (Appendix~\ref{sec:supervised_learning_iid}).

\subsection{Two-Layer Hierarchical Models}
\label{sec:hierarchical}
We apply \csymmpi to two-layer hierarchical data, a structure commonly arising in applications like meta-learning \citep{park2022pac} and cluster randomized trials \citep{wang2024conformal}. We show that \csymmpi extends prior marginal coverage results \citep{lee2023distribution, dobriban2024symmpipredictiveinferencedata} by providing a stronger near-conditional coverage guarantee. We next illustrate this methodology in supervised meta-learning and for estimating treatment effects in cluster-randomized trials.

\subsubsection{Setup: Data with Hierarchical Symmetries}
\label{sec:hierarchical_setup}
Let $Z = (Z_1, \dots, Z_K)$ consist of $K$ clusters. The $i$-th cluster $Z_i = (Z_1^{(i)}, \dots, Z_{N_i}^{(i)})^\top$ contains a random number $N_i$ of individuals, so it is a vector of variable length taking values in $\cZ_0^* := \bigcup_{m \ge 1} \cZ_0^m$, and the cluster sizes $N = (N_1, \dots, N_K)$ are part of the data. 

\begin{assumption}
    \label{assumption:two_layer_exchangeability}
    We assume that the data exhibit a two-layer exchangeability structure:

    \begin{itemize}
        \item[(i)] (Between clusters) The clusters are exchangeable: $(Z_1,\dots,Z_K)\deq(Z_{\pi(1)},\dots,Z_{\pi(K)})$ for all $\pi\in\rmS_K$.
        \item[(ii)] (Within clusters) For each $i\in[K]$ and each $m$ with $\PP(N_i=m)>0$, conditionally on $N_i=m$, the individuals in cluster $i$ are exchangeable: $Z\deq(Z_1,\dots,\sigma\cdot Z_i,\dots,Z_K)$ for all $\sigma\in\rmS_m$.
    \end{itemize}
\end{assumption}

Assumption~\ref{assumption:two_layer_exchangeability} holds under the hierarchical i.i.d.\ model of \citet{lee2023distribution}: independently for each cluster, one draws a distribution $P_i\sim P_\Pi$ and a size $N_i\mid P_i\sim P_{N\mid\Pi}$, and then $Z_1^{(i)},\dots,Z_{N_i}^{(i)}\stackrel{\mathrm{i.i.d.}}{\sim}P_i$ given $(P_i,N_i)$. In particular, the cluster sizes may depend on the cluster distributions.

\paragraph{Group action.}
Let $\cN := \{N_1, \dots, N_K\}$ be the set of realized cluster sizes. Since $\cN$ is unchanged by permuting individuals or clusters, we condition on it throughout, so that all guarantees hold conditionally on $\cN$ and hence also marginally. Permuting clusters can move a cluster of any size in $\cN$ into a given position, so we attach to each position one within-cluster permutation for every size in $\cN$. Formally, define the inner group $\rmS_\inn := \prod_{m\in\cN}\rmS_m$, whose element $\sigma = (\sigma_m)_{m\in\cN}$ acts on a cluster through the component that matches its size,
$$
\sigma\cdot Z_i := \bigl(Z_{\sigma_{N_i}^{-1}(1)}^{(i)}, \dots, Z_{\sigma_{N_i}^{-1}(N_i)}^{(i)}\bigr)^\top,
$$
and let $\rmS_\out := \rmS_K$ permute the clusters, $\pi\cdot Z := (Z_{\pi^{-1}(1)}, \dots, Z_{\pi^{-1}(K)})$. The group $\mathcal{G} := (\rmS_\inn)^K\rtimes\rmS_\out$ acts on $Z$ by
$$
g\cdot Z := \pi\cdot\bigl(\sigma^{(1)}\cdot Z_1, \dots, \sigma^{(K)}\cdot Z_K\bigr),\qquad g = \bigl((\sigma^{(1)},\dots,\sigma^{(K)}),\pi\bigr)\in\cG,
$$
where $\sigma^{(i)}\in\rmS_\inn$ is the within-cluster permutation applied to the $i$-th cluster and $\pi\in\rmS_\out$ permutes the clusters; that is, $g$ first shuffles the individuals within each cluster and then permutes the clusters.

For $G\sim U$, only the components $\sigma^{(i)}_{N_i}$ act on $Z$, so $G\cdot Z$ permutes the clusters uniformly at random and shuffles the individuals within each cluster uniformly at random. Hence, Assumption~\ref{assumption:two_layer_exchangeability} implies that $Z \deq G\cdot Z$ for $G\sim U$, given  $\cN$.

\subsubsection{Example: Supervised Learning with Two-Layer Hierarchical Data}
\label{sec:hierarchical_supervised_learning}
We now apply the \csymmpi framework to a supervised learning problem with two-layer hierarchical data, a setting also considered in \citet{lee2023distribution,dobriban2024symmpipredictiveinferencedata}. For realized cluster sizes $N = n = (n_1,\dots,n_K)$, let the complete data be
$$
z = (z_1,\dots,z_K), \qquad z_i = \left((x_1^{(i)}, y_1^{(i)}), \dots, (x_{n_i}^{(i)}, y_{n_i}^{(i)})\right)^\top \in (\cX \times \RR)^{n_i},
$$
which satisfies the two-layer exchangeability structure described in Section~\ref{sec:hierarchical_setup}. We consider the problem of predicting the response of the last individual in the final cluster, denoted by $y_{n_K}^{(K)}$. Accordingly, the observation function reveals all data except this target response:
$$
\obs(z) := (z_1,\dots, z_{K-1}, o_K), \quad \text{where } o_K := \left((x_1^{(K)}, y_1^{(K)}), \dots, (x_{n_K-1}^{(K)}, y_{n_K-1}^{(K)}), x_{n_K}^{(K)}\right)^\top.
$$

\paragraph{Transformation and adaptive threshold.}

Given a pretrained black-box predictor $\hat{\mu}:\cX\to\RR$, we map the data to absolute-residual non-conformity scores,
$$
V(z) := (s_1,\dots,s_K), \qquad s_i := \left(|y_1^{(i)} - \hat{\mu}(x_1^{(i)})|, \dots, |y_{n_i}^{(i)} - \hat{\mu}(x_{n_i}^{(i)})|\right)^\top \in \RR^{n_i}.
$$
Since $V$ acts on each individual separately, it commutes with permuting individuals and clusters, so $V$ is $\mathcal{G}$-deterministically equivariant. Let the test function be $\psi(\tilde z) := s_{n_K}^{(K)} = |y_{n_K}^{(K)} - \hat{\mu}(x_{n_K}^{(K)})|$, which extracts the non-conformity score of the target individual.

We use the flexible conditioning of Section~\ref{sec:flexible_conditioning}, which leads to the optimization problem in \eqref{eq:hat_t_eta}. In particular, we choose $\eta$ such that $\proj(z) = x_{n_K}^{(K)}$, so the threshold depends only on the covariate of the target individual. We take $\cW=\cX$ and optimize over a linear space $\cF_\cX\subseteq\{\cX\to\RR\}$ with a penalty $\cR:\cF_\cX\to\RR$ satisfying Assumption~\ref{assumption:regularization}.

Different group elements move different individuals to the target position, and with clusters of different sizes, individuals are moved there with different probabilities. The following corollary, proved in Section~\ref{sec:proof_hierarchical_threshold}, gives the resulting threshold explicitly.
\begin{corollary}
    \label{cor:hierarchical_threshold}
    For every $z$ with cluster sizes $n_1,\dots,n_K$, the threshold \eqref{eq:hat_t_eta} reduces to
    \begin{equation}
        \label{eq:hierarchical_hat_t}
        \hat{t}_{z}(\cdot) = \argmin_{t(\cdot)\in\cF_\cX}\frac{1}{K}\sum_{i=1}^{K}\frac{1}{n_i}\sum_{j=1}^{n_i}\ell_\alpha\left(t(x_j^{(i)}), s_j^{(i)}\right) + \cR(t).
    \end{equation}
\end{corollary}
This is a \emph{weighted} empirical pinball loss with regularization, in which each cluster receives equal weight and observations within a cluster are weighted inversely to the cluster size. Similar weights appear in prior work on marginal coverage for two-layer hierarchical data \citep{lee2023distribution, dobriban2024symmpipredictiveinferencedata,wang2024conformal}. With adaptive thresholds, our framework strengthens these results to near-conditional coverage.

\paragraph{C-SymmPI prediction region.}

Substituting $\hat{t}_z(\cdot)$ from \eqref{eq:hierarchical_hat_t} into the \csymmpi prediction set \eqref{eq:symmpi} and isolating the unobserved component $y_{n_K}^{(K)}$, we obtain
\begin{equation}
    \label{eq:hierarchical_prediction_set}
    T^{\symmpi}(z_\obs) = \cbr{y_{n_K}^{(K)}:s_{n_K}^{(K)}\leq \hat{t}_z(x_{n_K}^{(K)})}.
\end{equation}
By Corollary~\ref{cor:eta_coverage}, this prediction set satisfies a near-conditional coverage guarantee under two-layer exchangeability. The guarantee holds conditionally on the set of cluster sizes $\cN$, and hence also marginally, accounting for the joint randomness of the covariates, the responses, and the cluster sizes.

For a linear threshold class, the guarantee comes with an explicit rate. Since the weights in \eqref{eq:hierarchical_hat_t} are unequal when the cluster sizes differ, this rate does not follow directly from Theorem~\ref{thm:linear_class}; we prove it in Section~\ref{sec:proof_hierarchical_rate}. Let $\cF_\cX=\{\langle\theta,\varphi(\cdot)\rangle:\theta\in\RR^d\}$ for a feature map $\varphi:\cX\to\RR^d$ with $\|\varphi(\cdot)\|_2\leq b_\varphi$, take $\cR\equiv0$ with the minimum-norm selection of Section~\ref{sec:linear_function_class}, and write $S_j^{(i)}:=|Y_j^{(i)}-\hat\mu(X_j^{(i)})|$ for the scores.
\begin{corollary}
    \label{cor:hierarchical_rate}
    Let $n_{\min}:=\min\cN$, and suppose that, conditionally on $\cN$, the covariates, and the other scores, each score $S_j^{(i)}$ has a continuous distribution. Then, conditionally on $\cN$, for all $f(\cdot)=\langle\theta,\varphi(\cdot)\rangle$ with $\|\theta\|_2\leq b_\theta$, the prediction set \eqref{eq:hierarchical_prediction_set} satisfies
    $$
    \Bigl|\EE\Bigl[f(X^{(K)}_{N_K})\Bigl(\1\bigl\{Y^{(K)}_{N_K}\in T^{\symmpi}(Z_\obs)\bigr\}-(1-\alpha)\Bigr)\Bigm|\cN\Bigr]\Bigr|\leq\frac{b_\varphi b_\theta d}{Kn_{\min}}.
    $$
\end{corollary}
The individuals of the smallest cluster carry the largest weight $1/(Kn_{\min})$ and thus determine the rate. When all clusters have equal size $n$, the bound becomes $b_\varphi b_\theta d/(Kn)$, matching Theorem~\ref{thm:linear_class} with $[\cG:\cH]=Kn$.

An analogous result holds for an RKHS class, with the rate of Theorem~\ref{thm:RKHS}; its proof is given in Section~\ref{sec:proof_hierarchical_rkhs}.
\begin{corollary}
    \label{cor:hierarchical_rkhs}
    Let $\cF_\cX=\cF_K$ be the RKHS of a kernel $K$ on $\cX$ with $K(x,x)\leq\kappa^2$ for all $x\in\cX$, and take $\cR(t)=\lambda\|t\|_K^2$ with $\lambda=(Kn_{\min})^{-1/2}$. Suppose that, conditionally on $\cN$, the covariates, and the scores of all individuals other than the target, the target score $S^{(K)}_{N_K}$ has a density bounded by $b_\psi$, and that $\EE[\|\hat t_Z\|_K\mid\cN]\leq M_0$. Then, conditionally on $\cN$, for all $f\in\cF_K$ with $\|f\|_K\leq M$, the prediction set \eqref{eq:hierarchical_prediction_set} satisfies
    $$
    \Bigl|\EE\Bigl[f(X^{(K)}_{N_K})\Bigl(\1\bigl\{Y^{(K)}_{N_K}\in T^{\symmpi}(Z_\obs)\bigr\}-(1-\alpha)\Bigr)\Bigm|\cN\Bigr]\Bigr|\leq\frac{2MM_0+b_\psi\kappa^3M}{\sqrt{Kn_{\min}}}.
    $$
\end{corollary}
As in Corollary~\ref{cor:hierarchical_rate}, the rate is determined by the smallest cluster, and for clusters of equal size $n$ it becomes $\cO(1/\sqrt{Kn})$, matching Theorem~\ref{thm:RKHS} with $[\cG:\cH]=Kn$.

These results provide a stronger guarantee than the marginal coverage results established in \citet{lee2023distribution, dobriban2024symmpipredictiveinferencedata}. Appendix~\ref{sec:computational_considerations} presents a computationally efficient algorithm for constructing the prediction set. More generally, our framework can also be used to predict other components by specifying alternative forms of $\psi$.

\subsubsection{Example: Cluster Randomized Trials}
\label{sec:crt}
We now apply the \csymmpi framework to cluster randomized trials (CRTs), building on the two-layer hierarchical data structure introduced in Section~\ref{sec:hierarchical_setup} \citep{wang2024conformal}. We begin by introducing the setup and notation for CRTs.

Consider a trial with $K$ clusters indexed by $i = 1, \dots, K$, where a cluster $i$ contains $N_i$ individuals. All individuals within each cluster receive the same binary treatment assignment $A_i \in \{0, 1\}$, where $A_i = 1$ denotes treatment and $A_i = 0$ denotes control. For each individual $j = 1, \dots, N_i$ in cluster $i$, we denote the potential outcomes as $Y_{ij}(1)\in \mathcal{Y}$ (under treatment) and $Y_{ij}(0)\in \mathcal{Y}$ (under control). The available covariates are $C_i\in\mathcal{C}$ (cluster-level) and $X_{ij}\in\mathcal{X}$ (individual-level).

To embed this CRT setting within our framework, we define the entire trial as $Z = (Z_1^\top, \dots, Z_K^\top)^\top$. The complete data for each cluster is given by $Z_i = (Z^{(i)}_1, \dots, Z^{(i)}_{N_i})^\top$, and the complete data for each individual is $Z^{(i)}_j = (Y_{ij}(1), Y_{ij}(0), X_{ij}, C_i)$. The data is assumed to satisfy the $\cG$-distributional invariance discussed in Section~\ref{sec:hierarchical_setup}. The observed data $Z_\obs$ consist of all covariates and the outcomes $Y_{ij}(A_i)$ observed under each individual's assigned treatment $A = (A_1, \dots, A_K)$:
$$
\obs(Z) = \bigl((Y_{ij}(A_i), X_{ij}, C_i)\bigr)^\top_{1\leq i\leq K, 1\leq j\leq N_i}.
$$
We now demonstrate how to construct prediction sets for both individual- and cluster-level treatment effects.

\paragraph{Individual-level treatment effect.}
Our objective is to construct a prediction set for the individual treatment effect $\Delta_{m n} := Y_{m n}(1) - Y_{m n}(0)$ for individual $n$ in cluster $m$. Since we observe $Y_{mn}(A_m)$, this requires constructing a prediction set for the unobserved potential outcome $Y_{mn}(1 - A_m)$.

We begin by defining the non-conformity scores. Assume we are given two pretrained models, $\hat{\mu}_a :\cX \times \cC \to \cY$ for $a \in \{0,1\}$, which predict the potential outcome under each treatment assignment based on covariates. For the complete data vector $Z$, we define the transformation $V_a(Z)$ for each treatment assignment $a\in\{0,1\}$ as the vector of absolute residuals:
$$
V_a(Z) := \bigl(S_{ij}(a)\bigr)^\top_{1\leq i\leq K, 1 \leq j \leq N_i} := \bigl(\left| Y_{ij}(a) - \hat{\mu}_{a}(X_{ij}, C_i) \right| \bigr)^\top_{1\leq i\leq K, 1 \leq j \leq N_i}.
$$
Among these scores, $S_{ij}(A_i)$ is a known quantity computed from the observed data, whereas $S_{ij}(1 - A_i)$ is unknown.  Since $V_a$ acts on each individual separately, it is $\mathcal{G}$-deterministically equivariant for $a \in \{0,1\}$, preserving the group symmetry required by our framework. To focus on the individual $(m, n)$, we define the test function $\psi$ to extract the corresponding non-conformity score for $a\in\{0,1\}$:
$$
\psi(V_a(Z)) := S_{mn}(a) := \left| Y_{m n}(a) - \hat{\mu}_{a}(X_{m n}, C_{m}) \right|.
$$
To construct the adaptive threshold for the unobserved potential outcome $Y_{mn}(1 - A_m)$, we leverage the non-conformity scores from clusters that received treatment assignment $1 - A_m$. Let $a^\star = 1 - A_m$ be the unobserved treatment assignment for individual $(m,n)$, and let $\cI_{a^\star} = \cbr{1\leq i\leq K:A_i = a^\star}$ denote the set of clusters that received this treatment. Similar to the approach in Section~\ref{sec:hierarchical_supervised_learning}, we restrict the threshold function to the covariate space $\cX\times \cC$. For computational efficiency, we extract individual $n$ from cluster $m$ and treat it as a singleton cluster containing only that individual. This leads to the following optimization problem analogous to \eqref{eq:hierarchical_hat_t}:
$$
    \hat{t}_{Z}(\cdot) = \argmin_{t(\cdot)\in\{\cX\times\cC\to\RR\}}\frac{1}{|\cI_{a^\star}| + 1}\cbr{\sum_{i\in\cI_{a^\star}}\frac{1}{N_i}\sum_{j=1}^{N_i}\ell_\alpha\Bigl(t(X_{i j}, C_{i}), S_{ij}(a^\star)\Bigr) + \ell_\alpha\Bigl(t(X_{mn}, C_{m}), S_{mn}(a^\star)\Bigr)} + \cR(t).
$$
The scores $S_{ij}(a^\star)$ for $i \in \cI_{a^\star}$ are known (as $A_i = a^\star$), while the score $S_{mn}(a^\star)$ is the unknown quantity associated with individual $(m,n)$. This adaptive threshold allows us to define a prediction set for the potential outcome $Y_{m n}(1 - A_m)$:
$$
\hat{C}_{mn}(1 - A_m) = \left\{y\in\RR: \left| y - \hat{\mu}_{1 - A_m}(X_{m n}, C_{m}) \right| \leq \hat{t}_Z(X_{m n},C_{m}) \right\}.
$$

For $\Delta_{m n} = Y_{m n}(1) - Y_{m n}(0)$, we combine the observed outcome $Y_{mn}(A_m)$ with the prediction set $\hat{C}_{mn}(1-A_m)$ for the unobserved outcome. The \csymmpi prediction set for $\Delta_{m n}$ is therefore defined as
$$
T^{\symmpi}(Z_\obs) =
\begin{cases}
    \hat{C}_{mn}(1) - Y_{mn}(0), & \text{if } A_m = 0, \\
    Y_{mn}(1) - \hat{C}_{mn}(0), & \text{if } A_m = 1.
\end{cases}
$$
By Corollary~\ref{cor:eta_coverage}, this prediction set satisfies a corresponding near-conditional coverage guarantee, whose formal statement we omit for brevity. This is stronger than the marginal coverage ensured by existing prediction sets for treatment effects in CRTs \citep{lee2023distribution, dobriban2024symmpipredictiveinferencedata, wang2024conformal}. Appendix~\ref{sec:computational_considerations} provides a computationally efficient algorithm for constructing this prediction set.

\paragraph{Cluster-Level Treatment Effect.}

We now relax the within-cluster exchangeability in Assumption~\ref{assumption:two_layer_exchangeability}(ii) and construct a prediction set for the cluster-level treatment effect. This approach aggregates all information to the cluster level. First, we define the cluster-average potential outcome $\overline{Y}_i(a) = \sum_{j=1}^{N_i} Y_{ij}(a)/N_i$ for $a \in \{0, 1\}$ and the corresponding average covariate $\overline{X}_i = \sum_{j=1}^{N_i} X_{ij}/N_i$. The complete data $Z$ then consist of the cluster-level tuples $(\overline{Y}_i(0), \overline{Y}_i(1), \overline{X}_i, C_i)_{1\leq i\leq K}$. Consequently, the observation function reveals the observed tuples $(\overline{Y}_i(A_i), \overline{X}_i, C_i)_{1\leq i\leq K}$.

Our objective is to construct a prediction set for $\Delta_{m} = \overline{Y}_m(1) - \overline{Y}_m(0)$. The methodology proceeds analogously to the individual-level case. We first define cluster-level non-conformity scores $S_i(a) = |\overline{Y}_i(a) - \hat{\mu}_a(\overline{X}_i, C_i)|$ using pretrained models $\hat{\mu}_a$. An adaptive threshold $\hat{t}_Z(\cdot)$ is then constructed by solving the function-based quantile regression, using the observed scores $S_i(1-A_m)$ from clusters $i \in \mathcal{I}_{1-A_m}$. The final prediction set for $\Delta_m$ is formed by combining the observed outcome $\overline{Y}_m(A_m)$ with the prediction set for the unobserved outcome, $\hat{C}_m(1-A_m)$. By Theorem~\ref{thm:coverage}, the resulting \csymmpi prediction set for $\Delta_m$ satisfies a near-conditional coverage guarantee, which is stronger than existing marginal coverage results \citep{wang2024conformal}. Appendix~\ref{sec:computational_considerations} provides a computationally efficient algorithm for constructing this prediction set.

\subsection{Network-Structured Data}
\label{sec:network_structured_data_main}
\label{sec:network_structured_data}

We consider the setting where data possess a jointly exchangeable network structure, following the framework of \citet{lunde2023conformalpredictionnetworkassistedregression}. In Appendix~\ref{sec:network_beyond_joint_exchangeability}, we extend this setting to more general network models that do not satisfy joint exchangeability.

\paragraph{Problem setup.}

Let $Y_1, \dots, Y_{n+1} \in \cY\subseteq\RR$ denote response variables, $X_1, \dots, X_{n+1} \in \cX$ denote covariates, and let $A$ denote the $(n+1) \times (n+1)$ adjacency matrix, where $A_{ij}$ encodes the relationship between nodes $i$ and $j$.
For $1 \leq i, j \leq n+1$, define $V_{ij} = (Y_i, Y_j, X_i, X_j, A_{ij})$. We posit the following joint exchangeability assumption, which holds in many common network models, including sparse graphon models \citep{bickel2009nonparametric} and spatial autoregressive models \citep{manski1993identification}.
\begin{assumption}
    \label{assumption:joint_exchangeability}
    For any permutation $\sigma: [n+1] \to [n+1]$, we have
    $$
    (V_{\sigma^{-1}(i)\sigma^{-1}(j)})_{1 \leq i, j \leq n+1} \deq (V_{ij})_{1 \leq i, j \leq n+1}.
    $$
\end{assumption}

To leverage the network structure for prediction, we define local statistics $C_i \in \mathcal{C}$ for each node $i$. These statistics are functions of the adjacency matrix $A$ and the covariates $(X_1, \dots, X_{n+1})$. Many natural network summaries have inherent symmetry properties. Examples include the node degree $D_i = \sum_{j\neq i}A_{ij}$ and the neighborhood covariate average $\overline{X}_i = \sum_{j\neq i} A_{ij} X_j/D_i$. We formalize this requirement in the following exchangeability assumption on the network statistics:
\begin{assumption}
    \label{assumption:exchangeability_of_network_statistics}
    Let $(C_1, \dots, C_{n+1}) = \zeta(A, X_1, \dots, X_{n+1})$ for some function $\zeta$.
    For any permutation $\sigma: [n+1] \to [n+1]$, the network statistics $(C_1, \dots, C_{n+1})$ satisfy
    $$
    (C_{\sigma^{-1}(1)}, \dots, C_{\sigma^{-1}(n+1)}) = \zeta(A_{\sigma^{-1}}, X_{\sigma^{-1}(1)}, \dots, X_{\sigma^{-1}(n+1)}) \quad \text{a.s.}
    $$
    where $A_{\sigma^{-1}}$ is the permuted adjacency matrix with entries $A^{\sigma^{-1}}_{ij} = A_{\sigma^{-1}(i)\sigma^{-1}(j)}$ for $1 \leq i, j \leq n+1$.
\end{assumption}

Assumption~\ref{assumption:exchangeability_of_network_statistics} thus ensures that a permutation of the node labels induces a corresponding permutation of the network statistics vector. Under Assumptions~\ref{assumption:joint_exchangeability} and \ref{assumption:exchangeability_of_network_statistics}, \citet[Theorem 1]{lunde2023conformalpredictionnetworkassistedregression} shows that the triplets $(X_i, C_i, Y_i)$ for $1 \leq i \leq n+1$ are exchangeable.

This exchangeability allows us to embed the problem within our framework. We define the complete data space as $\cZ = (\cX \times \cC \times \cY)^{n+1}$ and the complete data vector as $Z = (X_i, C_i, Y_i)^\top_{1\leq i\leq n+1}\in\cZ$. The relevant symmetry group is the permutation group $\rmS_{n+1}$, acting on $Z$ via the permutation action $\sigma\cdot Z := (X_{\sigma^{-1}(i)}, C_{\sigma^{-1}(i)}, Y_{\sigma^{-1}(i)})^\top_{1\leq i\leq n+1}$. With $U$ the uniform distribution on $\rmS_{n+1}$, the exchangeability of the triplets implies that $Z$ is $\rmS_{n+1}$-distributionally invariant. In the standard predictive inference task, we observe all data except for the final response. For $z = (x_i, c_i, y_i)^\top_{1\leq i\leq n+1}\in \cZ$, the observation function is thus$$\obs(z) := \bigl((x_1, c_1, y_1), \dots, (x_{n}, c_{n}, y_{n}), (x_{n+1}, c_{n+1})\bigr)^\top,$$which reveals all components of $z$ except the response $y_{n+1}$.

\paragraph{C-SymmPI methodology.}

We construct non-conformity scores using a black-box predictor $\hat{\mu}:\cX\times \cC \to\RR$. This defines the transformation
$$
V(z) := (s_{i})^\top_{1\leq i\leq n+1} := \bigl(|y_i - \hat{\mu}(x_i, c_i)|\bigr)^\top_{1\leq i\leq n+1}\in\tilde{\cZ}:= \RR^{n+1}.
$$
The transformed data $\tilde z$ represent the non-conformity scores for all nodes. Since $\hat{\mu}$ is applied element-wise, $V$ is $\rmS_{n+1}$-deterministically equivariant. The test function isolates the score of the target node:
$$
\psi(\tilde z) := s_{n+1} := |y_{n+1} - \hat{\mu}(x_{n+1}, c_{n+1})|.
$$
We now construct the adaptive threshold function $\hat{t}_{z}(\cdot)$ using \csymmpi. Specifically, we define the threshold as a function over the space $\cX\times \cC$. This yields the following optimization problem, analogous to \eqref{eq:iid_hat_t}:
$$
\hat{t}_{z}(\cdot) = \argmin_{t\in\{\cX\times \cC\to\RR\}}\frac{1}{n+1}\sum_{i=1}^{n+1}\ell_\alpha(t(x_i, c_i), s_i) + \cR(t).
$$
The \csymmpi prediction set for $y_{n+1}$ is then given by
$$
T^{\symmpi}(z_\obs) = \cbr{y_{n+1}:s_{n+1}\leq \hat{t}_{z}(x_{n+1}, c_{n+1})}.
$$
By construction, this prediction set satisfies the conditional coverage guarantees of Theorem~\ref{thm:coverage}. This result strengthens the marginal coverage guarantee provided by standard network-assisted conformal prediction \citep{lunde2023conformalpredictionnetworkassistedregression}. Appendix~\ref{sec:computational_considerations} provides a computationally efficient algorithm for constructing this prediction set.

\section{Experiments}
\label{sec:experiments}

We evaluate \csymmpi through a simulation on a hierarchical model and two real-world applications, the PPACT cluster randomized trial and the Cora citation network. These experiments demonstrate the advantages of our method over existing benchmarks in constructing valid and adaptive prediction intervals. In all experiments, we use the flexible conditioning of Section~\ref{sec:flexible_conditioning}, with $\eta$ extracting covariates of the target unit. These are the covariate $X_{n_K}^{(K)}$ in the simulation, the baseline PEGS score in the PPACT trial, and the node covariates in the Cora network, which include network-derived features. The resulting guarantees are therefore relative to functions of these covariates.

\subsection{Numerical Simulation}
\label{sec:numerical_simulation}

\paragraph{Data-generating process.}

We consider a supervised learning task from a hierarchical model with $K=5$ clusters, as described in Section~\ref{sec:hierarchical_supervised_learning}. For each cluster $i \in \{1, \dots, K\}$, the number of samples $N_i$ is drawn from $\text{Poisson}(100)$. The covariates are drawn i.i.d. as $X_{j}^{(i)} \sim \text{Unif}([-0.5, 0.5])$. The response $Y_{j}^{(i)}$ is generated from a linear heterogeneous model:
$$
Y_{j}^{(i)} = X_{j}^{(i)} \cdot (\theta_i + \varepsilon_{j}^{(i)}),
$$
where $\theta_i \sim \mathcal{N}(0, \sigma_\theta^2 = 1.0^2)$ are the slopes for each cluster, and $\varepsilon_{j}^{(i)} \sim \mathcal{N}(0, \sigma_\varepsilon^2 = 0.5^2)$ are independent noise terms for each individual. This model exhibits covariate-dependent heterogeneity, since the conditional variance of the response is a function of the covariate:
$$
\Var\rbr{Y_{j}^{(i)} | X_{j}^{(i)}} = (X_{j}^{(i)})^2 (\sigma_\theta^2 + \sigma_\varepsilon^2).
$$
The task is to construct a prediction interval for a new response $Y_{n_K}^{(K)}$ from a test covariate $X_{n_K}^{(K)}$ in the $K$-th cluster.

\paragraph{Methods for comparison.}

We compare the performance of our proposed \csymmpi with several benchmarks. For all methods, we use linear regression predictors, and we set the target miscoverage level to $\alpha = 0.1$.

\begin{enumerate}[label=(\roman*)]
    \item \textbf{Standard Split-CP\footnote{Here, ``CP'' stands for ``conformal prediction''.}:}
    Since standard split conformal prediction does not account for hierarchical structure, we pool all training data from all clusters to fit a single predictor $\hat{\mu}$. We also pool all calibration data from all clusters into one set. We then compute a standard split conformal interval using the residuals from the pooled predictor as scores.
    \item \textbf{Standard Conditional-CP:} This method applies the conditional calibration technique of \citet{gibbs2025conformal} to the pooled data. It uses the residuals from the single pooled predictor as scores and computes an adaptive interval using function-based quantile regression.
    \item \textbf{Single-Tree Split-CP:} This benchmark uses only the training and calibration data from the target cluster $i=K$. It computes a standard split conformal interval using the cluster-specific predictor $\hat{\mu}_K$. While this respects the local data distribution, it suffers from a small effective sample size.
    \item \textbf{Single-Tree Conditional-CP:} This method is the conditional counterpart to the single-tree split conformal approach. It uses only the data from the target cluster $K$ and applies conditional calibration to compute an adaptive interval.
    \item \textbf{SymmPI:} Proposed by \citet{dobriban2024symmpipredictiveinferencedata}. This method computes residuals using per-cluster predictors and then pools these scores from all clusters. A constant-width interval is constructed based on the weighted quantile of the pooled scores.
    \item \textbf{\texttt{C-SymmPI}:} Our proposed method. We first compute non-conformity scores using per-cluster predictors trained on each cluster's data. Next, we pool the calibration scores from all clusters into a unified, symmetry-aware calibration set. Finally, we apply the conditional calibration technique to this set and compute an adaptive quantile via function-based quantile regression.
\end{enumerate}

\paragraph{Experimental details.}

We evaluate the proposed procedure over $N_{\text{trials}} = 40$ independent trials, each consisting of $N_{\text{reps}} = 100$ repetitions with independently generated datasets. For each of the $K$ clusters, the dataset $\cD^{(i)} = \{(X_j^{(i)}, Y_j^{(i)})\}_{j=1}^{N_i}$ is randomly split in a $50/50$ ratio into a training set $\cD_{\text{train}}^{(i)}$ and a calibration set $\cD_{\text{cal}}^{(i)}$. The training set is used to fit linear regression models. The calibration set is used to compute non-conformity scores and to construct prediction intervals for $Y_{n_K}^{(K)}$. For the function-based quantile regression, we employ an RKHS with a Gaussian kernel of length scale $L = 0.1$ and regularization parameter $\lambda = 0.005$.

We compute the \emph{average empirical coverage} and \emph{average interval length} for prediction intervals over the $100$ repetitions. The final table reports the mean and standard deviation of these $40$ trial-level averages. To assess conditional coverage, we report the empirical coverage and average interval length both marginally and conditionally on regions of the covariate space that represent different levels of variance:
$R_1 = \{x : |x| \leq 0.1\}$,
$R_2 = \{x : 0.1 < |x| \leq 0.3\}$, and
$R_3 = \{x : 0.3 < |x| \leq 0.5\}$.

\paragraph{Results and discussion.}

\begin{figure}[htbp]
    \centering
    \begin{minipage}[t]{0.32\textwidth}
        \centering
        \includegraphics[width=\linewidth]{figures/adaptive_symmpi.pdf}
    \end{minipage}
    \hfill
    \begin{minipage}[t]{0.32\textwidth}
        \centering
        \includegraphics[width=\linewidth]{figures/standard_conditional_conformal.pdf}
    \end{minipage}
    \hfill
    \begin{minipage}[t]{0.32\textwidth}
        \centering
        \includegraphics[width=\linewidth]{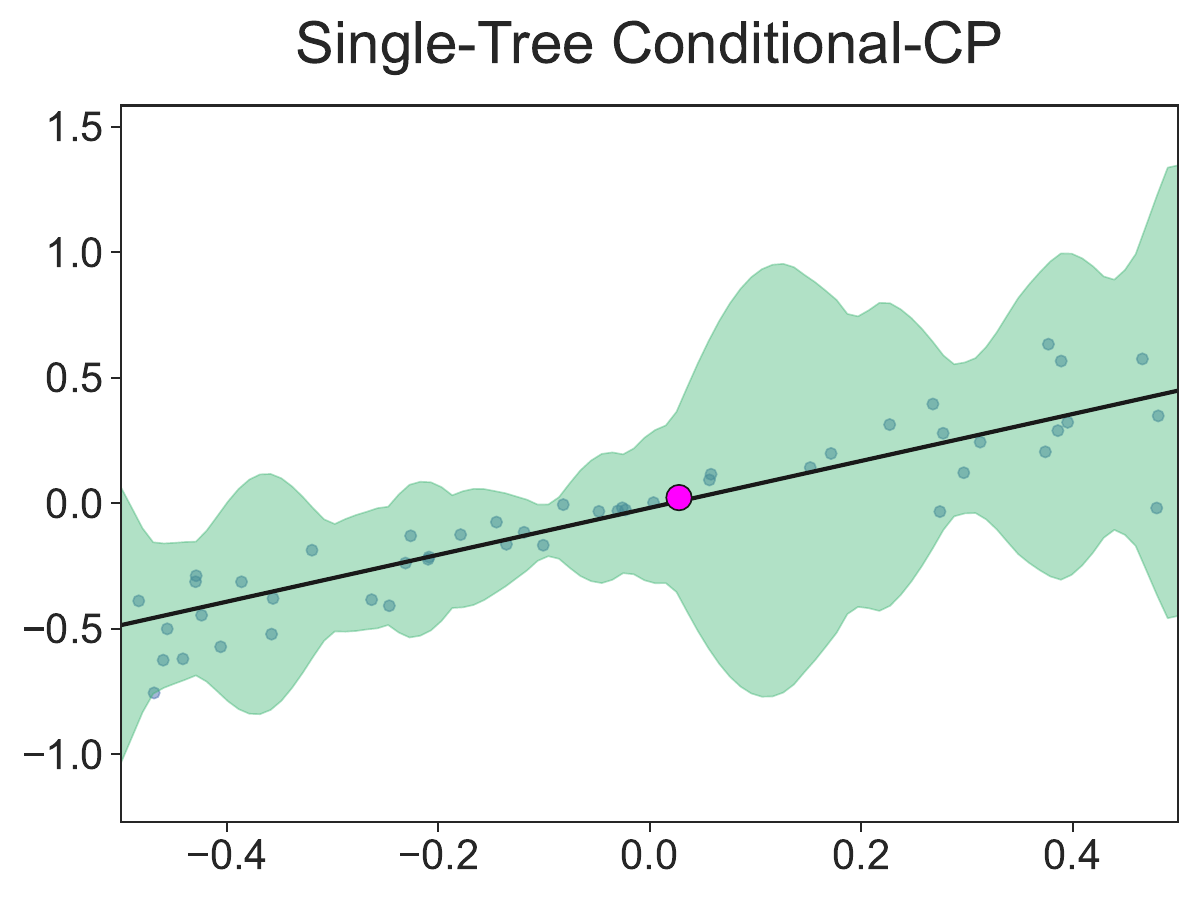}
    \end{minipage}

    \vspace{0.5em}

    \begin{minipage}[t]{0.32\textwidth}
        \centering
        \includegraphics[width=\linewidth]{figures/standard_symmpi.pdf}
    \end{minipage}
    \hfill
    \begin{minipage}[t]{0.32\textwidth}
        \centering
        \includegraphics[width=\linewidth]{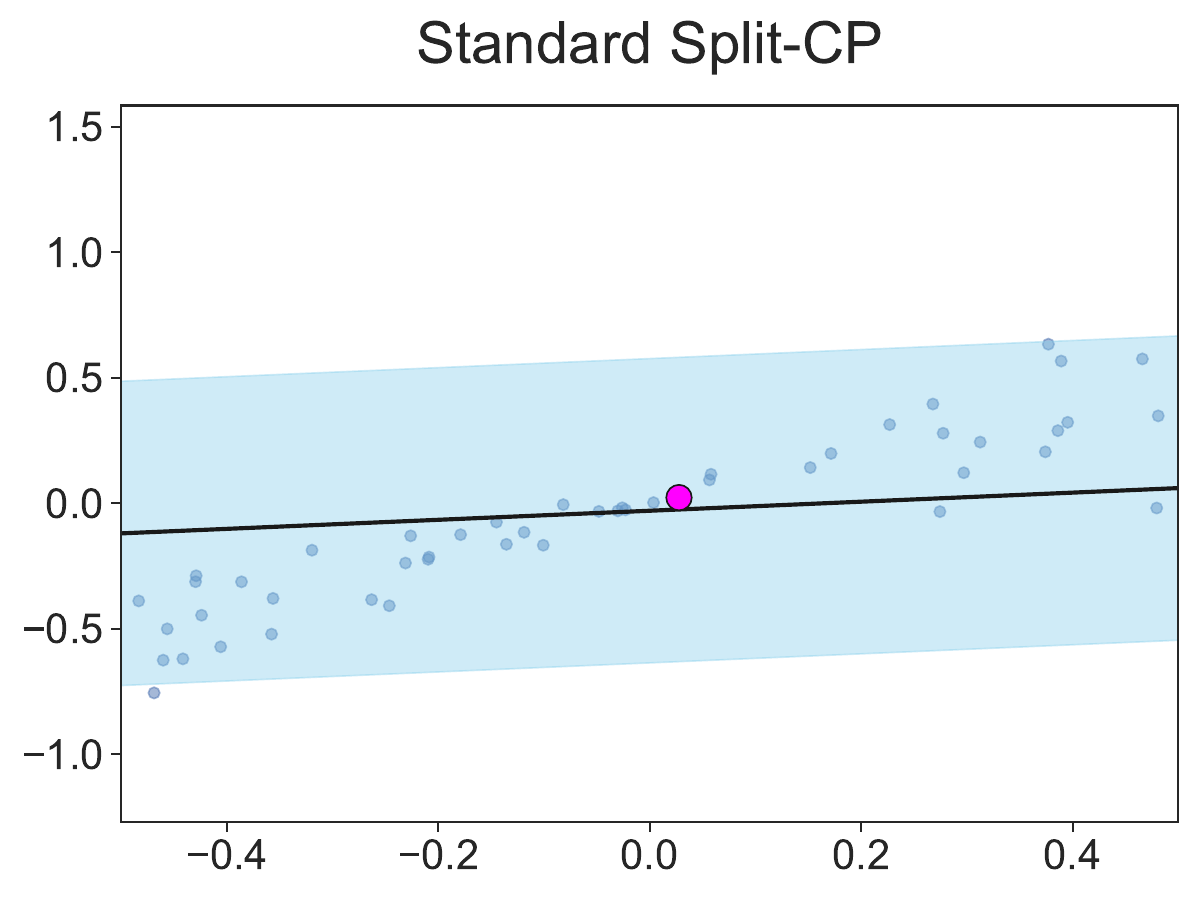}
    \end{minipage}
    \hfill
    \begin{minipage}[t]{0.32\textwidth}
        \centering
        \includegraphics[width=\linewidth]{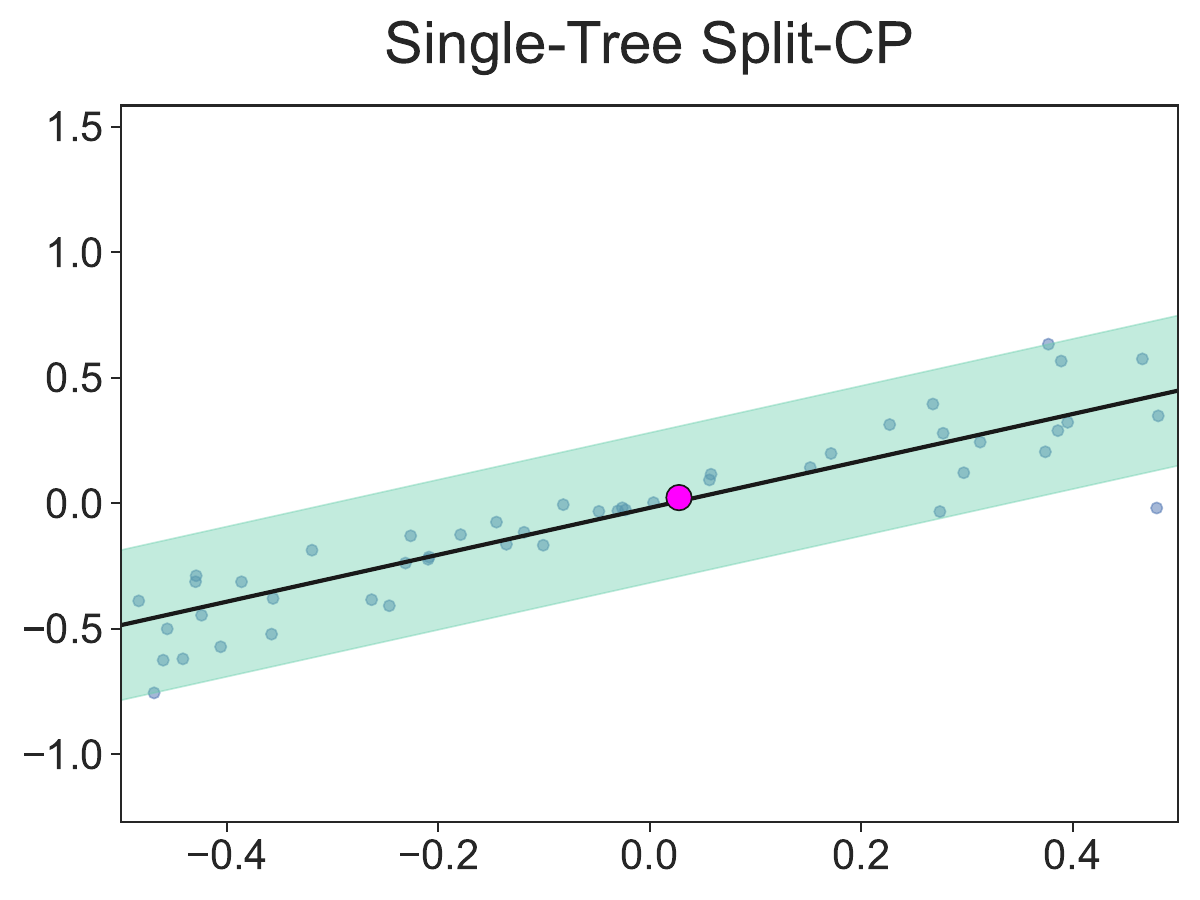}
    \end{minipage}
    \caption{
        Visual comparison of $90\%$ prediction intervals for a single test point (magenta dot) from one cluster of the simulated linear heterogeneous model. \textbf{Top row:} Methods adaptive to the covariate $X_{n_K}^{(K)}$. \textbf{Bottom row:} Constant-width methods. \textbf{Left column:} SymmPI methods that are adaptive to the hierarchical structure. \textbf{Center column:} Standard conformal methods that pool all data naively. \textbf{Right column:} Single-tree methods that use data only from the target cluster. \textbf{Top left:} \csymmpi produces a tight, adaptive interval by leveraging the hierarchical group symmetry and the full calibration set.
        }
    \label{fig:two_layer_experiments}
\end{figure}

Figure~\ref{fig:two_layer_experiments} compares the six methods visually on a single realization of the data, and Table~\ref{tab:two_layer_experiments} summarizes the average empirical coverage and interval length across all simulation trials.

\begin{table}[t]
\centering
\begin{tabular}{clcc}
\toprule
Region & Method & Interval Length & Coverage Probability \\
\midrule
\multirow{6}{*}{Overall}
& \csymmpi         & 0.46 (0.28) & 0.917 (0.056) \\
& SymmPI         & 0.59 (0.01) & 0.910 (0.111) \\
& Standard Conditional-CP   & 0.71 (0.44) & 0.909 (0.052) \\
& Standard Split-CP        & 1.02 (0.06) & 0.924 (0.100) \\
& Single-Tree Conditional-CP & 0.91 (0.29) & 0.982 (0.024) \\
& Single-Tree Split-CP       & 0.71 (0.03) & 0.939 (0.085) \\
\cmidrule(l){2-4}
\multirow{6}{*}{\textbf{$0.0\leq |x| \leq 0.1$}}
& \csymmpi         & 0.13 (0.01) & 0.932 (0.062) \\
& SymmPI         & 0.59 (0.01) & 1.000 (0.000) \\
& Standard Conditional-CP   & 0.20 (0.03) & 0.938 (0.056) \\
& Standard Split-CP      & 1.02 (0.08) & 1.000 (0.000) \\
& Single-Tree Conditional-CP & 0.63 (0.09) & 0.994 (0.016) \\
& Single-Tree Split-CP       & 0.70 (0.03) & 1.000 (0.000) \\
\cmidrule(l){2-4}
\multirow{6}{*}{\textbf{$0.1 < |x| \leq 0.3$}}
& \csymmpi         & 0.42 (0.02) & 0.920 (0.051) \\
& SymmPI         & 0.59 (0.01) & 0.967 (0.028) \\
& Standard Conditional-CP   & 0.67 (0.04) & 0.903 (0.041) \\
& Standard Split-CP       & 1.01 (0.05) & 0.980 (0.018) \\
& Single-Tree Conditional-CP & 0.81 (0.06) & 0.975 (0.024) \\
& Single-Tree Split-CP     & 0.71 (0.02) & 0.983 (0.022) \\
\cmidrule(l){2-4}
\multirow{6}{*}{\textbf{$0.3 < |x| \leq 0.5$}}
& \csymmpi         & 0.82 (0.02) & 0.898 (0.048) \\
& SymmPI         & 0.59 (0.01) & 0.764 (0.062) \\
& Standard Conditional-CP    & 1.27 (0.05) & 0.887 (0.045) \\
& Standard Split-CP         & 1.01 (0.04) & 0.792 (0.058) \\
& Single-Tree Conditional-CP & 1.29 (0.06) & 0.975 (0.025) \\
& Single-Tree Split-CP     & 0.71 (0.03) & 0.833 (0.066) \\
\bottomrule
\end{tabular}
\caption{Comparison of average prediction interval length and empirical coverage probability for the two-layer linear heterogeneous model with $90\%$ target coverage. Interval length and coverage probability are evaluated marginally and conditionally on three disjoint covariate regions. We display the average and the standard error over $40$ independent trials.}
\label{tab:two_layer_experiments}
\end{table}

As Table~\ref{tab:two_layer_experiments} shows, \csymmpi outperforms the marginal-only methods, which produce constant-width intervals and suffer from two deficiencies. First, they substantially under-cover in high-variance regions, with SymmPI \citep{dobriban2024symmpipredictiveinferencedata} achieving an empirical coverage of only $0.764$ in region $R_3$, below the target level $0.9$. Second, they are overly conservative in low-variance regions. In region $R_1$, for instance, SymmPI has an average interval length of $0.59$, whereas \csymmpi achieves $0.13$ while maintaining valid coverage of $0.932$.

\csymmpi also outperforms other conditional methods. The standard conditional approach ignores the hierarchical structure and produces wider intervals, with an average length of $0.71$ compared to $0.46$ for \symmpi. Conversely, the single-tree conditional method uses only local data, and its small sample size leads to overly conservative intervals, with an average length of $0.91$ compared to $0.46$ for \symmpi. Moreover, as seen in Figure~\ref{fig:two_layer_experiments}, the single-tree conditional intervals are visibly less stable than those from \symmpi.

\subsection{Real Data Applications}

\subsubsection{PPACT Cluster Randomized Trial}

The Pain Program for Active Coping and Training (PPACT) study is a CRT designed to evaluate a care-based cognitive behavioral therapy (CBT) intervention for long-term opioid users with chronic pain \citep{debar2022primary}. This dataset was previously analyzed by \citet{wang2024conformal}, who proposed the Split-CCI method.\footnote{Here, ``CCI'' stands for ``conformal causal inference''.} Following their setup, we use \symmpi to construct confidence intervals for both cluster-level and individual-level treatment effects, and we compare it with Split-CCI.

\paragraph{Experimental details.}
The PPACT study randomized $106$ primary care providers (clusters) to either the CBT intervention or usual care. The dataset includes $1$ to $10$ participants per cluster. Our analysis focuses on the primary outcome, the PEGS score at $12$ months. PEGS (pain intensity and interference with enjoyment of life, general activity, and sleep) is a continuous measure of pain intensity and interference on a scale from $1$ to $10$. We adjust for a set of $13$ individual-level baseline variables: baseline PEGS score ($Y_0$), age, gender, disability, smoking status, body mass index, alcohol abuse, drug abuse, comorbidity, depression, number of pain types, average morphine dose, and heavy opioid usage.

We conduct $100$ independent trials. In each trial, $20$ clusters are held out as a test set, and the remaining $86$ clusters are split evenly into training and calibration sets, stratified by the intervention arm $A$. We implement the \csymmpi method described in Section~\ref{sec:crt}, using the efficient computation strategy detailed in Appendix~\ref{sec:computational_considerations}. The predictor $\hat{\mu}_a$ is specified as an ensemble learner combining linear regression and random forest models. We condition on the baseline PEGS score $Y_0$, which serves as an important determinant of the 12-month outcome. We employ an RKHS with a Gaussian kernel of length scale $L = 0.1$ and regularization parameter $\lambda = 0.01$.

We report the average and standard error for two performance metrics: \emph{length of intervals} and \emph{fraction of negatives}. The fraction of negatives is the proportion of test intervals contained in $(-\infty, 0)$. Since negative values indicate treatment benefits, this metric reveals how many clusters or individuals are associated with beneficial treatment effects with probability $1-\alpha$. A higher fraction of negatives indicates a more effective method for identifying treatment benefits.

\paragraph{Results and discussion.}

Table~\ref{tab:CRT} summarizes the performance of \csymmpi and the Split-CCI benchmark for both cluster-level and individual-level treatment effects across four target coverage levels $1-\alpha$. At the cluster level, when $\alpha=0.10$, \csymmpi attains a higher fraction of negatives ($0.143$ vs.\ $0.089$) while keeping the interval length similar to that of the Split-CCI method ($3.968$ vs.\ $4.056$). At the individual level, \csymmpi attains both shorter intervals and a higher fraction of negatives compared to the Split-CCI method across all target coverage levels. The improvement is most pronounced at $\alpha=0.10$, where \csymmpi achieves an average interval length of $4.391$ (vs.\ $6.908$) and a negative fraction of $0.180$ (vs.\ $0.055$). In both cluster-level and individual-level analyses, \csymmpi attains a lower standard deviation for both metrics, indicating more stable performance across trials.

\begin{table}[t]
\centering
\begin{tabular}{cccccc}
\toprule
\multirow{2}{*}{$\alpha$} & \multirow{2}{*}{Method} & \multicolumn{2}{c}{Cluster-level Treatment Effect} & \multicolumn{2}{c}{Individual-level Treatment Effect} \\
\cmidrule(lr){3-4} \cmidrule(lr){5-6}
& & Interval Length & Negative Fraction & Interval Length & Negative Fraction \\
\midrule
\multirow{2}{*}{0.10}
& \csymmpi & 3.968(0.019) & 0.143(0.008) & 4.391(0.009) & 0.180(0.003) \\
& Split-CCI & 4.056(0.557) & 0.089(0.069) & 6.908(0.590) & 0.055(0.026) \\
\addlinespace
\multirow{2}{*}{0.20}
& \csymmpi & 3.354(0.023) & 0.204(0.009) & 3.816(0.009) & 0.225(0.004) \\
& Split-CCI & 2.874(0.412) & 0.173(0.092) & 4.800(0.345) & 0.132(0.044) \\
\addlinespace
\multirow{2}{*}{0.30}
& \csymmpi & 2.861(0.025) & 0.261(0.010) & 3.293(0.009) & 0.270(0.004) \\
& Split-CCI & 2.233(0.313) & 0.238(0.104) & 3.811(0.220) & 0.199(0.052) \\
\addlinespace
\multirow{2}{*}{0.40}
& \csymmpi & 2.443(0.024) & 0.319(0.010) & 2.800(0.009) & 0.314(0.004) \\
& Split-CCI & 1.799(0.255) & 0.304(0.117) & 3.108(0.189) & 0.257(0.055) \\
\bottomrule
\end{tabular}
\caption{Comparison of treatment effects for the \csymmpi and Split-CCI methods \citep{wang2024conformal}. For both the length of intervals and the fraction, we display the average and standard error over $100$ independent trials.}
\label{tab:CRT}
\end{table}

\begin{figure}[t]
    \centering
    \includegraphics[width=0.32\textwidth]{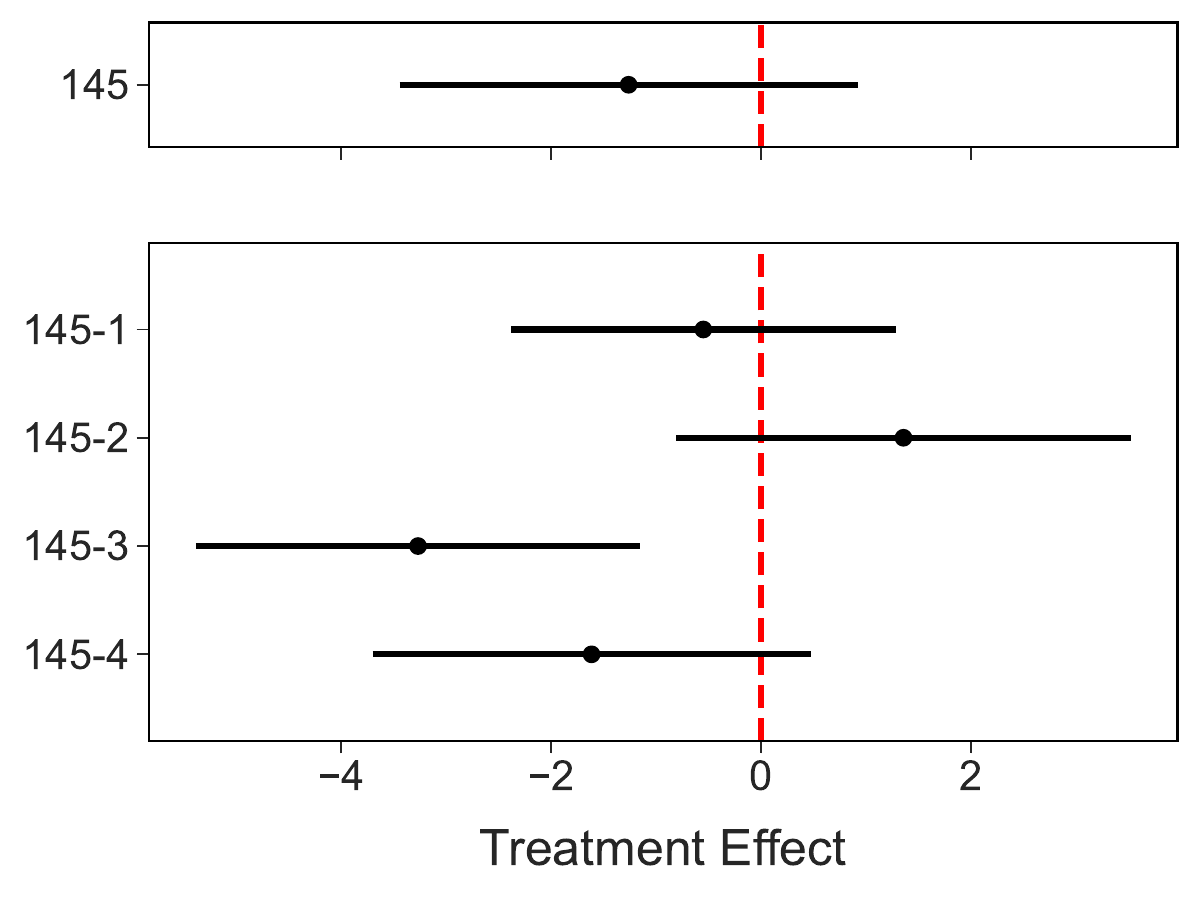}
    \hfill
    \includegraphics[width=0.32\textwidth]{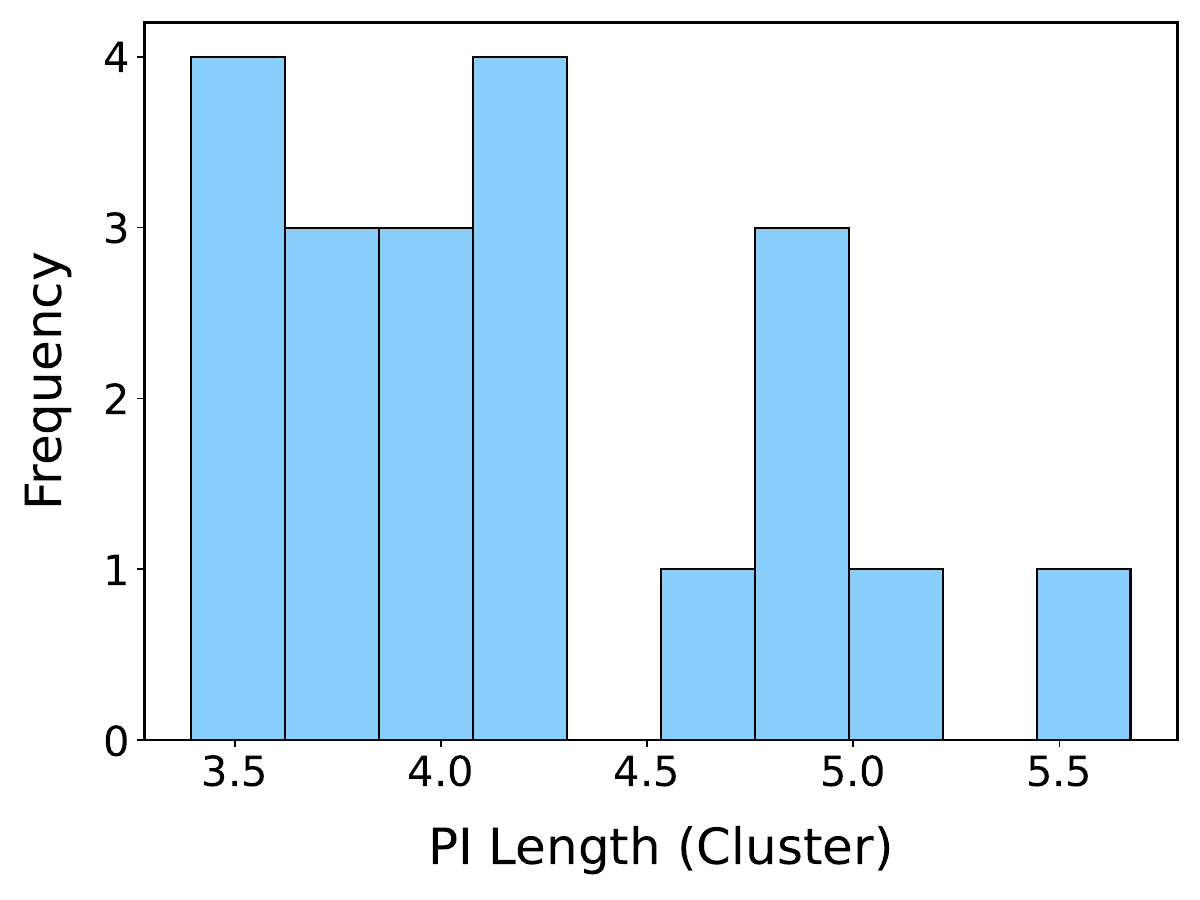}
    \hfill
    \includegraphics[width=0.32\textwidth]{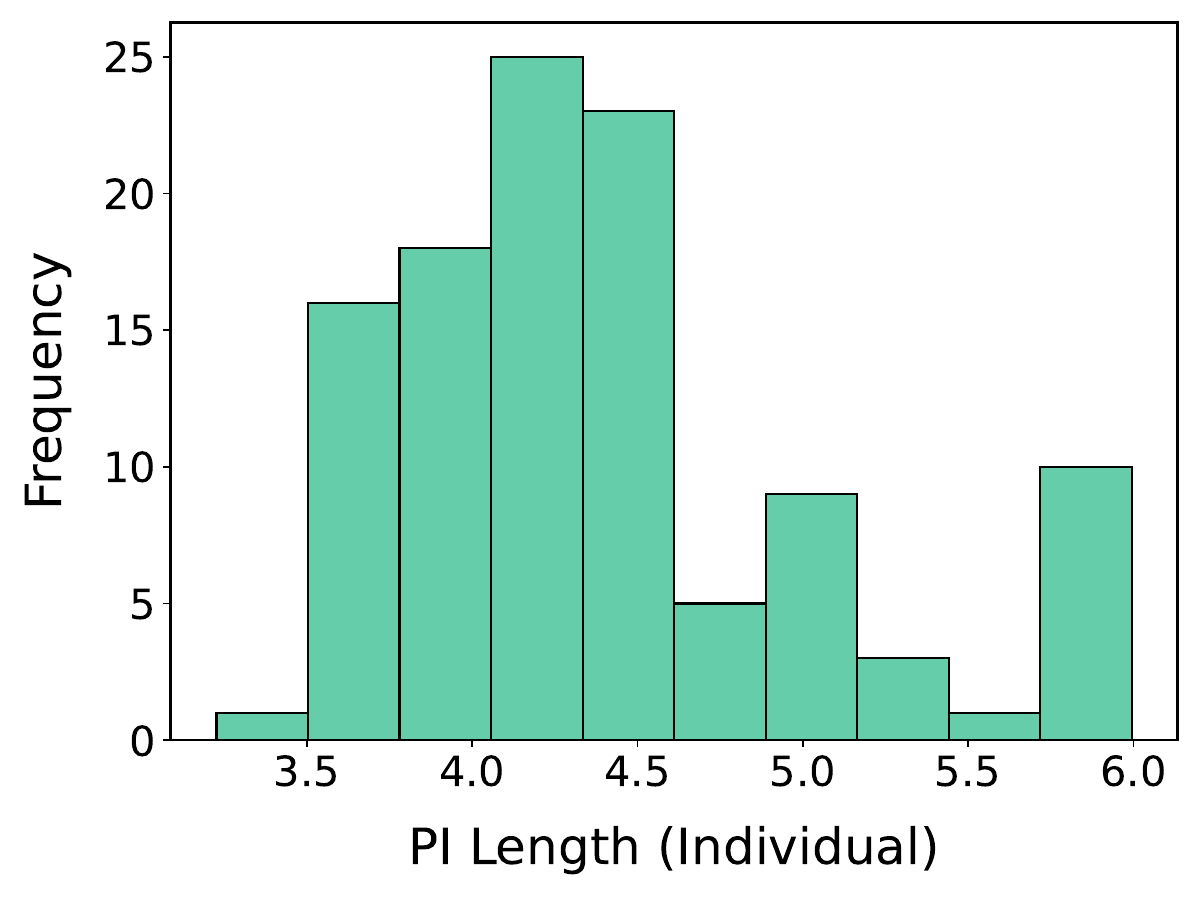}
    \caption{\textbf{Left:} 90\% prediction intervals for the treatment effects of cluster 145 and its individuals. \textbf{Middle:} Distribution of cluster-level treatment-effect interval lengths across all clusters. \textbf{Right:} Distribution of individual-level treatment-effect interval lengths across all individuals.
}
    \label{fig:CRT_experiments}
\end{figure}

The left panel of Figure~\ref{fig:CRT_experiments} presents the $90\%$ prediction intervals for cluster 145 and four of its individuals. The top panel shows that the cluster-level interval contains zero, indicating no statistically significant aggregate effect. The bottom panel reveals individual-level heterogeneity, as the interval for individual 145-3 lies entirely below zero, indicating a statistically significant beneficial treatment effect. This highlights the importance of analyzing individual-level effects, which \csymmpi is designed to capture more effectively. The middle and right panels of Figure~\ref{fig:CRT_experiments} display the distributions of interval lengths across all clusters and individuals, respectively. The interval lengths vary with the underlying uncertainty across individuals and clusters, illustrating the adaptivity of our method.

\subsubsection{Network-Assisted Classification on Cora Dataset}
\label{real_data_network}

The Cora dataset consists of $2708$ machine learning papers categorized into $7$ research topics \citep{mccallum2000automating}. It includes the contents of the papers and a citation network, which are represented by a document-word matrix and an adjacency matrix, respectively. Leveraging these features, we consider the task of predicting whether a paper belongs to the ``Neural Networks'' category. This classification problem was also studied by \citet{lunde2023conformalpredictionnetworkassistedregression}, who applied split conformal prediction to network-assisted models. We illustrate our \csymmpi method for this problem and compare its performance with standard split conformal prediction.\footnote{We do not include a direct comparison with the results in \citet{lunde2023conformalpredictionnetworkassistedregression}, as the evaluation metrics and score definitions are not directly comparable. Their approach measures predictive performance by the size of the prediction set, which takes only discrete values in $\{0, 1, 2\}$. In contrast, we evaluate performance using a continuous interval length on the outcome scale to better reflect adaptivity. Moreover, they adopt the score function from \citet{romano2020classification}, whereas our method consistently employs an absolute residual score.}

\paragraph{Experimental details.}

We construct the covariate $X$ using three groups of features: (i) the top $20$ principal components extracted from the document-word matrix; (ii) $3$-dimensional latent position embeddings and degree-related parameters obtained from the logit model of \citet{ma2020universal}; and (iii) the node degrees and neighborhood averages of the response variable $Y$. The dataset is divided into a training set ($n_{\text{train}} = 1254$), a calibration set ($n_{\text{cal}} = 1254$), and a test set ($n_{\text{test}} = 200$).
A logistic regression model is then fit on the training data to produce the predicted probability $\PP(Y=1)$, where $Y=1$ indicates that the paper belongs to the ``Neural Networks'' category. We implement the \csymmpi method described in Section~\ref{sec:network_structured_data}, using the efficient computation strategy detailed in Appendix~\ref{sec:computational_considerations}. We condition on the covariate set $X$ using an RKHS with a Gaussian kernel of length scale $L = 5.0$ and a regularization parameter $\lambda = 0.004$. We use the absolute residual score function for both \csymmpi and standard split conformal prediction. We evaluate the performance of both methods at target miscoverage levels $\alpha = 0.05$ and $\alpha = 0.10$.

\paragraph{Results and discussion.}

\begin{figure}[t]
    \centering
    \includegraphics[width=0.455\textwidth]{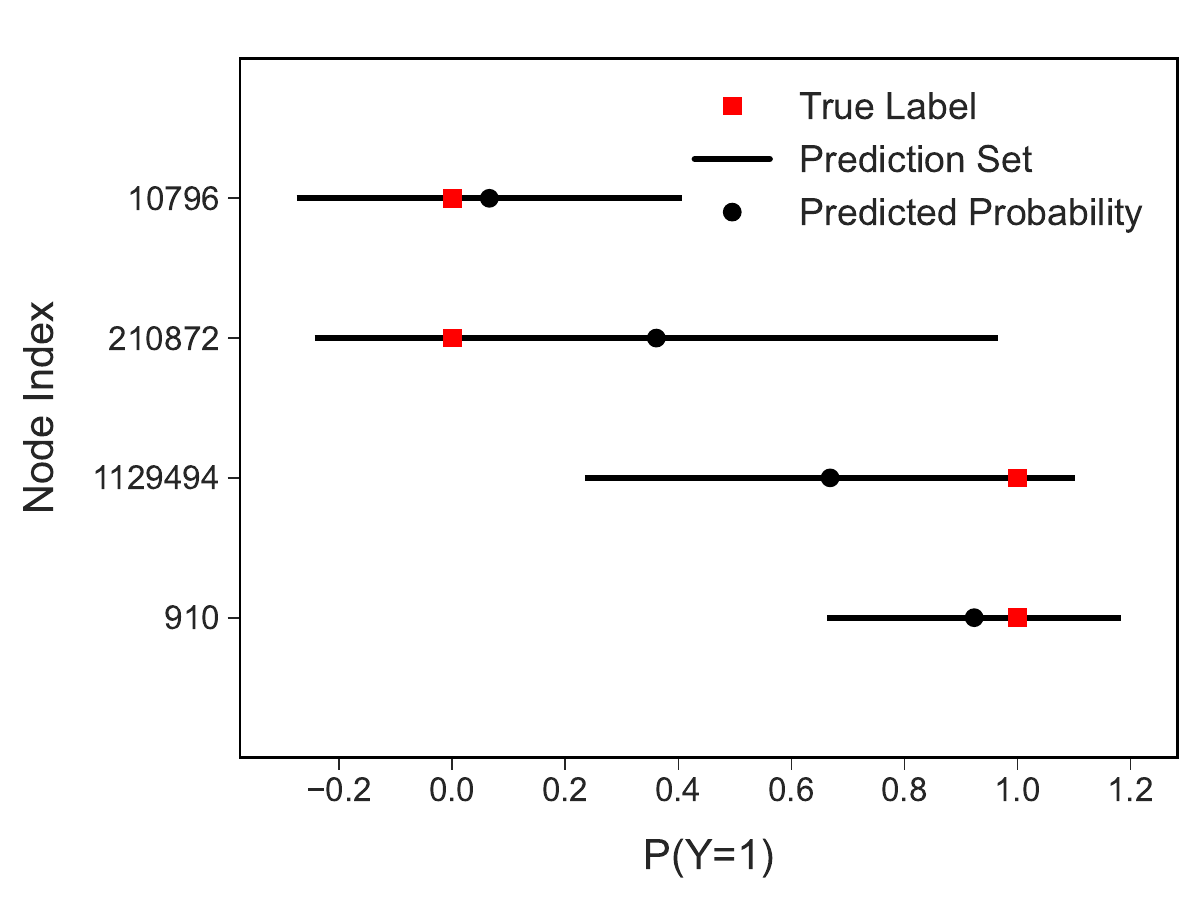}
    \hfill
    \includegraphics[width=0.44\textwidth]{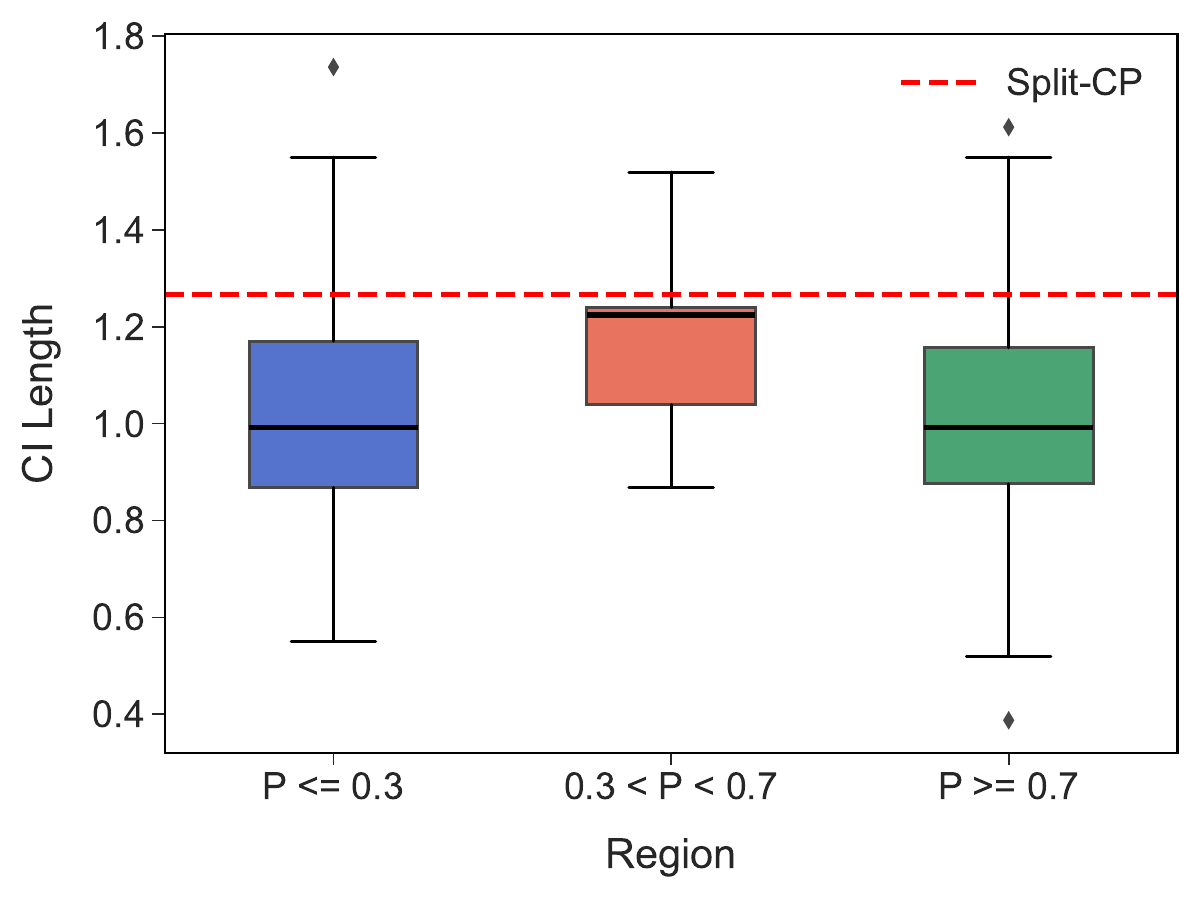}
    \caption{\textbf{Left:} Prediction intervals produced by \csymmpi for selected nodes at $\alpha = 0.10$. \textbf{Right:} Boxplots of interval lengths for \csymmpi across all test nodes, stratified by predicted-probability region.}
    \label{fig:network_experiments}
\end{figure}

Table~\ref{tab:network} reports the mean interval length and coverage probability of \csymmpi and split conformal prediction, both overall and within three regions of the predicted probability $P = \PP(Y=1)$. These are a low-confidence region ($0.3 < P < 0.7$) and two high-confidence regions ($0.0 \leq P \leq 0.3$ and $0.7 \leq P \leq 1.0$). Overall, \csymmpi produces shorter intervals than split conformal prediction while maintaining valid coverage. Moreover, Figure~\ref{fig:network_experiments} shows that the intervals are shorter in the high-confidence regions and longer in the low-confidence region, demonstrating the adaptivity of our method to prediction uncertainty.

\begin{table}[t]
\centering
\begin{tabular*}{\textwidth}{@{\extracolsep{\fill}}cccccc@{}}
\toprule
\multirow{2}{*}{$\alpha$} & \multirow{2}{*}{Prediction Region} & \multicolumn{2}{c}{Mean Interval Length} & \multicolumn{2}{c}{Coverage Probability} \\
\cmidrule(lr){3-4} \cmidrule(lr){5-6}
& & \csymmpi & Split-CP & \csymmpi & Split-CP \\
\midrule
\multirow{4}{*}{0.05}
& Overall & 1.4511 & 1.6121 & 0.9450 & 0.9600 \\
& $0.0\leq P \leq 0.3$ & 1.4513 & 1.6121 & 0.9407 & 0.9481 \\
& $0.3 < P < 0.7$ & 1.5575 & 1.6121 & 1.0000 & 1.0000 \\
& $0.7\leq P\leq 1.0$ & 1.4213 & 1.6121 & 0.9412 & 0.9804 \\
\midrule
\multirow{4}{*}{0.10}
& Overall & 1.0437 & 1.2667 & 0.9100 & 0.9100 \\
& $0.0\leq P \leq 0.3$ & 1.0319 & 1.2667 & 0.9111 & 0.9111 \\
& $0.3 < P < 0.7$ & 1.1871 & 1.2667 & 0.9286 & 0.9286 \\
& $0.7\leq P\leq 1.0$ & 1.0355 & 1.2667 & 0.9020 & 0.9020 \\
\bottomrule
\end{tabular*}
\caption{Comparison of mean interval length and coverage probability for \csymmpi and split conformal prediction at different target miscoverage levels and prediction regions.}
\label{tab:network}
\end{table}

\section{Discussion}
\label{sec:discussion}

We propose \symmpi, a distribution-free method that strengthens marginal guarantees to near-conditional guarantees for data with symmetries. This is achieved by learning an adaptive threshold that keeps a function-weighted average of coverage errors small. We prove fast convergence rates for linear function spaces and RKHSs. The framework allows conditioning on any chosen representation of the observed data, and a sampling-based approximation of the group average extends it to large or infinite groups. Our method recovers the globally exchangeable setting and extends to hierarchical and network data. In these settings, our studies show that it yields more adaptive and robust prediction intervals than prior methods. For future work, it would be interesting to learn a deterministic equivariant map $V$ beyond the non-conformity scores used in this paper, which could further improve adaptivity.
\bibliographystyle{apalike}
\bibliography{main}
\clearpage
\appendix
\section{Review of Group Theory}
\label{sec:group_theory_review}
A \emph{group} is a set $\mathcal{G}$ equipped with a binary operation $\cdot: \mathcal{G} \times \mathcal{G} \to \mathcal{G}$\footnote{For brevity, the operation $g \cdot g'$ is often written simply as $gg'$.} that satisfies three axioms: (i) Associativity: for all $g, g', g'' \in \mathcal{G}$, $(gg')g'' = g(g'g'')$; (ii) Identity element: there exists an element $e \in \mathcal{G}$ such that for every $g \in \mathcal{G}$, $eg = ge = g$; (iii) Inverse element: for each $g \in \mathcal{G}$, there exists an element $g^{-1} \in \mathcal{G}$ such that $gg^{-1} = g^{-1}g = e$. For example,
symmetric group $\rmS_n$ is the group of all permutations on a set of $n$ elements. The group operation is the composition of functions ($\circ$). The identity element is the identity map (i.e., the permutation that leaves all elements fixed), and the inverse of a permutation is its functional inverse.

A group $\mathcal{G}$ can act on a set $\mathcal{Z}$ through a \emph{group action}, which is a map $\mathcal{G} \times \mathcal{Z} \to \mathcal{Z}$, $(g, z)\mapsto g\cdot z$, satisfying two conditions for all $g, g' \in \mathcal{G}$ and $z \in \mathcal{Z}$: (i) $(gg')\cdot z = g\cdot(g'\cdot z)$; and (ii) $e\cdot z = z$. For example, the symmetric group $\rmS_n$ acts on a space $\mathcal{Z}_0^n$ by permuting the coordinates of its elements. For any $g \in \rmS_n$ and $z = (z_1, \dots, z_n)^\top\in \mathcal{Z}_0^n$, this permutation action is defined as
\begin{equation}
    \label{eq:permutation_action}
    g\cdot z := (z_{g^{-1}(1)}, \dots, z_{g^{-1}(n)})^{\top}.\footnote{By definition, we have $(g_1\cdot(g_2\cdot z))_i = (g_2\cdot z)_{g_1^{-1}(i)} = z_{g_2^{-1}(g_1^{-1}(i))} = z_{(g_1g_2)^{-1}(i)}$. The second condition is trivial.}
\end{equation}
Throughout this paper, we will use this permutation action whenever we consider an action of the symmetric group $\rmS_n$.

Given a group action ``$\cdot$'' of $\mathcal{G}$ on $\mathcal{Z}$, there is a natural induced action on a function $F:\cZ\to\cW$, defined by $(g\cdot F)(z) := F(g^{-1}\cdot z)$, where $\cW$ is another measurable space.\footnote{By definition, $((g_1g_2)\cdot F)(z) = F((g_1g_2)^{-1}\cdot z) = F(g_2^{-1}\cdot (g_1^{-1}\cdot z)) = (g_1\cdot (g_2\cdot F))(z)$. Also, $(e\cdot F)(z)=F(e^{-1}\cdot z)=F(z)$. When there is no ambiguity, we use ``$\cdot$'' for both actions on $\cZ$ and on functions.} The \emph{orbit} of the function $F$ is the set of all functions obtainable by acting on $F$, formally defined as $$O_F := \{g\cdot F: g \in \mathcal{G}\}.$$
For example, define $F:\cZ_0^n\to\cZ_0$ as $F(z) = z_{n}$. The orbit of $F$ under the permutation action of $\rmS_{n}$ is $O_F = \{F_1, \dots, F_n\}$, where $F_i(z) = z_i$ for $i = 1, \dots, n$. In other words, $O_F$ consists of all coordinate projection functions.

The \emph{stabilizer} of the function $F:\cZ\to\cW$ is the subgroup\footnote{A subset $\cH$ of $\cG$ is called a \emph{subgroup} of $\cG$, if $\cH$ also forms a group under the operation of $\cG$.} of $\mathcal{G}$ that leaves $F$ unchanged, formally defined as $\stab{F}:=\{g\in\cG:g\cdot F = F\}$, or equivalently,
$$
\stab{F}=\cbr{g\in\cG: F(g\cdot z) = F(z),\ \forall z\in\cZ}.
$$
Continuing the example above, for the function $F(z) = z_{n}$, the stabilizer of $F$ under the permutation action of $\rmS_{n}$ is $\stab{F} = \{g\in\rmS_{n}: g(n) = n\}$, which is isomorphic to $\rmS_{n-1}$. Equivalently, $\stab{F}$ is the set of all permutations that fix the last coordinate.

There is a fundamental relationship between the orbit $O_F$ and the stabilizer $\stab{F}$. To state it, we first introduce some additional concepts. Given $g\in\cG$ and the subgroup $\stab{F}$, the (left) \emph{coset} of $\stab{F}$ associated with $g$ is defined as
$g\stab{F} := \{gh: h \in \stab{F}\}$. Intuitively, a coset is a ``bundle'' of group elements that are indistinguishable in their effect on $F$, since all elements of the coset $g\stab{F}$ produce the same element of the orbit.\footnote{Indeed, for any $h \in \stab{F}$, we have $(gh)\cdot F = g\cdot(h\cdot F) = g\cdot F$.} The set of all distinct cosets of $\stab{F}$ in $\cG$ is denoted by $\cG/\stab{F} := \{g\stab{F}: g \in \cG\}$.
We are now ready to state the \emph{orbit--stabilizer theorem}:
$$
O_F \cong \cG/\stab{F}.
$$
Here, ``$\cong$'' means that there exists a natural bijection. This theorem states that there is a one-to-one correspondence between the elements of the orbit and the cosets of the stabilizer subgroup. In particular, when $\cG$ is finite, it implies that
$$
|O_F| = \frac{|\cG|}{|\stab{F}|},
$$
where $|\cdot|$ denotes the cardinality of a set. Continuing the previous example, since $|\rmS_n| = n!$ and $|\stab{F}| = (n-1)!$, the orbit $O_F$ has size $|O_F| = n!/(n-1)! = n$, which aligns with our earlier observation that $O_F = \{F_1, \dots, F_n\}$.

Finally, the \emph{uniform distribution} $U$ on a group $\mathcal{G}$, known in group theory as its \emph{Haar probability measure}, formalizes the notion of drawing a group element uniformly at random. Its key property is invariance under the group operation: if $G \sim U$\footnote{The expression ``$G\sim U$'' means that $G$ is a random variable taking values in $\cG$ whose distribution is the uniform distribution $U$.}, then for any fixed $g \in \mathcal{G}$, we have $gG\sim U$ and $Gg\sim U$. Such a distribution exists for all finite groups and many common infinite groups (e.g., the rotation group $\mathrm{SO}(p)$).\footnote{Formally, every compact (Hausdorff) topological group admits a unique Haar probability measure, and it is both left- and right-invariant \citep{diestel2014joys}.} For any finite group, $U$ is the discrete uniform distribution over its elements. Throughout this paper, all groups we consider are equipped with their uniform distribution $U$.

To illustrate these concepts with a structure beyond global exchangeability, we consider a ``nested'' symmetric group \citep{dixon1996permutation}, which is relevant to two-layer hierarchical settings discussed in Section~\ref{sec:hierarchical}.

\begin{example}
    Suppose the data is a vector $Z = (Z_1^\top, \dots, Z_K^\top)^\top$ composed of $K$ distinct clusters. Each cluster $Z_i = (Z^{(i)}_1, \dots, Z^{(i)}_{n})^\top$ in turn contains $n$ data points for all $i=1, \dots, K$.

    The symmetries of this structure arise from a two-layer permutation mechanism: first, permuting the elements within each cluster, and second, permuting the clusters themselves. This hierarchical symmetry is formally represented by a ``nested'' symmetric group $\mathcal{G} := \left(\mathrm{S}_{n}\right)^K \rtimes \mathrm{S}_K$, which can be described as follows:
    \begin{itemize}
        \item[(i)] Each group element $g \in \mathcal{G}$ is a pair $((\sigma_1, \dots, \sigma_K), \pi)$, where $\sigma_i \in \mathrm{S}_{n}$ permutes the elements within the $i$-th cluster, and $\pi \in \mathrm{S}_K$ permutes the clusters themselves.
        \item[(ii)] The group operation between $g_1 = ((\sigma_1, \dots, \sigma_K), \pi)$ and $g_2 = ((\tau_1, \dots, \tau_K), \eta)$ is defined as
        $g_1g_2 = ((\sigma_{\eta(1)}\tau_1, \dots, \sigma_{\eta(K)}\tau_K), \pi\eta)$.
        In other words, the cluster permutations are composed directly, while $g_2$ permutes the $i$-th cluster by $\tau_i$ and moves it to position $\eta(i)$, where $g_1$ then permutes it by $\sigma_{\eta(i)}$. This is the operation that makes \eqref{eq:nested_symmetric_group_action} below a group action.
    \end{itemize}

    The group action of $g = ((\sigma_1, \dots, \sigma_K), \pi) \in \mathcal{G}$ on the data vector $Z$ proceeds in two stages: first, the elements within each cluster are permuted according to $\sigma_i$, and then the clusters are permuted according to $\pi$. Formally, this action is expressed as
    \begin{equation}
        \label{eq:nested_symmetric_group_action}
        g\cdot Z := \pi \cdot (\sigma_1 \cdot Z_1^\top, \dots, \sigma_K \cdot Z_K^\top)^\top,
    \end{equation}
    where the symbol ``$\cdot$'' denotes the permutation action defined in \eqref{eq:permutation_action}.

    Suppose we are interested in the function $F:\cZ\to\cZ_0$ that extracts the last data point from the last cluster, i.e., $F(Z) = Z^{(K)}_{n}$. Because $\pi$ can send any cluster $i\in[K]$ to the last cluster $K$, and within that cluster $\sigma_i$ can send any element $j\in[n]$ to the last element $n$, the orbit consists of all coordinate projections:
    $O_F = \{F_j^{(i)}\}_{1\leq i\leq K,1\leq j\leq n}$, where $F_j^{(i)}(Z) = Z^{(i)}_{j}$.
    The stabilizer $\stab{F}$ consists of all group elements that fix the last data point in the last cluster. Specifically, $\pi$ must fix cluster $K$, and $\sigma_K$ must fix element $n$. The other permutations $\sigma_1, \dots, \sigma_{K-1}$ can be arbitrary. Therefore,
    $\stab{F} \cong (\mathrm{S}_n)^{K-1} \times \mathrm{S}_{n-1}\times \mathrm{S}_{K-1}$.
    As a check, we can verify the orbit--stabilizer theorem:
    $$
    \frac{|\mathcal{G}|}{|\stab{F}|} = \frac{(n!)^K \cdot K!}{(n!)^{K-1} \cdot (n-1)! \cdot (K-1)!} = Kn = |O_F|.
    $$

    Finally, since $\mathcal{G}$ is a finite group, its uniform distribution $U$ is the discrete uniform distribution over its elements. This allows us to define a probability space $(\cG, 2^{\cG}, U)$, which enables probabilistic analysis on the group.\footnote{Here, $2^{\cG}$ denotes the power set of $\cG$, representing the sigma-algebra of all subsets of $\cG$.}
    \qed
\end{example}

\section{Classical Examples: Exchangeable Data}
\label{sec:supervised_learning_iid}

In this section, we study near-conditional predictive inference under the exchangeability assumption. In particular, we demonstrate that \csymmpi recovers the conditional conformal prediction method of \citet{gibbs2025conformal}, together with its coverage guarantees (their Theorems~2 and~3).

\paragraph{Problem setup.}

Let $Z = (X_i, Y_i)^\top_{1\leq i\leq n+1}\in \cZ := \rbr{\cX\times\RR}^{n+1}$ be a random vector consisting of $n+1$ exchangeable pairs $(X_i, Y_i)$. Let $\rmS_{n+1}$ be the permutation group on the index set $[n+1]$, which acts on $Z$ by the permutation action defined in \eqref{eq:permutation_action}. The exchangeability assumption ensures $Z\deq G\cdot Z$ for $G\sim U$, where $U$ denotes the uniform distribution on $\rmS_{n+1}$. For all $z = (x_i, y_i)^\top_{1\leq i\leq n+1}\in\cZ$, the observation function is $\obs(z) := \left((x_1, y_1), \dots, (x_{n}, y_{n}), x_{n+1}\right)^\top\in\cO := \rbr{\cX\times\RR}^{n}\times \cX$, which reveals all the data points (both features and labels) except for $y_{n+1}$.

\paragraph{Transformations.}

We focus primarily on split conformal prediction due to its computational efficiency \citep{papadopoulos2002inductive,angelopoulos2021gentle}, and our method can be extended  to full conformal prediction easily. We assume access to an arbitrary black-box predictor $\hat{\mu}:\cX\to\RR$ \footnote{One can always partition the sample into two disjoint subsets, using $n_0$ observations to train the predictor and the remaining $n_1 := n-n_0$ observations to construct the prediction set. Without loss of generality, we treat $n_1$ as the full calibration size $n$ and leave the $n_0$ training sample implicit. Throughout, $\hat\mu$ is treated as fixed, i.e., all probabilities and assumptions below are conditional on the training sample.} Define
$$
V(z) := (s_1, \dots, s_{n+1})^\top := \bigl(|y_1 - \hat{\mu}(x_1)|, \dots, |y_{n+1} - \hat{\mu}(x_{n+1})|\bigr)^\top \in \tilde{\cZ}.
$$
We obtain that for all $g \in \rmS_{n+1}$ and $z\in\cZ$, the transformed data satisfy $V(g\cdot z) = g\cdot\tilde z$, and hence $V$ is $\rmS_{n+1}$-deterministically equivariant. Additionally, $\tilde z$ collects the non-conformity scores of the original sample $z$ with respect to $\hat{\mu}$.

\paragraph{Adaptive threshold.}

Let the test function be $\psi(\tilde z) := s_{n+1} = |y_{n+1} - \hat{\mu}(x_{n+1})|$, which extracts the last component of $\tilde z$. We now construct an adaptive threshold by applying a threshold function to the scores $\{\psi(g\cdot\tilde z) : g \in \rmS_{n+1}\}$ over the group orbit.

To recover this method, we use the flexible conditioning of Section~\ref{sec:flexible_conditioning}, in which the threshold function is restricted to depend only on the covariates in $\cX$, rather than the full observed data $\cO$. Formally, we define the projection map $\proj:\cZ \to \cX$ by $\proj(z) = \eta(\obs(z))= x_{n+1}$, which yields the optimization problem in \eqref{eq:hat_t_eta} with $\cW=\cX$. Throughout this section, $\cF_\cX\subseteq\{\cX\to\RR\}$ denotes a linear space of threshold functions and $\cR:\cF_\cX\to\RR$ a penalty satisfying Assumption~\ref{assumption:regularization} with $\cF$ replaced by $\cF_\cX$; for instance, $\cF_\cX$ can be the linear class or the RKHS considered below.

Next, we simplify the calculation of \eqref{eq:hat_t_eta} by partitioning the group. For any $i \in [n+1]$, the set of permutations that map the $i$-th element to the last position has size $n!$. For any such permutation $g$, we have $\proj(g\cdot z) = x_i$ and $\psi(g\cdot\tilde z) = s_i$. Therefore, the optimization problem reduces to
\begin{equation}
    \label{eq:iid_hat_t}
    \begin{split}
        \hat{t}_{z}(\cdot) &= \argmin_{t(\cdot)\in\cF_\cX}\frac{1}{|\rmS_{n+1}|}\sum_{g\in\rmS_{n+1}}\ell_\alpha\bigl(t(\proj(g\cdot z)), \psi(g\cdot\tilde z)\bigr) + \cR(t)\\
        &= \argmin_{t(\cdot)\in\cF_\cX}\frac{1}{(n+1)!}\sum_{i=1}^{n+1}\bigl|\left\{g\in\rmS_{n+1}:\proj(g\cdot z)=x_i, \psi(g\cdot\tilde z)=s_i\right\}\bigr|\cdot\ell_\alpha(t(x_i), s_i) + \cR(t)\\
        &= \argmin_{t(\cdot)\in\cF_\cX}\frac{1}{(n+1)!}\sum_{i=1}^{n+1}n!\cdot\ell_\alpha(t(x_i), s_i) + \cR(t)\\
        & = \argmin_{t(\cdot)\in\cF_\cX}\frac{1}{n+1}\sum_{i=1}^{n+1}\ell_\alpha(t(x_i), s_i) + \cR(t).
    \end{split}
\end{equation}
This is exactly the conditional conformal formulation, where the adaptive threshold is learned by minimizing the empirical pinball loss over the calibration scores $\{s_i\}_{i=1}^{n+1}$ with a regularization term.\footnote{Throughout this section, we omit the ``proj'' subscript and superscript from the threshold function and the prediction region for simplicity.}

\paragraph{C-SymmPI prediction region.}

Substituting $\hat{t}_z(\cdot)$ from \eqref{eq:iid_hat_t} into the \csymmpi prediction region \eqref{eq:symmpi} and isolating the unobserved component $y_{n+1}$, we obtain
\begin{equation}
    \label{eq:iid_prediction_set}
    T^{\symmpi}(z_\obs) = \cbr{y_{n+1}: s_{n+1}\leq \hat{t}_z(x_{n+1})}.
\end{equation}
This prediction region coincides with the conditional conformal prediction set. The following corollary, which is directly implied by Theorem~\ref{thm:coverage} and Corollary~\ref{cor:eta_coverage}, establishes the near-conditional coverage guarantee for $Y_{n+1}$ given $X_{n+1}$ for the prediction set in \eqref{eq:iid_prediction_set}. Its proof closely parallels the arguments of Theorem~\ref{thm:coverage} and  Corollary~\ref{cor:eta_coverage} and is therefore omitted for brevity.
\begin{corollary}
    \label{cor:supervised_learning_iid}
    Under the split conformal prediction setup, for any $f\in\cF_\cX$ with $\EE|f(X_{n+1})|<\infty$ and $\EE|D_f\cR(\hat t_Z)|<\infty$, the prediction region $T^{\symmpi}$ defined in \eqref{eq:iid_prediction_set} satisfies
    $$
        \abr{\EE_Z\left[f(X_{n+1})\Bigl(\1\{Y_{n+1}\in T^{\symmpi}(Z_\obs)\} - (1-\alpha)\Bigr) \right]} \leq \varepsilon_\pen + \varepsilon_{\inte},
    $$
    where the penalty error and interpolation error are defined as
    $$
    \varepsilon_\pen = \abr{\EE_{Z}\left[D_{f}\cR(\hat{t}_{Z})\right]},\quad\varepsilon_{\inte} = \mathbb{E}_Z\left[|f(X_{n+1})| \1\left\{S_{n+1} = \hat{t}_Z(X_{n+1})\right\}\right].$$
\end{corollary}

\begin{remark}
    While our derivation employs the absolute residual score, $S_i = \abr{Y_i - \hat{\mu}(X_i)}$, the framework readily accommodates other non-conformity scores through appropriate choices of the transformation $V$ and the test function $\psi$.
\end{remark}

\paragraph{Specialization to a linear function class.}
Let $\cF_{L} = \{f:\cX\to\RR \mid f(\cdot) = \langle\theta, \varphi(\cdot)\rangle,\ \theta\in\RR^d\}$, where $\|\varphi\|_2\leq b_\varphi$ as specified in Assumption~\ref{assumption:bounded_feature_map}. By taking $\cF_\cX=\cF_L$ in \eqref{eq:iid_hat_t} (without any norm constraint on the threshold) and setting the penalty $\cR \equiv 0$, we obtain the corresponding prediction region $T^{\symmpi}_L(Z_{\obs})$. Here the stabilizer $\cH$ consists of the permutations fixing $n+1$, and the right coset $\cH g$ is determined by $g^{-1}(n+1)$, so that $[\cG:\cH]=n+1$ and the distinct orbit elements are $(X_i,S_i)$, $i=1,\dots,n+1$. Corollary~\ref{cor:linear_class_iid} establishes a convergence rate of $\cO(d/n)$ for the conditional coverage error, which follows from Theorem~\ref{thm:linear_class}.

\begin{corollary}
    \label{cor:linear_class_iid}
    Suppose that $(X_i,Y_i)_{1\leq i\leq n+1}$ are i.i.d.\ and that the conditional distribution of the score $S$ given $X$ is continuous. Then, in the setting above, for any $f=\langle\theta,\varphi\rangle\in \cF_L$ with $\|\theta\|_2\leq b_\theta$, the prediction region $T^{\symmpi}_L(Z_{\obs})$ satisfies
    $$
        \left|\EE_Z\left[f(X_{n+1})\Bigl(\1\{Y_{n+1}\in T^{\symmpi}_L(Z_{\obs})\} - (1-\alpha)\Bigr) \right]\right| \leq \frac{b_{\varphi}b_\theta d}{n+1}.
    $$
\end{corollary}
A continuous marginal distribution of $S$ is not sufficient: if, for instance, $S=X$ with $X$ continuous and $\varphi(x)=x$, then the threshold $t(x)=x$ interpolates every score.

\paragraph{Specialization to an RKHS.}

Let $\cF_K$ be the RKHS with norm $\|\cdot\|_K$ induced by a positive semidefinite kernel $K:\cX\times\cX\to\RR$ satisfying Assumption~\ref{assumption:bounded_kernel}. By taking $\cF_\cX=\cF_K$ in \eqref{eq:iid_hat_t} and choosing the penalty $\cR(t) = \lambda\|t\|_{K}^2$, we obtain the corresponding prediction region $T^{\symmpi}_K(Z_{\obs})$. Corollary~\ref{cor:RKHS_iid} follows from Theorem~\ref{thm:RKHS} with $[\cG:\cH]=n+1$.

\begin{corollary}
\label{cor:RKHS_iid}
Suppose that $(X_i,Y_i)_{1\leq i\leq n+1}$ are i.i.d., and that the conditional distribution of the score $S$ given $X=x$ admits a density bounded above by a constant $b > 0$ for every $x$. Choose $\lambda=(n+1)^{-1/2}$ and suppose $\EE_Z\|\hat t^{\,K}_Z\|_K\leq M_0$. Then, in the setting above, for any $f \in \cF_K$ with $\|f\|_K\leq M$, the prediction region $T^{\symmpi}_K(Z_{\obs})$ satisfies
$$
    \left|\EE_Z\left[f(X_{n+1})\Bigl(\1\{Y_{n+1}\in T_K^{\symmpi}(Z_\obs)\} - (1-\alpha)\Bigr) \right]\right| \leq \frac{2MM_0+b\kappa^3M}{\sqrt{n+1}}.
$$
\end{corollary}

\begin{remark}
    \citet{gibbs2025conformal} considered a richer function class $\cF = \{f = f_1 + f_2 :\cX\to\RR \mid f_1\in\cF_L, f_2\in\cF_K\}$, defined as the sum of a $d$-dimensional linear component and an RKHS component. A direct consequence of their Proposition~1 is a statistical rate of $\cO(\sqrt{d\log n/n})$ for the conditional coverage error. Our analysis considers the RKHS function class only; the leave-one-out bound and the resulting rate in Corollary~\ref{cor:RKHS_iid} are derived in this paper.
\end{remark}

\section{Additional Examples: Graph Automorphism Symmetries}
\label{sec:additional_examples_symmpi}

The SymmPI framework introduced in Section~\ref{sec:symmpi_framework} is highly general and accommodates a wide range of data structures and group actions, including network-structured data, rotationally invariant data, and two-layer hierarchical data. In this section, we first present an example in which the symmetries are given by the automorphism group of a graph (Section~\ref{sec:symmpi_graph_example}), and then use it to extend the network model of Section~\ref{sec:network_structured_data_main} beyond joint exchangeability (Section~\ref{sec:network_beyond_joint_exchangeability}). For further examples and a more detailed discussion, readers are referred to \citet{dobriban2024symmpipredictiveinferencedata}.

\subsection{SymmPI under Graph Automorphisms}
\label{sec:symmpi_graph_example}

\begin{figure}[ht]
    \centering
    \includegraphics[width=0.6\textwidth]{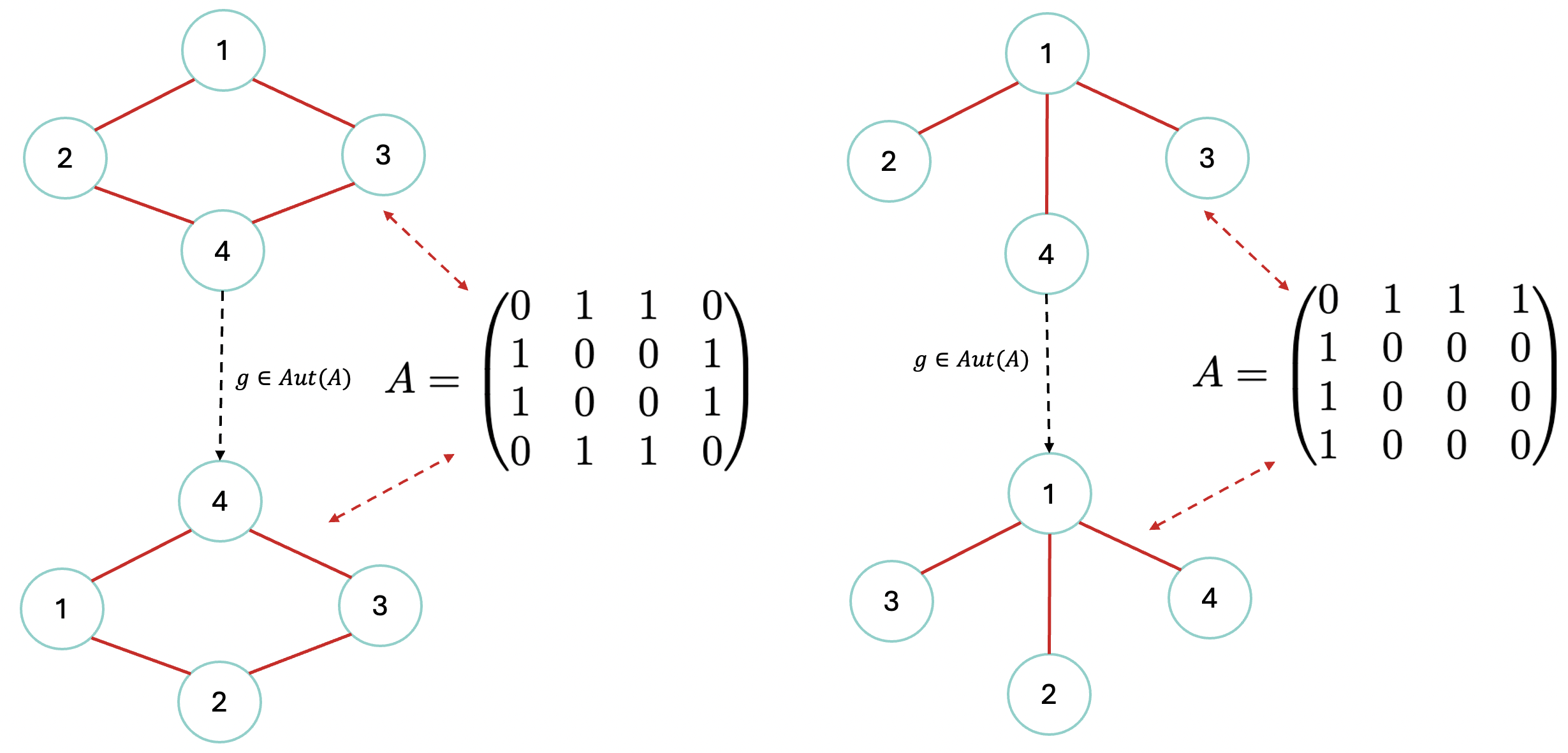}
    \caption{Examples of undirected graphs along with their adjacency matrices. Relabeling the vertices of a graph according to its automorphism group leaves the adjacency matrix unchanged. Adapted from \citet{dobriban2024symmpipredictiveinferencedata}.}
    \label{fig:symmPI}
\end{figure}

\paragraph{Problem setup.} We consider an illustration of the SymmPI framework for random variables whose symmetries are characterized by a graph structure. Let $Z = (Z_{1}, \ldots, Z_{n+1})^{\top} \in \mathcal{Z}_0^{n+1} =: \mathcal{Z}$. In this setting, we are given an undirected graph with adjacency matrix $A \in [0,\infty)^{(n+1)\times(n+1)}$. We assume the data $Z$ is distributionally invariant under the automorphism group $\mathrm{Aut}(A) \subseteq \rmS_{n+1}$ with the permutation action defined in \eqref{eq:permutation_action}. The elements of $\mathrm{Aut}(A)$ are permutations $g$ that satisfy $gAg^{\top} = A$ when viewed as linear maps $\mathbb{R}^{n+1} \rightarrow \mathbb{R}^{n+1}$. See Figure~\ref{fig:symmPI} for examples of the adjacency matrices of undirected graphs and their corresponding automorphism group actions. This framework generalizes the classical notion of exchangeability, which is recovered by taking the identity matrix $A = I_{n+1}$.

\paragraph{Transformation.} In machine learning, numerous graph neural network (GNN) architectures exhibiting deterministic equivariance have been developed, and a prominent class among them is the message-passing GNN \citep{gilmer2017neural, xu2018representation}. For a chosen depth $L$, we define the layers sequentially as $z^{0} := z$ and $z^{1}, \ldots, z^{L}$. For any $l \ge 0$ and any $i \in [n+1]$, the $i$-th coordinate of $z^{l+1}$ is computed by first aggregating information from the neighborhood $N(i)$ of node $i$ via a message function $\lambda_1$, and then applying a node update function $\lambda_0$:
$$
z_{i}^{l+1}
= \lambda_{0}\left(z_{i}^{l}, \sum_{j \in N(i)} \lambda_{1}(z_{i}^{l}, z_{j}^{l})\right)
:= F_{l}(z^{l})_{i}.
$$
The message-passing GNN is then defined as $\mathrm{MPGNN}_{L}(z) := F_{L} \circ F_{L-1} \circ \cdots \circ F_{1}(z)$ for all $z\in\cZ$. Since all nodes apply the same functions and aggregate neighbor information using a permutation-invariant operation, relabeling the nodes only relabels the outputs without changing their values. Therefore, any message-passing GNN is $\mathrm{Aut}(A)$-deterministically equivariant under the permutation action defined in \eqref{eq:permutation_action}, i.e., $\mathrm{MPGNN}_{L}(g \cdot z) = g \cdot \mathrm{MPGNN}_{L}(z)$, for all $g \in \mathrm{Aut}(A)$ and $z \in \mathcal{Z}$.

\paragraph{Threshold.}

In general, determining the automorphism group of a graph is a hard problem, and therefore we do not provide the details of the threshold calculation here. Instead, we refer readers to \citet{dobriban2024symmpipredictiveinferencedata} for a comprehensive discussion of how to compute the SymmPI threshold in this setting.

\subsection{Network Data Beyond Joint Exchangeability}
\label{sec:network_beyond_joint_exchangeability}

In this subsection, we consider a generalized setting for network-structured data that extends beyond the joint exchangeability assumption in \citet{lunde2023conformalpredictionnetworkassistedregression}. In this setting, we assume that the edges of the network are observed and that the adjacency matrix $A \in \RR^{(n+1)\times(n+1)}$ is fixed and known. For $1\leq i,j\leq n+1$, define $V_{ij} = (Y_i, Y_j, X_i, X_j)$, where $X_i \in \cX$ and $Y_i \in \RR$ denote the covariate and response associated with node $i$, respectively. We assume that the array $(V_{ij})_{1\leq i,j\leq n+1}$ is distributionally invariant under the automorphism group $\mathrm{Aut}(A)\subseteq\rmS_{n+1}$ of the graph with adjacency matrix $A$. This assumption is formally stated as follows.

\begin{assumption}
    \label{assumption:network_automorphism}
    For any $g \in \mathrm{Aut}(A)$, i.e., for any permutation $g$ satisfying $gAg^\top = A$, we have
    $$
    (V_{g^{-1}(i)g^{-1}(j)})_{1\leq i,j\leq n+1} \deq (V_{ij})_{1\leq i,j\leq n+1}.
    $$
\end{assumption}
Heuristically, Assumption~\ref{assumption:network_automorphism} posits that relabeling the nodes of the graph according to its automorphism group does not affect the joint distribution of the edge variables (see Figure~\ref{fig:symmPI} for an illustration). This assumption is less restrictive than the joint exchangeability condition in Assumption~\ref{assumption:joint_exchangeability} of Section~\ref{sec:network_structured_data_main}. For network statistics $(C_1, \dots, C_{n+1})$, we also consider a relaxation of Assumption~\ref{assumption:exchangeability_of_network_statistics}.
\begin{assumption}
    \label{assumption:network_statistics_automorphism}
    Let $(C_1, \dots, C_{n+1}) = \zeta(X_1, \dots, X_{n+1})$ for some function $\zeta$. For any $g\in\mathrm{Aut}(A)$, the network statistics $(C_1, \dots, C_{n+1})$ satisfy
    $$
    (C_{g^{-1}(1)}, \dots, C_{g^{-1}(n+1)}) = \zeta(X_{g^{-1}(1)}, \dots, X_{g^{-1}(n+1)}) \quad \text{a.s.}
    $$
\end{assumption}
Under Assumptions~\ref{assumption:network_automorphism} and~\ref{assumption:network_statistics_automorphism}, we can conclude that the triple $(X_i, C_i, Y_i)_{1\leq i\leq n+1}$ is distributionally invariant under the automorphism group $\mathrm{Aut}(A)$. This is a direct consequence of \citet[Proposition 2]{lunde2023conformalpredictionnetworkassistedregression}, and we omit the details here for brevity. Consequently, we can directly apply the SymmPI framework introduced in Section~\ref{sec:symmpi_framework} to construct valid predictive inference for the unobserved response $Y_{n+1}$ based on the observed data $\{(X_i, C_i, Y_i): 1\leq i\leq n\}$ and $(X_{n+1}, C_{n+1})$.

\section{Additional Details for Section~\ref{sec:main_results}}
\label{sec:method_details}

\subsection{RKHS Function Class}
\label{sec:rkhs_details}
This subsection collects details on the RKHS function class of Section~\ref{sec:rkhs}.

\paragraph{Basic properties.} Under Assumption~\ref{assumption:bounded_kernel}, the reproducing property and the Cauchy--Schwarz inequality give $|t(o)|=|\langle t,K(\cdot,o)\rangle_K|\leq\kappa\|t\|_K$ for all $t\in\cF_K$ and $o\in\cO$; in particular, $\|f\|_\infty\leq\kappa M$ for $f\in\cF_K$ with $\|f\|_K\leq M$. The objectives in \eqref{eq:hat_t_RKHS} and \eqref{eq:hat_t_RKHS_loo} are continuous, coercive and strongly convex on $\cF_K$, so their minimizers exist and are unique (Lemma~\ref{lem:RKHS_technical}).

\paragraph{Finite-dimensional reformulation.} Write $o_i:=o_i(z)$ and $s_i:=s_i(z)$ for the distinct orbit elements of Section~\ref{sec:distinct_orbit_elements}, and let $\bm{K}\in \RR^{[\cG:\cH]\times[\cG:\cH]}$ denote the Gram matrix with entries $\bm{K}_{ij}=K(o_i, o_j)$. By the representer theorem \citep{scholkopf2001generalized}, the minimizer $\hat{t}_z^{\,K}(\cdot)$ admits a finite expansion of the form $\hat{t}_z^{\,K}(o_i) = (\bm{K}c)_i$ for some coefficient vector $c \in \RR^{[\cG:\cH]}$, and its squared RKHS norm is given by $\|\hat{t}_z^{\,K}\|_K^2 = c^\top \bm{K} c$. Consequently, \eqref{eq:hat_t_RKHS} reduces to
$$
    \min_{c\in\RR^{[\cG:\cH]}} \ \frac{1}{[\cG:\cH]}\sum_{i=1}^{[\cG:\cH]}\ell_\alpha\bigl((\bm{K}c)_i,s_i\bigr)
    +\lambda\, c^\top \bm{K} c.
$$
This is a standard convex optimization problem that can be solved efficiently using existing algorithms, e.g., through the dual formulation in Section~\ref{sec:computational_considerations}.

\paragraph{Leave-one-out threshold and the density condition.} To control the interpolation error, we compare $\hat t_z^{\,K}$ with the threshold fitted without the test point. Since $g_1=e$, the $i=1$ term in \eqref{eq:hat_t_RKHS} is the test term $\ell_\alpha(t(z_\obs),\psi(\tilde z))$. We define the leave-one-out threshold
\begin{equation}
    \label{eq:hat_t_RKHS_loo}
    \hat{t}_z^{-e}(\cdot) = \argmin_{t(\cdot) \in \cF_{K}} \frac{1}{[\cG:\cH]} \sum_{i=2}^{[\cG:\cH]} \ell_\alpha\bigl(t(o_i(z)), s_i(z)\bigr) + \lambda\|t\|_K^2,
\end{equation}
which removes the test term while keeping the weights of the remaining terms unchanged; it is only used in the analysis. The penalty makes the threshold stable: by Lemma~\ref{lem:loo_stability}, $\|\hat t_z^{\,K}-\hat t_z^{-e}\|_\infty\leq\kappa^2/(2\lambda[\cG:\cH])$. Hence the test statistic rarely hits $\hat t_Z^{\,K}(Z_\obs)$ as long as it does not concentrate around $\hat t_Z^{-e}(Z_\obs)$, which is guaranteed by the following condition.
\begin{assumption}
    \label{assumption:conditional_density}
    There exists a $\sigma$-field $\cA$ such that $\hat t^{-e}_Z(Z_\obs)$ is $\cA$-measurable and, conditionally on $\cA$, the test statistic $\psi(\tilde Z)$ admits a density bounded above by a constant $b_\psi>0$ almost surely.
\end{assumption}
A natural choice is $\cA=\sigma\bigl(Z_\obs,\{(o_i(Z),s_i(Z))\}_{i\geq2}\bigr)$: since $\hat t^{-e}_Z$ is fitted only from the terms with $i\geq2$, $\hat t^{-e}_Z(Z_\obs)$ is $\cA$-measurable by construction. In the exchangeable setting, this gives $\cA=\sigma(X_{1:n+1},S_{1:n})$, and Assumption~\ref{assumption:conditional_density} reduces to a bounded conditional density of the score given the covariate (Corollary~\ref{cor:RKHS_iid}).

\subsection{Sampled C-SymmPI for Large or Infinite Groups}
\label{sec:sampled_adaptive_symmpi}

When $\cG$ is large or infinite (e.g., the rotation group $\mathrm{SO}(p)$), evaluating the group average in \eqref{eq:hat_t} becomes computationally costly or infeasible. To improve tractability, we approximate the expectation over $G\sim U$ by a Monte Carlo average. Specifically, we draw $G_1,\dots,G_N\simiid U$ independently of $Z$ and set $G_0:=e$. For each completion $z$ of $z_\obs$, the sampled threshold is
\begin{equation}
    \label{eq:hat_t_sampled}
    \hat{t}_{z,\samp}(\cdot) = \argmin_{t(\cdot)\in\cF}\frac{1}{N+1}\sum_{i=0}^{N}\ell_\alpha\bigl(t(\obs(G_i\cdot z)), \psi(G_i\cdot\tilde z)\bigr) + \cR(t),
\end{equation}
where the same draw $G_{1:N}=(G_1,\dots,G_N)$ is used for all completions, and, if the minimizer is not unique, it is selected by a rule that does not depend on the order of the $N+1$ terms. The \texttt{Sampled C-SymmPI} prediction set is
\begin{equation}
    \label{eq:symmpi_sampled}
    T_s^{\symmpi}(z_\obs) = \cbr{z \in \cZ : \psi(\tilde z) \leq \hat{t}_{z,\samp}(z_\obs),\ \obs(z) = z_\obs}.
\end{equation}
The following theorem establishes its near-conditional coverage; the proof is deferred to Section~\ref{sec:proof_sampled_coverage}.

\begin{theorem}
    \label{thm:sampled_coverage}
    Let $\cG$ be a compact metrizable group equipped with its Borel $\sigma$-field and the uniform distribution $U$, whose action on $\cZ$ is jointly measurable, and suppose $Z\deq G\cdot Z$ for $G\sim U$. Under the remaining conditions of Theorem~\ref{thm:coverage}, for every $N\geq0$ and every $f\in\cF$ with $\EE|f(Z_\obs)|<\infty$ and $\EE|D_f\cR(\hat t_{Z,\samp})|<\infty$, the prediction set \eqref{eq:symmpi_sampled} satisfies
    $$
        \abr{\EE_{Z, G_{1:N}}\Bigl[f(Z_{\obs})\Bigl(\1\{Z\in T_s^{\symmpi}(Z_{\obs})\} - (1 - \alpha)\Bigr)\Bigr]} \leq  \varepsilon_{\pen}^{\samp} + \varepsilon_{\inte}^{\samp},
    $$
    where
    $$
        \varepsilon_\pen^{\samp} = \abr{\EE_{Z, G_{1:N}}\left[D_{f}\cR(\hat{t}_{Z,\samp})\right]}, \quad
        \varepsilon_\inte^{\samp} = \mathbb{E}_{Z, G_{1:N}}\left[|f(Z_\obs)| \1\{\psi(\tilde Z) = \hat{t}_{Z,\samp}(Z_\obs)\}\right].
    $$
\end{theorem}

Theorem~\ref{thm:sampled_coverage} has the same form as Theorem~\ref{thm:coverage}, holds for every finite $N$, and does not require $\cG$ to be finite; the expectation is taken over both the data and the sampled group elements. The key ingredient is the term $i=0$ in \eqref{eq:hat_t_sampled}: because the completion itself is one of the $N+1$ terms, the test point is exchangeable with the terms used to fit the threshold. For the RKHS class, Theorem~\ref{thm:sampled_approximation} below further shows that the sampled prediction set agrees with the full one at the test point with probability tending to one as $N\to\infty$, under a density condition on the orbit.

\paragraph{Comparison with the full threshold.} We now compare \texttt{Sampled C-SymmPI} with the full method \eqref{eq:hat_t} for the RKHS class, where the group may be infinite. Let $\cG$ be a compact metrizable group with its Borel $\sigma$-field and uniform distribution $U$. The full threshold reads
\begin{equation}
    \label{eq:hat_t_RKHS_full}
    \hat t^{\,K}_z(\cdot)=\argmin_{t(\cdot)\in\cF_K}\EE_{G\sim U}\left[\ell_\alpha\bigl(t(\obs(G\cdot z)),\psi(G\cdot\tilde z)\bigr)\right]+\lambda\|t\|_K^2.
\end{equation}
We assume $\int|\psi(g\cdot\tilde z)|\,\diff U(g)<\infty$ for every $z\in\cZ$, so that the objective in \eqref{eq:hat_t_RKHS_full} is finite. Let $T^{\symmpi}_K(z_\obs)$ be the prediction set based on \eqref{eq:hat_t_RKHS_full}, and let $\hat t^{\,K}_{z,\samp}$ and $T^{\symmpi}_{K,s}(z_\obs)$ be the sampled threshold and prediction set with the same RKHS class and penalty. For a weighting function $f$, let
$$
\Delta_f:=\EE_Z\left[f(Z_\obs)\left(\1\{Z\in T^{\symmpi}_K(Z_\obs)\}-(1-\alpha)\right)\right],
$$
and let $\Delta_{f,\samp}(G_{1:N})$ be the same quantity with $T^{\symmpi}_K$ replaced by $T^{\symmpi}_{K,s}$, computed from a given draw $G_{1:N}$ (the expectation is over $Z$ only). Finally, for $u\geq0$ and $\delta\in(0,1]$, define the mass of the full residual near zero and a threshold radius,
$$
\beta_f(u):=\EE_Z\left[|f(Z_\obs)|\,\1\bigl\{|\psi(\tilde Z)-\hat t^{\,K}_Z(Z_\obs)|\leq u\bigr\}\right],\qquad
u_N(\delta):=\frac{\kappa^2}{2\lambda(N+1)}\Bigl(\sqrt N+2+\sqrt{2N\log(1/\delta)}\Bigr).
$$

For infinite groups, we control $\beta_f$ through the following condition on the orbit distribution.
\begin{assumption}
    \label{assumption:orbit_density}
    For $z\in\cZ$, let $P_z$ be the distribution of $\bigl(\obs(G\cdot z),\psi(G\cdot\tilde z)\bigr)$ for $G\sim U$, and write $P_z(A\times B)=\int_A q_z(o,B)\,\diff P_z^{\obs}(o)$ for Borel sets $A,B$, where $P_z^{\obs}$ is the first marginal of $P_z$ and $q_z$ is a probability kernel. There exists a measurable $b_{\rm orb}:\cZ\to[0,\infty)$ with $\bar b_{\rm orb}:=\EE\, b_{\rm orb}(Z)<\infty$ such that, for every $z$ and $o$, $q_z(o,\cdot)$ admits a Lebesgue density bounded by $b_{\rm orb}(z)$.
\end{assumption}
Assumption~\ref{assumption:orbit_density} requires that, along the orbit of each $z$, the score retains randomness given the observed part. It cannot hold for finite groups, but it may hold for continuous groups such as $\mathrm{SO}(p)$.

\begin{theorem}
    \label{thm:sampled_approximation}
    Let $f:\cO\to\RR$ be bounded and measurable, and let $B_f:=\sup_{o\in\cO}|f(o)|$. Then, without requiring the distributional invariance of $Z$, for every $N\geq1$ and $\delta\in(0,1]$,
    \begin{equation}
        \label{eq:sampled_disagreement}
        \EE_{Z,G_{1:N}}\left[|f(Z_\obs)|\,\bigl|\1\{Z\in T^{\symmpi}_K(Z_\obs)\}-\1\{Z\in T^{\symmpi}_{K,s}(Z_\obs)\}\bigr|\right]\leq\beta_f\bigl(u_N(\delta)\bigr)+B_f\delta.
    \end{equation}
    Consequently, $\bigl|\Delta_f-\EE_{G_{1:N}}\Delta_{f,\samp}(G_{1:N})\bigr|\leq\beta_f(u_N(\delta))+B_f\delta$, and for every $\eta\in(0,1)$, with probability at least $1-\eta$ over $G_{1:N}$, $|\Delta_f-\Delta_{f,\samp}(G_{1:N})|\leq\beta_f(u_N(\delta))+B_f\delta/\eta$. Moreover:
    \begin{itemize}
        \item[(a)] if $Z\deq G\cdot Z$ for $G\sim U$ and Assumption~\ref{assumption:orbit_density} holds, then $\beta_f(u)\leq 2\bar b_{\rm orb}B_f\,u$;
        \item[(b)] if $\cG$ is finite and Assumption~\ref{assumption:conditional_density} holds, then $\beta_f(u)\leq 2B_fb_\psi\bigl(u+\kappa^2/(2\lambda[\cG:\cH])\bigr)$.
    \end{itemize}
\end{theorem}
The proof is provided in Section~\ref{sec:proof_sampled_approximation}. Theorem~\ref{thm:sampled_approximation} bounds how often the full and the sampled prediction sets make different decisions at the test point, through the distance between the two thresholds and the mass of the full residual near zero. In case (a), taking $\delta=1/(N+1)$ gives a bound of order $\bar b_{\rm orb}B_f\kappa^2\lambda^{-1}\sqrt{\log N/N}+B_f/(N+1)$, so the two procedures agree at the test point with probability tending to one as $N\to\infty$. In case (b), the bound does not vanish as $N\to\infty$ for fixed $[\cG:\cH]$.

\subsection{Dual Formulation for Efficient Computation}
\label{sec:computational_considerations}

In this subsection, we present a dual formulation of the optimization problem in \eqref{eq:hat_t} to enable efficient computation of the prediction region. This type of duality-based approach has been previously explored by \citet{gibbs2025conformal}, who considered quantile regression with regularization under uniform weights. Our approach generalizes their framework to accommodate \emph{arbitrary} weights, thereby extending it to the broader setting of weighted quantile regression with regularization.

\paragraph{Dual formulation.}

Consider a dataset $z = ((x_1, y_1), \dots, (x_n, y_n)) \in (\cX \times \RR)^n$, where the covariates $x_1, \dots, x_n$ are fully observed, the responses $y_1, \dots, y_{n-1}$ are observed, and $y_n$ is unobserved. Define the score for each observation by $s_i = s(x_i, y_i)$ for some score function $s(\cdot, \cdot)$. Accordingly, $s_1, \dots, s_{n-1}$ are fixed, while $s_n$ is treated as a parameter. We then formulate the following regularized, weighted quantile regression problem, whose decision variable is a function $t(\cdot) \in \{\cX \to \RR\}$:

\begin{equation}
    \label{eq:opt_main}
    \minimize \quad \sum_{i=1}^{n} w_i\cdot \ell_{\alpha}(t(x_i), s_i) + \cR(t).
\end{equation}
Here, $w = [w_i]_{1\leq i\leq n}$ is a vector of non-negative weights, which we assume are normalized such that $\sum_{i=1}^{n} w_i = 1$. The following proposition presents the dual formulation of this optimization problem, whose proof is provided in Section~\ref{sec:proof_dual_problem}.

\begin{proposition}
    \label{prop:dual_problem}
    The dual problem of the optimization problem in (\ref{eq:opt_main}) is given by
    \begin{equation}
        \label{eq:dual_problem}
        \begin{aligned}
        &\maximize \quad (w\odot\lambda)^\top s - \cR^*(w\odot\lambda) \\
        &\st \quad -\alpha\one \preceq \lambda \preceq (1-\alpha)\one,
        \end{aligned}
    \end{equation}
    where $\lambda\in \RR^{n}$ is the dual variable, and $\cR^*(\alpha) := \sup_{t} \cbr{\sum_{i=1}^{n} \alpha_i t(x_i) - \cR(t)}$ is the Fenchel conjugate of $\cR$.\footnote{$w\odot\lambda:=[w_i\lambda_i]_{i=1}^{n}$ denotes the Hadamard product of vectors $w$ and $\lambda$, $\one$ denotes the vector of all ones, and ``$\preceq$'' denotes element-wise inequality.}
\end{proposition}

\begin{remark}
The Fenchel conjugate $\cR^*(\cdot)$ often admits a closed-form expression for many commonly used regularizers. As an illustrative example, consider the RKHS setting. Let $K \in \RR^{n\times n}$ be the Gram matrix with entries $K_{ij} = K(x_i, x_j)$, where common choices of the kernel function $K$ include the linear, polynomial, and Gaussian RBF kernels. By the representer theorem \citep{scholkopf2001generalized}, any function $t$ in the RKHS admits a finite expansion of the form $t(x_i) = (K c)_i$ for some coefficient vector $c = (c_1, \dots, c_{n})^\top \in \RR^{n}$, and its squared RKHS norm is given by $\|t\|_K^2 = c^\top K c$. Consequently, the conjugate is expressed as $
\cR^*(\alpha) = \sup_{c \in \RR^{n}} \{\alpha^\top K c - \lambda c^\top K c\}$, and solving this quadratic maximization problem yields
$$
\cR^*(\alpha) = \frac{1}{4\lambda}\alpha^\top K \alpha.
$$
In the experiments presented in Section~\ref{sec:experiments}, we employ this RKHS kernel to enable efficient computation.
\end{remark}

Let $\hat{t}_z(\cdot)\in\{\cX\to\RR\}$ and $\hat{\lambda} = [\hat{\lambda}_i]_{1\leq i\leq n}$ denote the primal and dual optimal solutions, respectively. Recall that the scores $s_1,\dots,s_{n-1}$ are fixed, while $s_n$ is treated as a parameter. Consequently, the optimal dual solution $\hat{\lambda}$ is a function of $s_n$. Among the dual variables, our primary interest lies in the $n$-th component $\hat{\lambda}_n = \hat{\lambda}_n(s_n)$. The following proposition identifies the precise form of $\hat{\lambda}_n(s_n)$, whose proof is provided in Section~\ref{sec:proof_primal_dual_solution}.

\begin{proposition}
    \label{prop:primal_dual_solution}
    Let $\hat{t}_z:\cX\to\RR$ and $\hat{\lambda}(s_n)=[\hat{\lambda}_i(s_n)]_{1\leq i\leq n}$ denote the optimal primal and dual solutions to \eqref{eq:opt_main}. Then $\hat{\lambda}_n(s_n)$ satisfies:
    $$
    \hat{\lambda}_{n}(s_n)\in\begin{cases}
    -\alpha & \mathrm{if\ } s_n < \hat{t}_z(x_n), \\
    [-\alpha, 1-\alpha] & \mathrm{if\ } s_n = \hat{t}_z(x_n), \\
    1-\alpha & \mathrm{if\ } s_n > \hat{t}_z(x_n).
    \end{cases}
    $$
\end{proposition}

\begin{remark}
    For the prediction set $\hat{C}(x_n) = \{y_{n}: s_n\leq \hat{t}_{z}(x_n)\}$, Proposition~\ref{prop:primal_dual_solution} reveals that evaluating the condition $s_n\leq \hat{t}_{z}(x_n)$ is nearly equivalent to verifying whether $\hat{\lambda}_n(s_n)<1-\alpha$. This motivates the following dual representation of the prediction region:
    \begin{equation}
        \label{eq:dual_prediction_region}
        \hat{C}^{*}(x_n) = \{y_{n}: \hat{\lambda}_n(s_n)<1-\alpha\}.
    \end{equation}
    Under strong duality, \citet{gibbs2025conformal} show that $\hat{C}^{*}(x_n)$ attains the same near-conditional coverage guarantee as $\hat{C}(x_n)$.
\end{remark}

\paragraph{Efficient computation.}

To enable efficient computation of $\hat{C}^{*}(x_n)$ in \eqref{eq:dual_prediction_region}, the following proposition establishes a key monotonicity property that facilitates fast search procedures. Its proof is provided in Section~\ref{sec:proof_dual_solution_monotone}.

\begin{proposition}
    \label{prop:dual_solution_monotone}
        The function $\hat{\lambda}_n(\cdot)$ is a monotonically increasing function.
\end{proposition}

Leveraging this monotonicity, we can recover the prediction region by identifying the largest value of $s_n$ for which the dual feasibility condition holds. Specifically, we compute $s^* = \sup\{s_n : \hat{\lambda}_n(s_n) < 1-\alpha\}$ using binary search and return the prediction region $\hat{C}^{*}(x_n) = \{y_n : s_n \leq s^*\}$. The detailed algorithmic steps are provided in Algorithm~\ref{alg:dual_prediction_region}.

\begin{algorithm}[ht]
\caption{Computation of the prediction region via dual formulation}
\label{alg:dual_prediction_region}
\begin{algorithmic}[1]
\Require Observed data $z = ((x_1, y_1), \dots, (x_{n-1}, y_{n-1}), x_n)$, score function $s(\cdot, \cdot)$, miscoverage level $\alpha$, weights $w = [w_i]_{1\leq i\leq n}$, regularizer $\cR(\cdot)$, tolerance $\varepsilon > 0$.
\State Initialize $s_{\min}$ and $s_{\max}$ as lower and upper bounds for $s_n$.
\While{$s_{\max} - s_{\min} > \varepsilon$} \Comment{Binary search.}
    \State Set $s_n = (s_{\min} + s_{\max})/2$.
    \State Solve the dual problem in \eqref{eq:dual_problem} to obtain $\hat{\lambda}_n(s_n)$.
    \If{$\hat{\lambda}_n(s_n) < 1 - \alpha$}
        \State Set $s_{\min} = s_n$.
    \Else
        \State Set $s_{\max} = s_n$.
    \EndIf
\EndWhile
\State Set $s^* = s_{\min}$. \Comment{An approximation of $\sup\{s_n : \hat{\lambda}_n(s_n) < 1-\alpha\}$.}
\Ensure Prediction region $\hat{C}^{*}(x_n) = \{y_n : s(x_n, y_n) \le s^*\}$.
\end{algorithmic}
\end{algorithm}

\begin{remark}
    In the final step of Algorithm~\ref{alg:dual_prediction_region}, the dual formulation enables efficient recovery of the prediction region when $s_n = s(x_n, y_n)$ is defined as a non-conformity score, such as $s_n = |\hat{\mu}(x_{n+1}) - y_{n+1}|$, where $\hat{\mu}(x_{n+1})$ denotes the predicted response. In this case, once $s^*$ is obtained, the prediction region immediately reduces to the closed-form interval $\sbr{\hat{\mu}(x_{n+1}) - s^*, \hat{\mu}(x_{n+1}) + s^*}$.
\end{remark}

\section{Proofs}
\label{sec:proofs}

\subsection{Proofs for Section~\ref{sec:preliminaries}}
\label{sec:proofs_sec2}

\subsubsection{Proof of Proposition~\ref{prop:conditional_coverage}}
\label{sec:proof_conditional_coverage}
For (i) implies (ii), taking conditional expectation with respect to $Z_{\obs}$ on (ii), we have
\begin{align*}
    &\quad\ \EE\left[f(Z_{\obs})\left(\1\{Z\in \hat{C}(Z_{\obs})\} - (1-\alpha)\right)\right]\\
    &= \EE\left[f(Z_{\obs})\EE\left[\1\{Z\in \hat{C}(Z_{\obs})\} - (1-\alpha)\mid Z_{\obs}\right]\right]\\
    &= \EE\left[f(Z_{\obs})\rbr{\PP\left(Z\in \hat{C}(Z_{\obs})\mid Z_{\obs}\right) -\rbr{1-\alpha}}\right] = 0,
\end{align*}
where the last equality follows from assumption (i).

For (ii) implies (i), the above calculation implies that for all bounded non-negative $f:\cO\to\RR$, we have
$$
\EE\left[f(Z_{\obs})\rbr{\PP\left(Z\in \hat{C}(Z_{\obs})\mid Z_{\obs}\right) -\rbr{1-\alpha}}\right] = 0.
$$
Take $f = \1_B$ for any measurable set $B\subseteq \cO$ and define $g(Z_{\obs}) = \PP\left(Z\in \hat{C}(Z_{\obs})\mid Z_{\obs}\right) -\rbr{1-\alpha}$.
Then for any measurable set $B\subseteq \cO$, we have
$$
\EE\left[\1_B(Z_{\obs})g(Z_{\obs})\right] = \int_{B}g(z_{\obs})\diff \PP_{Z_{\obs}}(z_{\obs}) = 0.
$$
Suppose there is a measurable set $B_+\subseteq \cO$ with positive measure such that $g(z_{\obs})>0$ on $B_+$. Then choosing $B = B_+$ gives
$$\int_{B_+}g(z_{\obs})\diff \PP_{Z_{\obs}}(z_{\obs}) > 0,$$
which contradicts the above equation. Similar results can be shown for the case when $g(z_{\obs})<0$ on some measurable set $B_-$.
Therefore, we conclude that $g(z_{\obs}) = 0$ almost surely, which is equivalent to (i).
This completes the proof.
\qed

\subsection{Proofs for Section~\ref{sec:theoretical_properties}}
\label{sec:proofs_sec32}

\subsubsection{Proof of Theorem~\ref{thm:coverage}}
\label{sec:proof_coverage}

Recall that $\tilde{Z}=V(Z)$, and let $|\cG|$ denote the cardinality of the finite group $\cG$. For notational convenience, define $Z_i := g_i\cdot Z$, $\tilde{Z}_i := g_i\cdot\tilde{Z}$, and $Z_i^\obs := \obs(Z_i)$ for each $i = 1, \dots, |\cG|$; by deterministic equivariance, $\tilde Z_i=V(Z_i)$. With these definitions, \eqref{eq:hat_t} can be equivalently rewritten as

\begin{equation}
    \label{eq:hat_t_rewrite}
    \hat{t}_{Z}(\cdot) = \argmin_{t(\cdot) \in \cF} \frac{1}{|\cG|}\sum_{i=1}^{|\cG|} \ell_\alpha\bigl(t(Z_i^\obs), \psi(\tilde{Z}_i)\bigr) + \cR(t).
\end{equation}

It is easy to verify that the pinball loss $\ell_\alpha(\theta, S) := \alpha[\theta - S]_+ + (1 - \alpha)[S - \theta]_+$ is convex in its first argument. Since $t\mapsto t(Z_i^\obs)$ is linear, each loss term in \eqref{eq:hat_t_rewrite} is convex in $t$, and by Assumption~\ref{assumption:regularization}, so is $\cR$. Therefore, \eqref{eq:hat_t_rewrite} is a convex optimization problem over the linear space $\cF$.

\paragraph{First-order condition.} Fix any $f\in\cF$. Since $\cF$ is a linear space, $\hat t_Z+\varepsilon f\in\cF$ for all $\varepsilon\in\RR$. Hence the function
$$
H(\varepsilon):=\frac{1}{|\cG|}\sum_{i=1}^{|\cG|}\ell_\alpha\bigl(\hat t_Z(Z_i^\obs)+\varepsilon f(Z_i^\obs), \psi(\tilde{Z}_i)\bigr)+\cR(\hat t_Z+\varepsilon f)
$$
is a finite convex function on $\RR$ that is minimized at $\varepsilon=0$, so its one-sided derivatives satisfy $H'_-(0)\leq 0\leq H'_+(0)$. One-sided derivatives of finite convex functions on $\RR$ are additive, and by Assumption~\ref{assumption:regularization}(ii) the regularizer contributes $D_f\cR(\hat t_Z)$ to both of them. Let $A := \{1 \leq i \leq |\cG| : \psi(\tilde{Z}_i) = \hat{t}_{Z}(Z_i^\obs)\}$ be the index set of interpolation points. For $i\notin A$, the $i$-th loss term is differentiable at $\varepsilon=0$ with derivative $f(Z_i^\obs)\bigl(\alpha-\1\{\psi(\tilde Z_i)>\hat t_Z(Z_i^\obs)\}\bigr)$. For $i\in A$, the $i$-th loss term equals $\alpha[\varepsilon f(Z_i^\obs)]_++(1-\alpha)[-\varepsilon f(Z_i^\obs)]_+$, whose left and right derivatives at $\varepsilon=0$ are the endpoints of the interval $\{s_if(Z_i^\obs):s_i\in[\alpha-1,\alpha]\}$, regardless of the sign of $f(Z_i^\obs)$. Consequently,
\begin{align*}
[H'_-(0),H'_+(0)]=\Biggl\{&\frac{1}{|\cG|}\left(\sum_{i\in A^C} f(Z_i^\obs)\left(\alpha - \1\{\psi(\tilde{Z}_i) > \hat{t}_{Z}(Z_i^\obs)\}\right) + \sum_{i\in A}s_if(Z_i^\obs)\right)\\
&+D_f\cR(\hat t_Z): s_i\in[\alpha-1, \alpha]\text{ for } i\in A\Biggr\}.
\end{align*}
Since this interval contains $0$, there exist coefficients $s_i^* \in [\alpha - 1, \alpha]$, $i\in A$, for which the expression inside the braces vanishes. Noting that $\1\{\psi(\tilde{Z}_i) > \hat{t}_{Z}(Z_i^\obs)\}=0$ for $i\in A$, we obtain
\begin{equation}
    \label{eq:optimality_condition}
    \frac{1}{|\cG|}\sum_{i=1}^{|\cG|} f(Z_i^\obs)\left(\alpha - \1\{\psi(\tilde{Z}_i) > \hat{t}_{Z}(Z_i^\obs)\}\right) = - D_f\cR(\hat{t}_{Z}) + \frac{1}{|\cG|} \sum_{i\in A}(\alpha - s_i^*)f(Z_i^\obs).
\end{equation}

\paragraph{Invariance.} We claim that for all $1\leq i\leq |\cG|$,
\begin{equation}
    \label{eq:distributional_invariance}
        \EE_{Z}\left[f(Z_i^\obs)\left(\alpha - \1\{\psi(\tilde{Z}_i) > \hat{t}_{Z}(Z_i^\obs)\}\right)\right] = \EE_{Z}\left[f(Z_\obs)\left(\alpha - \1\{\psi(\tilde{Z}) > \hat{t}_{Z}(Z_\obs)\}\right)\right].
\end{equation}

To prove this claim, we first show that $f(Z_i^\obs)\left(\alpha - \1\{\psi(\tilde{Z}_i) > \hat{t}_{Z}(Z_i^\obs)\}\right)$ is a function of $Z_i$. Indeed, $\tilde{Z}_i = V(Z_i)$. Moreover, recall the definition of $\hat{t}_Z(\cdot)$ in \eqref{eq:hat_t}, and it holds that
\begin{align*}
    \hat{t}_{g\cdot Z}(\cdot) &= \argmin_{t(\cdot)\in\cF}\EE_{G\sim U}\left[\ell_\alpha\bigl(t(\obs(G\cdot(g\cdot Z))), \psi(G\cdot V(g\cdot Z))\bigr)\right] + \cR(t) \\
    &= \argmin_{t(\cdot)\in\cF}\EE_{G\sim U}\left[\ell_\alpha\bigl(t(\obs((Gg)\cdot Z)), \psi((Gg)\cdot\tilde Z)\bigr)\right] + \cR(t) \\
    &= \argmin_{t(\cdot)\in\cF}\EE_{G\sim U}\left[\ell_\alpha\bigl(t(\obs(G\cdot Z)), \psi(G\cdot\tilde Z)\bigr)\right] + \cR(t) \\
    &= \hat{t}_{Z}(\cdot),
\end{align*}
where the first equality follows from the definition of $\hat{t}_Z(\cdot)$, the second equality follows from the definition of the action and the $\cG$-deterministic equivariance property of $V$, the third equality follows from the invariance property of the uniform distribution $U$, and the last equality follows from the definition of $\hat{t}_Z(\cdot)$ again. In the last step, the objective functions for $g\cdot Z$ and $Z$ coincide as functions on $\cF$, so the selected minimizers coincide as well (if the minimizer is not unique, this uses the selection rule in Section~\ref{sec:prediction_region}). Therefore, we write
$$f(Z_i^\obs)\left(\alpha - \1\{\psi(\tilde{Z}_i) > \hat{t}_{Z}(Z_i^\obs)\}\right) = f(\obs(Z_i))\left(\alpha - \1\{\psi(V(Z_i)) > \hat{t}_{Z_i}(\obs(Z_i))\}\right):=\varphi(Z_i),$$
which shows that the randomness in this expression lies only in $Z_i$.

Next, notice that for all $g\in \cG$, we have
$$
    g\cdot Z \stackrel{d}{=} g\cdot(G\cdot Z) = (gG)\cdot Z \stackrel{d}{=} G\cdot Z \stackrel{d}{=} Z,
$$
where the first equality follows from distributional invariance property of $Z$, the second equality follows from the definition of the action, the third equality follows from the invariance property of the uniform distribution $U$,
and the last equality also follows from the distributional invariance. Therefore, for the function $\varphi(\cdot)$ defined above, we have $\varphi(Z_i) \deq \varphi(Z)$; since $|\varphi|\leq|f\circ\obs|$ is integrable, this directly implies \eqref{eq:distributional_invariance}.

\paragraph{Coverage error.} We next proceed to compute coverage error terms:
\begin{equation}\label{eq:main_step}
    \begin{split}
        &\quad\ \EE_{Z}\left[f(Z_\obs)\left(\1\{Z\in T^{\symmpi}(Z_{\obs})\}-(1-\alpha)\right)\right]\\
        &= \EE_{Z}\left[f(Z_\obs)\left(\1\{\psi(\tilde{Z})\leq \hat{t}_{Z}(Z_\obs)\} - (1-\alpha)\right)\right] \\
        &= \EE_{Z}\left[f(Z_\obs)\left(\alpha - \1\cbr{\psi(\tilde{Z}) > \hat{t}_{Z}(Z_\obs)}\right)\right]\\
        &= \EE_{Z}\left[\frac{1}{|\cG|}\sum_{i=1}^{|\cG|} f(Z_i^\obs)\left(\alpha - \1\{\psi(\tilde{Z}_i) > \hat{t}_{Z}(Z_i^\obs)\}\right)\right],
    \end{split}
\end{equation}
where the first equality follows from Algorithm~\ref{alg:symmpi}, and the third equality follows from \eqref{eq:distributional_invariance}. Since $\alpha - s_i^* \in [0, 1]$, \eqref{eq:optimality_condition} implies the pointwise bound
\begin{equation}
    \label{eq:pointwise_bound}
    \left|\frac{1}{|\cG|}\sum_{i=1}^{|\cG|} f(Z_i^\obs)\left(\alpha - \1\{\psi(\tilde{Z}_i) > \hat{t}_{Z}(Z_i^\obs)\}\right)+D_f\cR(\hat{t}_{Z})\right|\leq\frac{1}{|\cG|}\sum_{i\in A} |f(Z_i^\obs)|,
\end{equation}
which holds for every realization of $Z$. Moreover, by the same reasoning as \eqref{eq:distributional_invariance} applied to the function $z\mapsto|f(\obs(z))|\1\{\psi(\tilde z)=\hat t_z(\obs(z))\}$,
\begin{equation}\label{eq:interpolation}
\begin{split}
    \EE_{Z}\left[\frac{1}{|\cG|}\sum_{i\in A} |f(Z_i^\obs)|\right] &= \EE_{Z}\left[\frac{1}{|\cG|}\sum_{i=1}^{|\cG|} |f(Z_i^\obs)|\1\{\psi(\tilde{Z}_i) = \hat{t}_{Z}(Z_i^\obs)\}\right] \\
    &= \EE_{Z}\left[|f(Z_\obs)|\1\left\{\psi(\tilde{Z}) = \hat{t}_{Z}(Z_\obs)\right\}\right].
\end{split}
\end{equation}

Finally, combining \eqref{eq:main_step}, \eqref{eq:pointwise_bound} and \eqref{eq:interpolation}, and using that $D_f\cR(\hat t_Z)$ is integrable, we obtain
$$
    \left|\EE_{Z}\left[f(Z_\obs)\left(\1\{Z\in T^{\symmpi}(Z_{\obs})\}-(1-\alpha)\right)\right]+\EE_Z\left[D_f\cR(\hat{t}_{Z})\right]\right| \leq \EE_{Z}\left[|f(Z_\obs)|\1\left\{\psi(\tilde{Z}) = \hat{t}_{Z}(Z_\obs)\right\}\right].
$$
The triangle inequality completes the proof.
\qed

\subsubsection{Proof of Theorem~\ref{thm:distribution_shift}}
\label{sec:proof_distribution_shift}
Recall that $Z_G := G\cdot Z$, $\tilde{Z}_G := V(Z_G)$, and $Z_G^\obs := \obs(Z_G)$, where $G\sim U$ is independent of $Z$. Following the same reasoning as in the first two steps of \eqref{eq:main_step}, we have
$$\EE_{Z}\left[f(Z_\obs)\left(\1\{Z\in T^{\symmpi}(Z_{\obs})\}-(1-\alpha)\right)\right] = \EE_{Z}\left[f(Z_\obs)\left(\alpha - \1\cbr{\psi(\tilde{Z}) > \hat{t}_{Z}(Z_\obs)}\right)\right].$$
Define
$$
\varphi(z) = f(\obs(z))\left(\alpha - \1\{\psi(\tilde z) > \hat{t}_z(\obs(z))\}\right),
$$
which is bounded since $f$ is bounded. We now decompose the coverage error into two terms:
\begin{equation}
    \label{main_estimate}
    \EE_{Z}\left[f(Z_\obs)\left(\1\{Z\in T^{\symmpi}(Z_{\obs})\}-(1-\alpha)\right)\right]
    = \underbrace{\EE_{Z}\left[\varphi(Z)\right] - \EE_{G,Z}\left[\varphi(Z_G)\right]}_{(\text{I})}
    + \underbrace{\EE_{G,Z}\left[\varphi(Z_G)\right]}_{(\text{II})}.
\end{equation}

We begin by analyzing term (II). Although $Z$ may not satisfy the distributional invariance property, $Z_G$ always does. Indeed, let $G'\sim U$ be independent of $(G,Z)$. Then
$$
G'\cdot Z_G = (G'G)\cdot Z \stackrel{d}{=} G\cdot Z = Z_G,
$$
where the equality follows from the definition of the action, and the distributional equality holds because, conditionally on $G'=g'$, the element $g'G$ is distributed according to $U$ and independent of $Z$, by the invariance property of the uniform distribution $U$. Moreover, by the definition of $\varphi$,
$$
(\text{II}) = \EE_{G,Z}\left[f(Z_G^\obs)\left(\1\{Z_G\in T^{\symmpi}(Z_G^\obs)\}-(1-\alpha)\right)\right],
$$
which is exactly the coverage error in Theorem~\ref{thm:coverage} with $Z$ replaced by $Z_G$. Since $f$ is bounded and $\EE_{G,Z}|D_f\cR(\hat t_{Z_G})|<\infty$, Theorem~\ref{thm:coverage} applies to $Z_G$ and yields
\begin{equation}
    \label{eq:distribution_shift_II}
    |(\text{II})| \leq \abr{\EE_{G,Z}\left[D_{f}\cR(\hat{t}_{Z_G})\right]} + \EE_{G,Z}\left[|f(Z_G^\obs)|\1\{\psi(\tilde{Z}_G) = \hat{t}_{Z_G}(Z_G^\obs)\}\right] = \varepsilon_{\pen} + \varepsilon_{\inte}.
\end{equation}

We now turn to term (I). Recall that $\PP$ and $\QQ$ denote the distributions of $Z$ and $Z_G$, respectively, and write $\mu:=\PP\circ\xi_f^{-1}$ and $\nu:=\QQ\circ\xi_f^{-1}$ for the distributions of $\xi_f(Z)$ and $\xi_f(Z_G)$ on $\RR\times\{0,1\}$. The key observation is that $\varphi$ depends on $z$ only through $\xi_f(z)$: defining $h(a,b):=a(\alpha-b)$ for $(a,b)\in\RR\times\{0,1\}$, we have $\varphi=h\circ\xi_f$, and hence
$$
(\text{I}) = \EE_Z\bigl[h(\xi_f(Z))\bigr]-\EE_{G,Z}\bigl[h(\xi_f(Z_G))\bigr] = \int h\,\diff\mu-\int h\,\diff\nu.
$$
To bound this difference, note that the first coordinate of $\xi_f(z)$ is $f(\obs(z))\in[-\|f\|_\infty,\|f\|_\infty]$ and the second is $0$ or $1$. Hence $h(\xi_f(z))$ lies, for every $z\in\cZ$, in the interval
$$
[a_-,a_+]=
\begin{cases}
[(\alpha-1)\|f\|_\infty,\ \alpha\|f\|_\infty], & \text{if } f\geq0,\\
[-\alpha\|f\|_\infty,\ (1-\alpha)\|f\|_\infty], & \text{if } f\leq0,\\
[-c\,\|f\|_\infty,\ c\,\|f\|_\infty] \text{ with } c:=\max(\alpha,1-\alpha), & \text{otherwise},
\end{cases}
$$
whose length satisfies $a_+-a_-\leq C_f\|f\|_\infty$ in each case. We can therefore compare the two expectations through the layer-cake formula: any random variable $W$ with values in $[a_-,a_+]$ satisfies
$$
\EE[W]=a_-+\int_{a_-}^{a_+}\PP(W>s)\,\diff s.
$$
Let $B_s:=\{x\in\RR\times\{0,1\}:h(x)>s\}$, which is a Borel set since $h$ is continuous in its first argument, so that $\PP(h(\xi_f(Z))>s)=\mu(B_s)$ and $\PP(h(\xi_f(Z_G))>s)=\nu(B_s)$. Applying the formula to $h(\xi_f(Z))$ and $h(\xi_f(Z_G))$ and subtracting, the constant $a_-$ cancels, and
$$
(\text{I}) = \int_{a_-}^{a_+}\bigl(\mu(B_s)-\nu(B_s)\bigr)\,\diff s.
$$
Since $|\mu(B_s)-\nu(B_s)|\leq\TV(\mu,\nu)$ for every $s$, we conclude that
\begin{equation}
    \label{eq:distribution_shift_I}
    |(\text{I})| \leq (a_+-a_-)\,\TV(\mu,\nu)\leq C_f\|f\|_\infty\,\TV\bigl(\PP\circ\xi_f^{-1},\QQ\circ\xi_f^{-1}\bigr).
\end{equation}

Finally, taking the absolute value of both sides of \eqref{main_estimate}, applying the triangle inequality, and combining \eqref{eq:distribution_shift_I} and \eqref{eq:distribution_shift_II}, we complete the proof.
\qed

\subsection{Proofs for Section~\ref{sec:choice_of_function_class}}
\label{sec:proofs_sec33}

\subsubsection{Proof of Theorem~\ref{thm:linear_class}}
\label{sec:proofs_linear_class}

Throughout the proof, we write $o_i:=o_i(Z)$ and $s_i:=s_i(Z)$ for the distinct orbit elements of Section~\ref{sec:distinct_orbit_elements}, and $\varphi_i:=\varphi(o_i)$ for their features, $i=1,\dots,[\cG:\cH]$, so that $\Phi$ has rows $\varphi_i^\top$ and $\Psi$ has entries $s_i$.

Since $\cF_L$ is a linear space containing $f$ and $\cR\equiv0$, so that $\varepsilon_\pen=0$, Theorem~\ref{thm:coverage} bounds the coverage error by the interpolation error $\varepsilon_\inte$ alone. By \eqref{eq:interpolation}, this error is an average over $g\sim U$ in which all elements of a right coset $\cH g_i$ produce the same term. Each coset has probability $1/[\cG:\cH]$ under $U$, so grouping the group elements by cosets as in \eqref{eq:hat_t_distinct} gives
\begin{equation}
    \label{eq:main_step_linear}
    \left|\EE_{Z}\left[f(Z_\obs)\left(\1\{Z\in T^{\symmpi}_L(Z_{\obs})\}-(1-\alpha)\right)\right]\right| \leq \frac{1}{[\cG:\cH]}\,\EE_{Z}\Bigl[\sum_{i\in A}|f(o_i)|\Bigr],
\end{equation}
where $A$ is the set of orbit elements that lie exactly on the fitted threshold,
$$
A:=\bigl\{i:\ s_i=\langle\hat\theta_Z,\varphi_i\rangle\bigr\}.
$$

The right-hand side is small because each orbit element carries weight $1/[\cG:\cH]$ and only few of them are interpolated. Indeed, $|f|\leq b_\theta b_\varphi$ by the Cauchy--Schwarz inequality and Assumption~\ref{assumption:bounded_feature_map}, so
$$
\frac{1}{[\cG:\cH]}\,\EE_{Z}\Bigl[\sum_{i\in A}|f(o_i)|\Bigr]\leq\frac{b_\theta b_\varphi}{[\cG:\cH]}\,\EE_Z|A|,
$$
and it remains to show that at most $d$ orbit elements are interpolated, i.e., $|A|\leq d$ almost surely.

Suppose that some $d+1$ orbit elements, indexed by a set $J$, are interpolated. Writing $\Phi_J$ and $\Psi_J$ for the rows of $\Phi$ and the entries of $\Psi$ indexed by $J$, we then have
$$
\Psi_J=\Phi_J\hat\theta_Z,
$$
so the scores $\Psi_J$ lie in the column space of $\Phi_J$. Since there are only finitely many sets $J$, it suffices to show that, for each fixed $J$ with $|J|=d+1$, this event has probability zero.

Fix such a $J$. The matrix $\Phi_J$ has $d+1$ rows but only $d$ columns, so some nonzero vector $a\in\RR^{d+1}$, chosen measurably in $\Phi$, is orthogonal to all its columns. Multiplying the display above by $a^\top$, the event therefore forces
$$
\sum_{i\in J}a_is_i=0.
$$
Pick $j\in J$ with $a_j\neq0$. Given $\Phi$ and the other scores, this equation pins $s_j$ down to the single value
$$
s_j=-\frac{1}{a_j}\sum_{i\in J\setminus\{j\}}a_is_i,
$$
which has conditional probability zero by the continuity condition of the theorem. Hence the event has probability zero, so $|A|\leq d$ almost surely, and the bound $b_\theta b_\varphi d/[\cG:\cH]$ follows.
\qed

\subsubsection{Proof of Theorem~\ref{thm:RKHS}}
\label{sec:proofs_rkhs}
Throughout the proof, write $\hat t_Z:=\hat t_Z^{\,K}$, and recall the distinct orbit elements $(o_i(Z),s_i(Z))$ of Section~\ref{sec:distinct_orbit_elements}, with $(o_1(Z),s_1(Z))=(Z_\obs,\psi(\tilde Z))$.

The class $\cF_K$ is a linear space containing $f$, and $\cR(t)=\lambda\|t\|_K^2$ satisfies Assumption~\ref{assumption:regularization} with
\begin{equation}
\label{eq:RKHS_penalty_term}
D_f\cR(\hat t_Z)=\lambda\frac{\diff}{\diff \varepsilon}\left(\langle \hat{t}_{Z}, \hat{t}_{Z} \rangle_K + 2\varepsilon\langle \hat{t}_{Z}, f \rangle_K + \varepsilon^2\langle f, f \rangle_K\right)\bigg|_{\varepsilon=0}=2\lambda\langle \hat{t}_{Z}, f \rangle_K .
\end{equation}
By Lemma~\ref{lem:RKHS_technical}, the minimizer in \eqref{eq:hat_t_RKHS} is unique, the quantities $\hat t_Z(o_i(Z))$ and $\langle\hat t_Z,f\rangle_K$ are measurable, and $\|\hat t_Z\|_K\leq\kappa/(2\lambda)$, so that $|D_f\cR(\hat t_Z)|\leq\kappa\|f\|_K$ is integrable. Moreover, by the reproducing property,
\begin{equation}
\label{eq:RKHS_h_infty_norm}
    \|f\|_{\infty} = \sup_{o \in \cO}|\langle f, K(\cdot, o)\rangle_K| \leq \sup_{o \in \cO}\|f\|_K \cdot \sqrt{K(o,o)} \leq \kappa M.
\end{equation}
Therefore, Theorem~\ref{thm:coverage} with $\cF=\cF_K$ yields
\begin{equation}
\label{eq:main_step_RKHS}
\begin{split}
    &\left|\EE_{Z}\left[f(Z_\obs)\left(\1\{Z\in T_K^{\symmpi}(Z_{\obs})\}-(1-\alpha)\right)\right]\right| \leq \underbrace{2\lambda\bigl|\EE_{Z}\langle \hat{t}_{Z}, f \rangle_K\bigr|}_{{(\text{I})}} \\
    &\qquad+ \underbrace{\EE_{Z}\left[|f(Z_\obs)|\1\left\{\psi(\tilde Z) = \hat{t}_{Z}(Z_\obs)\right\}\right]}_{(\text{II})}.
\end{split}
\end{equation}

By Lemma~\ref{lem:loo_stability} applied with weights $w_i=1/[\cG:\cH]$, the threshold \eqref{eq:hat_t_RKHS} and the leave-one-out threshold \eqref{eq:hat_t_RKHS_loo} satisfy
\begin{equation}
\label{eq:RKHS_infty_norm_bound}
    \|\hat{t}_{z} - \hat{t}_{z}^{-e}\|_{\infty} \leq \frac{\kappa^2}{2\lambda[\cG:\cH]}:=\delta
\end{equation}
for every $z\in\cZ$. Hence $\hat{t}_{Z}^{-e}(Z_\obs) - \delta\leq \hat{t}_{Z}(Z_\obs) \leq \hat{t}_{Z}^{-e}(Z_\obs) + \delta$, and by Assumption~\ref{assumption:conditional_density},
\begin{equation}
\label{eq:RKHS_probability_term}
\PP_{Z}\rbr{\psi(\tilde Z) = \hat{t}_{Z}(Z_\obs)\,\big|\,\cA} \leq \PP_{Z}\rbr{\hat{t}_{Z}^{-e}(Z_\obs) - \delta\leq \psi(\tilde Z)
\leq \hat{t}_{Z}^{-e}(Z_\obs) + \delta\,\big|\,\cA} \leq 2\delta b_{\psi}
= \frac{b_{\psi}\kappa^2}{\lambda[\cG:\cH]},
\end{equation}
almost surely, where we used that $\hat{t}_{Z}^{-e}(Z_\obs)$ is $\cA$-measurable and that the conditional density of $\psi(\tilde Z)$ given $\cA$ is bounded by $b_\psi$. Combining \eqref{eq:RKHS_h_infty_norm} and \eqref{eq:RKHS_probability_term},
\begin{equation}
\label{eq:RKHS_interpolation_2}
(\text{II})\leq \|f\|_{\infty}\,\PP_{Z}\rbr{\psi(\tilde Z) = \hat{t}_{Z}(Z_\obs)}\leq \frac{b_{\psi}\kappa^3 M}{\lambda[\cG:\cH]}.
\end{equation}
Together with \eqref{eq:main_step_RKHS}, this proves \eqref{eq:RKHS_bound}.

By the Cauchy--Schwarz inequality and the assumption $\EE_Z\|\hat t_Z\|_K\leq M_0$, $(\text{I})\leq 2\lambda M\,\EE_Z\|\hat t_Z\|_K\leq2\lambda MM_0$. Choosing $\lambda=[\cG:\cH]^{-1/2}$ in \eqref{eq:RKHS_bound} gives the second claim of Theorem~\ref{thm:RKHS}.
\qed

\subsection{Proofs for Section~\ref{sec:hierarchical}}
\label{sec:proofs_hierarchical}

\subsubsection{Proof of Corollary~\ref{cor:hierarchical_threshold}}
\label{sec:proof_hierarchical_threshold}

Fix $z$ with cluster sizes $n_1,\dots,n_K$, and let $g=((\sigma^{(1)},\dots,\sigma^{(K)}),\pi)\in\cG$. By the definition of the action, $g$ moves cluster $i:=\pi^{-1}(K)$ to the last position and, within it, moves individual $j:=(\sigma^{(i)}_{n_i})^{-1}(n_i)$ to the last position. Hence $\proj(g\cdot z)=x_j^{(i)}$ and $\psi(g\cdot\tilde z)=s_j^{(i)}$, and the objective in \eqref{eq:hat_t_eta} becomes
$$
\EE_{G\sim U}\Bigl[\ell_\alpha\bigl(t(\proj(G\cdot z)),\psi(G\cdot\tilde z)\bigr)\Bigr]
=\sum_{i=1}^K\sum_{j=1}^{n_i}p_{ij}\,\ell_\alpha\bigl(t(x_j^{(i)}),s_j^{(i)}\bigr),
\qquad
p_{ij}:=\PP_{G\sim U}\bigl(\pi^{-1}(K)=i,\ \sigma^{(i)}_{n_i}(j)=n_i\bigr).
$$
Under $U$, the cluster permutation $\pi$ and the component $\sigma^{(i)}_{n_i}$ are independent and uniformly distributed on $\rmS_K$ and $\rmS_{n_i}$, respectively. Therefore $p_{ij}=(1/K)(1/n_i)$, which gives \eqref{eq:hierarchical_hat_t}.
\qed

\subsubsection{Proof of Corollary~\ref{cor:hierarchical_rate}}
\label{sec:proof_hierarchical_rate}

Throughout the proof, we condition on $\cN$, so that $\cG$ is a fixed finite group and $Z\deq G\cdot Z$ for $G\sim U$ (Section~\ref{sec:hierarchical_setup}); all probabilities and expectations below are conditional on $\cN$. Write $\hat t:=\hat t_Z=\langle\hat\theta_Z,\varphi(\cdot)\rangle$ for the threshold \eqref{eq:hierarchical_hat_t} with $\cR\equiv0$.

Since $\cF_\cX$ is a linear space and $\cR\equiv0$, Corollary~\ref{cor:eta_coverage} bounds the coverage error by the interpolation error
$$
\varepsilon_\inte=\EE\Bigl[\bigl|f(X^{(K)}_{N_K})\bigr|\,\1\bigl\{S^{(K)}_{N_K}=\hat t(X^{(K)}_{N_K})\bigr\}\Bigr].
$$
By the same invariance argument as in \eqref{eq:interpolation}, $\varepsilon_\inte$ equals the average over $g\sim U$ of $|f(\proj(g\cdot Z))|\,\1\{\psi(g\cdot\tilde Z)=\hat t(\proj(g\cdot Z))\}$. By the proof of Corollary~\ref{cor:hierarchical_threshold}, a uniformly random $g$ moves individual $j$ of cluster $i$ to the target position with probability $1/(KN_i)$, in which case this quantity equals $|f(X_j^{(i)})|\,\1\{S_j^{(i)}=\hat t(X_j^{(i)})\}$. Therefore,
$$
\varepsilon_\inte=\EE\Bigl[\sum_{(i,j)\in A}\frac{|f(X_j^{(i)})|}{KN_i}\Bigr],
\qquad
A:=\bigl\{(i,j):\ S_j^{(i)}=\hat t(X_j^{(i)})\bigr\}.
$$
Since $|f|\leq b_\theta b_\varphi$ and $N_i\geq n_{\min}$ for all $i$, we obtain $\varepsilon_\inte\leq b_\theta b_\varphi\,\EE|A|/(Kn_{\min})$, and it remains to show that $|A|\leq d$ almost surely.

This follows from the argument in the proof of Theorem~\ref{thm:linear_class}, applied to the $\sum_iN_i$ pairs $(X_j^{(i)},S_j^{(i)})$ in place of the distinct orbit elements. If some $d+1$ of these pairs were interpolated, their scores would lie in the column space of the corresponding $(d+1)\times d$ matrix of features $\varphi(X_j^{(i)})^\top$. Given the covariates and the other scores, this pins one of the scores down to a single value, which has probability zero by the continuity condition of the corollary. Hence $|A|\leq d$ almost surely, and the bound $b_\theta b_\varphi d/(Kn_{\min})$ follows.
\qed

\subsubsection{Proof of Corollary~\ref{cor:hierarchical_rkhs}}
\label{sec:proof_hierarchical_rkhs}

As in the proof of Corollary~\ref{cor:hierarchical_rate}, we condition on $\cN$ throughout, and we write $\hat t:=\hat t_Z$ for the threshold \eqref{eq:hierarchical_hat_t} with $\cF_\cX=\cF_K$ and $\cR(t)=\lambda\|t\|_K^2$. By Lemma~\ref{lem:RKHS_technical}, applied with $\obs$ replaced by $\proj$, this threshold is well defined, satisfies $\|\hat t\|_K\leq\kappa/(2\lambda)$, and the relevant quantities are measurable. Corollary~\ref{cor:eta_coverage} therefore applies, and since $D_f\cR(\hat t)=2\lambda\langle\hat t,f\rangle_K$ as in \eqref{eq:RKHS_penalty_term}, it gives
$$
\Bigl|\EE\Bigl[f(X^{(K)}_{N_K})\Bigl(\1\bigl\{Y^{(K)}_{N_K}\in T^{\symmpi}(Z_\obs)\bigr\}-(1-\alpha)\Bigr)\Bigr]\Bigr|
\leq 2\lambda\bigl|\EE\langle\hat t,f\rangle_K\bigr|+\EE\Bigl[\bigl|f(X^{(K)}_{N_K})\bigr|\,\1\bigl\{S^{(K)}_{N_K}=\hat t(X^{(K)}_{N_K})\bigr\}\Bigr].
$$

It remains to bound the second term, which we do by comparing $\hat t$ with the threshold fitted without the target individual. Let $\hat t^{-}$ be the minimizer of the objective in \eqref{eq:hierarchical_hat_t} with the term of the target individual removed and all other weights unchanged. The removed term has weight $1/(KN_K)\leq1/(Kn_{\min})$, so Lemma~\ref{lem:loo_stability} gives
$$
\|\hat t-\hat t^{-}\|_\infty\leq\frac{\kappa^2}{2\lambda Kn_{\min}}=:\delta.
$$
The value $\hat t^{-}(X^{(K)}_{N_K})$ depends only on $\cN$, the covariates, and the scores of the other individuals. Given this information, the target score has a density bounded by $b_\psi$, so
$$
\PP\Bigl(S^{(K)}_{N_K}=\hat t(X^{(K)}_{N_K})\,\Big|\,\cN,\ \text{covariates},\ \text{other scores}\Bigr)
\leq\PP\Bigl(\bigl|S^{(K)}_{N_K}-\hat t^{-}(X^{(K)}_{N_K})\bigr|\leq\delta\,\Big|\,\cN,\ \text{covariates},\ \text{other scores}\Bigr)
\leq2b_\psi\delta.
$$
Combining this with $\|f\|_\infty\leq\kappa M$ from \eqref{eq:RKHS_h_infty_norm}, the second term is at most $\kappa M\cdot2b_\psi\delta=b_\psi\kappa^3M/(\lambda Kn_{\min})$. By the Cauchy--Schwarz inequality and $\EE\|\hat t\|_K\leq M_0$, the first term is at most $2\lambda MM_0$. Substituting $\lambda=(Kn_{\min})^{-1/2}$ into both bounds completes the proof.
\qed

\subsection{Proofs for Appendix~\ref{sec:sampled_adaptive_symmpi}}
\label{sec:proofs_sampled}

\subsubsection{Proof of Theorem~\ref{thm:sampled_coverage}}
\label{sec:proof_sampled_coverage}
The proof follows that of Theorem~\ref{thm:coverage}, with the invariance argument replaced by an exchangeability argument. For notational convenience, define $Z_i := G_i\cdot Z$, $\tilde{Z}_i := G_i\cdot\tilde{Z}$, and $Z_i^\obs := \obs(Z_i)$ for $i = 0, \dots, N$; in particular, $(Z_0,\tilde Z_0,Z_0^\obs)=(Z,\tilde Z,Z_\obs)$, and $\tilde Z_i=V(Z_i)$ by deterministic equivariance. Since the completion entering $\1\{Z\in T_s^{\symmpi}(Z_\obs)\}$ is $Z$ itself, \eqref{eq:hat_t_sampled} gives
\begin{equation}
    \label{eq:hat_t_sampled_rewrite}
    \hat{t}_{Z,\samp}(\cdot) = \argmin_{t(\cdot) \in \cF} \frac{1}{N+1}\sum_{i=0}^{N} \ell_\alpha\bigl(t(Z_i^\obs), \psi(\tilde{Z}_i)\bigr) + \cR(t).
\end{equation}

Applying first-order argument \eqref{eq:optimality_condition} to \eqref{eq:hat_t_sampled_rewrite} in the direction $f\in\cF$, and letting $A := \{0 \leq i \leq N : \psi(\tilde{Z}_i) = \hat{t}_{Z,\samp}(Z_i^\obs)\}$ be the index set of interpolation points, we obtain the pointwise bound
\begin{equation}
    \label{eq:sampled_pointwise_bound}
    \left|\frac{1}{N+1}\sum_{i=0}^{N} f(Z_i^\obs)\left(\alpha - \1\{\psi(\tilde{Z}_i) > \hat{t}_{Z,\samp}(Z_i^\obs)\}\right)+D_f\cR(\hat{t}_{Z,\samp})\right|\leq\frac{1}{N+1}\sum_{i\in A} |f(Z_i^\obs)|,
\end{equation}
which holds for every realization of $(Z,G_{1:N})$.

We claim that for all $0\leq i\leq N$,
\begin{equation}
    \label{eq:sampled_exchange}
    \begin{split}
    \EE\left[f(Z_i^\obs)\left(\alpha - \1\{\psi(\tilde{Z}_i) > \hat{t}_{Z,\samp}(Z_i^\obs)\}\right)\right] &= \EE\left[f(Z_\obs)\left(\alpha - \1\{\psi(\tilde{Z}) > \hat{t}_{Z,\samp}(Z_\obs)\}\right)\right],\\
    \EE\left[|f(Z_i^\obs)|\1\{\psi(\tilde{Z}_i) = \hat{t}_{Z,\samp}(Z_i^\obs)\}\right] &= \EE\left[|f(Z_\obs)|\1\{\psi(\tilde{Z}) = \hat{t}_{Z,\samp}(Z_\obs)\}\right],
    \end{split}
\end{equation}
where $\EE=\EE_{Z,G_{1:N}}$. To prove this claim, we first show that the threshold depends on $(Z_0,\dots,Z_N)$ symmetrically. For $w=(w_0,\dots,w_N)\in\cZ^{N+1}$, let $\hat t_w$ denote the minimizer, selected as in Section~\ref{sec:prediction_region}, of
$$
t\mapsto\frac{1}{N+1}\sum_{i=0}^{N}\ell_\alpha\bigl(t(\obs(w_i)),\psi(V(w_i))\bigr)+\cR(t).
$$
Then $\hat t_{Z,\samp}=\hat t_{(Z_0,\dots,Z_N)}$. As for Theorem~\ref{thm:coverage}, the maps $w\mapsto\hat t_w(\obs(w_i))$, $0\leq i\leq N$, and $w\mapsto D_f\cR(\hat t_w)$ are assumed to be measurable; for the RKHS class this follows from the proof of Lemma~\ref{lem:RKHS_technical}(iii), applied with the $N+1$ points $\obs(w_i)$, the scores $\psi(V(w_i))$ and weights $1/(N+1)$. For a permutation $\pi$ of $\{0,\dots,N\}$, write $w_\pi:=(w_{\pi(0)},\dots,w_{\pi(N)})$. The objective for $w_\pi$ is the same function on $\cF$ as that for $w$, and the selection depends only on the objective, so $\hat t_{w_\pi}=\hat t_w$. Now define
$$
\varphi(w):=f(\obs(w_0))\left(\alpha-\1\{\psi(V(w_0))>\hat t_w(\obs(w_0))\}\right),
$$
and let $\tau_i$ be the transposition of $0$ and $i$. Then
$$
\varphi\bigl((Z_0,\dots,Z_N)_{\tau_i}\bigr)=f(Z_i^\obs)\left(\alpha - \1\{\psi(\tilde{Z}_i) > \hat{t}_{Z,\samp}(Z_i^\obs)\}\right).
$$
By Lemma~\ref{lem:sampled_exchangeability}(ii), $(Z_0,\dots,Z_N)_{\tau_i}\deq(Z_0,\dots,Z_N)$, which gives the first identity in \eqref{eq:sampled_exchange}; here $|\varphi(w)|\leq|f(\obs(w_0))|$ and $Z_i\deq Z$, so all expectations are finite. The second identity follows in the same way with $\varphi$ replaced by $w\mapsto|f(\obs(w_0))|\1\{\psi(V(w_0))=\hat t_w(\obs(w_0))\}$.

Following the same reasoning as in \eqref{eq:main_step},
\begin{equation}
    \label{eq:sampled_main_step}
    \begin{split}
        &\quad\ \EE_{Z,G_{1:N}}\left[f(Z_\obs)\left(\1\{Z\in T^{\symmpi}_s(Z_{\obs})\}-(1-\alpha)\right)\right] \\
        &= \EE_{Z,G_{1:N}}\left[f(Z_\obs)\left(\alpha - \1\{\psi(\tilde{Z}) > \hat{t}_{Z,\samp}(Z_\obs)\}\right)\right] \\
        &= \EE_{Z,G_{1:N}}\left[\frac{1}{N+1}\sum_{i=0}^N f(Z_i^\obs)\left(\alpha - \1\{\psi(\tilde{Z}_i) > \hat{t}_{Z,\samp}(Z_i^\obs)\}\right)\right],
    \end{split}
\end{equation}
where the first equality follows from \eqref{eq:symmpi_sampled}, and the second equality follows from the first identity in \eqref{eq:sampled_exchange}. Moreover, by the second identity in \eqref{eq:sampled_exchange},
\begin{equation}
    \label{eq:sampled_interpolation}
    \EE_{Z,G_{1:N}}\left[\frac{1}{N+1}\sum_{i\in A}|f(Z_i^\obs)|\right] = \EE_{Z,G_{1:N}}\left[\frac{1}{N+1}\sum_{i=0}^N |f(Z_i^\obs)|\1\{\psi(\tilde{Z}_i) = \hat{t}_{Z,\samp}(Z_i^\obs)\}\right] = \varepsilon_\inte^{\samp}.
\end{equation}
Finally, combining \eqref{eq:sampled_pointwise_bound}, \eqref{eq:sampled_main_step} and \eqref{eq:sampled_interpolation}, and using that $D_f\cR(\hat t_{Z,\samp})$ is integrable, we obtain
$$
\left|\EE_{Z,G_{1:N}}\left[f(Z_\obs)\left(\1\{Z\in T^{\symmpi}_s(Z_{\obs})\}-(1-\alpha)\right)\right]+\EE_{Z,G_{1:N}}\left[D_f\cR(\hat{t}_{Z,\samp})\right]\right| \leq \varepsilon_\inte^{\samp}.
$$
The triangle inequality completes the proof.
\qed

\subsubsection{Proof of Theorem~\ref{thm:sampled_approximation}}
\label{sec:proof_sampled_approximation}

For notational convenience, write $\hat t_Z:=\hat t^{\,K}_Z$ and $\hat t_{Z,\samp}:=\hat t^{\,K}_{Z,\samp}$, and define
$$
R:=\psi(\tilde Z)-\hat t_Z(Z_\obs),\qquad R_\samp:=\psi(\tilde Z)-\hat t_{Z,\samp}(Z_\obs),\qquad d_\samp:=\bigl|\hat t_{Z,\samp}(Z_\obs)-\hat t_Z(Z_\obs)\bigr|,
$$
which are measurable by Lemmas~\ref{lem:RKHS_technical} and~\ref{lem:full_measurability}. Note that $\hat t_{Z,\samp}$ depends on the random samples $G_1,\dots,G_N$, and so do $R_\samp$ and $d_\samp$; in contrast, $\hat t_Z$ and $R$ do not depend on $G_{1:N}$.

By the definitions of the two prediction sets,
$$
\1\{Z\in T^{\symmpi}_K(Z_\obs)\}=\1\{R\leq0\},\qquad\1\{Z\in T^{\symmpi}_{K,s}(Z_\obs)\}=\1\{R_\samp\leq0\}.
$$ We claim that the two indicators differ only if $|R|\leq d_\samp$. Indeed, if $R>0\geq R_\samp$, then $0<R\leq R-R_\samp\leq d_\samp$; and if $R\leq0<R_\samp$, then $0\leq-R<R_\samp-R\leq d_\samp$. Therefore, for every $u>0$,
\begin{equation}
    \label{eq:indicator_split}
    \bigl|\1\{R\leq0\}-\1\{R_\samp\leq0\}\bigr|\leq\1\{|R|\leq u\}+\1\{d_\samp>u\}.
\end{equation}
Now fix a draw $G_{1:N}=g_{1:N}$. Multiplying \eqref{eq:indicator_split} by $|f(Z_\obs)|$ and taking the expectation over $Z$, we estimate the disagreement as follows:
\begin{equation}
    \label{eq:fixed_draw}
    \begin{split}
        \EE_Z\left[|f(Z_\obs)|\,\bigl|\1\{R\leq0\}-\1\{R_\samp\leq0\}\bigr|\right]
        &\leq\underbrace{\EE_Z\left[|f(Z_\obs)|\,\1\{|R|\leq u\}\right]}_{(\text{I})}+\underbrace{\EE_Z\left[|f(Z_\obs)|\,\1\{d_\samp>u\}\right]}_{(\text{II})}\\
        &\leq\beta_f(u)+B_f\,Q(g_{1:N}),
    \end{split}
\end{equation}
where $Q(g_{1:N}):=\PP_Z\bigl(d_\samp>u\bigr)$ is computed with $G_{1:N}=g_{1:N}$, and the second inequality follows from the definition of $\beta_f$ and $|f|\leq B_f$.

We first bound the expectation of $Q(G_{1:N})$. Take $u=u_N(\delta)$. Since $G_{1:N}$ is independent of $Z$, Fubini's theorem gives
\begin{equation}
    \label{eq:Q_bound}
    \EE_{G_{1:N}}\bigl[Q(G_{1:N})\bigr]=\int\PP_{G_{1:N}}\bigl(d_\samp(z,G_{1:N})>u_N(\delta)\bigr)\,\diff \PP_Z(z)\leq\sup_{z\in\cZ}\PP_{G_{1:N}}\Bigl(\bigl\|\hat t^{\,K}_{z,\samp}-\hat t^{\,K}_z\bigr\|_\infty>u_N(\delta)\Bigr)\leq\delta,
\end{equation}
where $d_\samp(z,g_{1:N})$ denotes $d_\samp$ evaluated at $Z=z$ and $G_{1:N}=g_{1:N}$, the first inequality follows from $d_\samp(z,g_{1:N})\leq\|\hat t^{\,K}_{z,\samp}-\hat t^{\,K}_z\|_\infty$, and the second inequality follows from Lemma~\ref{lem:sampled_stability}(b).

We now derive the three claimed bounds. Taking the expectation over $G_{1:N}$ in \eqref{eq:fixed_draw} with $u=u_N(\delta)$, and combining with \eqref{eq:Q_bound}, proves \eqref{eq:sampled_disagreement}. Next, by the definitions of $\Delta_f$ and $\Delta_{f,\samp}(g_{1:N})$,
\begin{align*}
\bigl|\Delta_f-\Delta_{f,\samp}(g_{1:N})\bigr|&=\Bigl|\EE_Z\left[f(Z_\obs)\bigl(\1\{R\leq0\}-\1\{R_\samp\leq0\}\bigr)\right]\Bigr|\\
&\leq\EE_Z\left[|f(Z_\obs)|\,\bigl|\1\{R\leq0\}-\1\{R_\samp\leq0\}\bigr|\right].
\end{align*}
Since $|\Delta_f-\EE_{G_{1:N}}\Delta_{f,\samp}(G_{1:N})|\leq\EE_{G_{1:N}}|\Delta_f-\Delta_{f,\samp}(G_{1:N})|$, the bound on the averaged coverage difference follows from \eqref{eq:sampled_disagreement}. Moreover, by Markov's inequality and \eqref{eq:Q_bound},
$$
\PP_{G_{1:N}}\bigl(Q(G_{1:N})>\delta/\eta\bigr)\leq\frac{\eta}{\delta}\,\EE_{G_{1:N}}\bigl[Q(G_{1:N})\bigr]\leq\eta,
$$
so with probability at least $1-\eta$ over $G_{1:N}$, \eqref{eq:fixed_draw} gives $|\Delta_f-\Delta_{f,\samp}(G_{1:N})|\leq\beta_f(u_N(\delta))+B_f\delta/\eta$.

It remains to bound $\beta_f(u)$. For (a), apply Lemma~\ref{lem:sampled_exchangeability}(i) to the function $h(z):=|f(\obs(z))|\1\{|\psi(\tilde z)-\hat t^{\,K}_z(\obs(z))|\leq u\}$. Since $\hat t^{\,K}_{g\cdot z}=\hat t^{\,K}_z$ by Lemma~\ref{lem:full_threshold}(ii), we have
\begin{align*}
\beta_f(u)&=\EE_Z\int|f(\obs(g\cdot Z))|\,\1\bigl\{|\psi(g\cdot\tilde Z)-\hat t_Z(\obs(g\cdot Z))|\leq u\bigr\}\,\diff U(g)\\
&=\EE_Z\int|f(o)|\,q_Z\bigl(o,[\hat t_Z(o)-u,\hat t_Z(o)+u]\bigr)\,\diff P_Z^{\obs}(o)\\
&\leq2\bar b_{\rm orb}B_f\,u,
\end{align*}
where the first equality follows from Lemma~\ref{lem:sampled_exchangeability}(i), the second equality follows from the definition of $P_z$ and the disintegration in Assumption~\ref{assumption:orbit_density} (for fixed $z$, $\hat t_z$ is a fixed function), and the inequality follows from the density bound in Assumption~\ref{assumption:orbit_density} and $|f|\leq B_f$. For (b), Lemma~\ref{lem:loo_stability} gives $|\hat t_Z(Z_\obs)-\hat t^{-e}_Z(Z_\obs)|\leq\delta_\cG:=\kappa^2/(2\lambda[\cG:\cH])$, as in \eqref{eq:RKHS_infty_norm_bound}. Hence
$$
\PP_Z\bigl(|R|\leq u\,\big|\,\cA\bigr)\leq\PP_Z\Bigl(\bigl|\psi(\tilde Z)-\hat t^{-e}_Z(Z_\obs)\bigr|\leq u+\delta_\cG\,\Big|\,\cA\Bigr)\leq2b_\psi(u+\delta_\cG)
$$
almost surely, where the last inequality follows from Assumption~\ref{assumption:conditional_density}. Multiplying by $B_f$ and taking expectations gives (b), which completes the proof.
\qed

\subsection{Proofs for Appendix~\ref{sec:computational_considerations}}
\label{sec:proofs_dual}

\subsubsection{Proof of Proposition~\ref{prop:dual_problem}}
\label{sec:proof_dual_problem}
Recall that the primal problem is
$$
    \minimize \quad \sum_{i=1}^{n} w_i\cdot\ell_{\alpha}(t(x_i), s_i) + \cR(t),
$$
where $\ell_\alpha(t(x_i), s_i) = (1-\alpha)\sbr{s_i - t(x_i)}_+ + \alpha\sbr{t(x_i) - s_i}_+$. Define $p_i = \sbr{s_i - t(x_i)}_+$ and $q_i = \sbr{t(x_i) - s_i}_+$, which satisfy $p_i, q_i \geq 0$ and the equality $s_i - t(x_i) - p_i + q_i = 0$. Let $p = [p_i]_{1\leq i\leq n}$ and $q = [q_i]_{1\leq i\leq n}$. With these decision variables $p, q \in \RR^{n}$ and $t \in \{\cX\to\RR\}$, the optimization problem can be reformulated as follows:
\begin{equation}
    \label{eq:primal_problem_reformulated}
    \begin{aligned}
    & \text{minimize}
    && \sum_{i=1}^{n} w_i\cdot \rbr{(1-\alpha) p_i + \alpha q_i } + \cR(t), \\
    & \text{subject to}
    && s_i - t(x_i) - p_i + q_i = 0, \quad i = 1, \dots, n, \\
    & && p_i, q_i \geq 0, \quad i = 1, \dots, n. \\
    \end{aligned}
\end{equation}
The Lagrangian of this problem is given by
\begin{align*}
    L(p, q, t, \lambda, \mu, \nu) &= \sum_{i=1}^{n} w_i\cdot\rbr{(1-\alpha) p_i + \alpha q_i } + \cR(t) \\
    &\qquad + \sum_{i=1}^{n} w_i\cdot \lambda_i \rbr{s_i - t(x_i) - p_i + q_i} - \sum_{i=1}^{n}w_i\cdot \mu_i p_i - \sum_{i=1}^{n} w_i\cdot\nu_i q_i \\
    &= \sum_{i=1}^{n}w_i\cdot\rbr{(1-\alpha) - \lambda_i - \mu_i}p_i + \sum_{i=1}^{n}w_i\cdot\rbr{\alpha + \lambda_i - \nu_i}q_i \\
    &\qquad  + \sum_{i=1}^{n}w_i\cdot \lambda_i s_i + \cR(t) - \sum_{i=1}^{n}w_i\cdot \lambda_i t(x_i).
\end{align*}
where $\lambda_i\in \RR$ and $\mu_i, \nu_i \geq 0$ for $i=1,\dots,n$. Therefore, the dual function is given by
$$\begin{aligned}
    g(\lambda, \mu, \nu) &= \inf_{p,q,t} L(p, q, t, \lambda, \mu, \nu) \\
    &=\underbrace{\inf_{p\in \RR^{n}}\cbr{\sum_{i=1}^{n}w_i\cdot\rbr{(1-\alpha) - \lambda_i - \mu_i}p_i}}_{\text{(I)}} + \underbrace{\inf_{q\in \RR^{n}}\cbr{\sum_{i=1}^{n}w_i\cdot\rbr{\alpha + \lambda_i - \nu_i}q_i}}_{(\text{II})} \\
    &\qquad + \sum_{i=1}^{n} w_i\cdot\lambda_i s_i + \underbrace{\inf_{t\in\{\cX\to\RR\}}\cbr{\cR(t) - \sum_{i=1}^{n} w_i\cdot \lambda_i t(x_i)}}_{(\text{III})}.
\end{aligned}$$
For term (I), it is easy to see that
$$\inf_{p\in \RR^{n}}\cbr{\sum_{i=1}^{n}w_i\cdot\rbr{(1-\alpha) - \lambda_i - \mu_i}p_i} =
\begin{cases}
    0, & \text{if } (1-\alpha) - \lambda_i - \mu_i = 0, \text{ for all } i=1,\dots,n,\\
    -\infty, & \text{otherwise.}
\end{cases}$$
Since $\mu_i\geq 0$, the condition $(1-\alpha) - \lambda_i - \mu_i = 0$ simplifies to $\lambda_i \leq 1-\alpha$ for all $i=1,\dots,n$. Similarly, for term (II), we have
$$\inf_{q\in \RR^{n}}\cbr{\sum_{i=1}^{n}w_i\cdot\rbr{\alpha + \lambda_i - \nu_i}q_i} =
\begin{cases}
    0, & \text{if } \lambda_i \geq -\alpha, \text{ for all } i=1,\dots,n,\\
    -\infty, & \text{otherwise.}
\end{cases}$$
For term (III), notice that $$\inf_{t\in\{\cX\to\RR\}}\cbr{\cR(t) - \sum_{i=1}^{n} w_i\cdot\lambda_i t(x_i)}  = - \cR^*(w\odot\lambda),$$
where $\cR^*(\alpha) := \sup_{t} \cbr{\sum_{i=1}^{n} \alpha_i t(x_i) - \cR(t)}$ is the Fenchel conjugate of $\cR$, and $w\odot \lambda := [w_i \lambda_i]_{1\leq i\leq n}$ denotes the Hadamard product of vectors $w, \lambda\in\RR^n$. Combining these results, the dual function simplifies to
$$\begin{aligned}
    g(\lambda, \mu, \nu) &=
    \begin{cases}
        \displaystyle\sum_{i=1}^{n} w_i\cdot\lambda_i s_i - \cR^*(w\odot\lambda), & \text{if } -\alpha\leq \lambda_i\leq 1-\alpha, \text{ for all } i=1,\dots,n,\\
        -\infty, & \text{otherwise.}
    \end{cases}
\end{aligned}$$
Therefore, the dual problem is to maximize the dual function subject to the derived constraints. We can express the dual problem in the following compact vector form:
$$\begin{aligned}
&\maximize \quad (w\odot\lambda)^\top s - \cR^*(w\odot\lambda) \\
&\st \quad -\alpha\one \preceq \lambda \preceq (1-\alpha)\one,
\end{aligned}$$
where $s = [s_i]_{1\leq i\leq n}$ is the vector of scores, and ``$\preceq$'' denotes element-wise inequality. This completes the proof.
\qed

\subsubsection{Proof of Proposition~\ref{prop:primal_dual_solution}}
\label{sec:proof_primal_dual_solution}

For the reformulated primal problem \eqref{eq:primal_problem_reformulated}, we denote an optimal primal solution as $(\hat{p}, \hat{q}, \hat{t}_z)$, where $\hat{p}=[\hat{p}_i]_{1\leq i\leq n}$ and $\hat{q}=[\hat{q}_i]_{1\leq i\leq n}$. Let $\hat{\lambda}=[\hat{\lambda}_i]_{1\leq i\leq n}$, $\hat{\mu}=[\hat{\mu}_i]_{1\leq i\leq n}$, and $\hat{\nu}=[\hat{\nu}_i]_{1\leq i\leq n}$ be the corresponding optimal dual solutions.\footnote{For notational simplicity, we keep the dependence of the optimal solutions on $s_n$ implicit.}

Recall the definition $\hat{p}_n = [s_n - \hat{t}_z(x_n)]_+$ and $\hat{q}_n = [\hat{t}_z(x_n) - s_n]_+$. Therefore,
\begin{itemize}
    \item[(i)] $\hat{p}_n = 0$ and $\hat{q}_n > 0$ if and only if $s_n < \hat{t}_z(x_n)$;
    \item[(ii)] $\hat{p}_n = 0$ and $\hat{q}_n = 0$ if and only if $s_n = \hat{t}_z(x_n)$;
    \item[(iii)] $\hat{p}_n > 0$ and $\hat{q}_n = 0$ if and only if $s_n > \hat{t}_z(x_n)$.
\end{itemize}

By the complementary slackness conditions, we have $\hat{\mu}_n \hat{p}_n = ((1-\alpha) - \hat{\lambda}_n)\hat{p}_n = 0$ and $\hat{\nu}_n \hat{q}_n = (\alpha + \hat{\lambda}_n)\hat{q}_n = 0$. Therefore, we have the following three cases:
\begin{itemize}
    \item[(i)] if $\hat{p}_n = 0$ and $\hat{q}_n > 0$, then $\hat{\lambda}_n = -\alpha$;
    \item[(ii)] if $\hat{p}_n = 0$ and $\hat{q}_n = 0$, then $\hat{\lambda}_n$ can take any value in the feasible region $[-\alpha, 1-\alpha]$;
    \item[(iii)] if $\hat{p}_n > 0$ and $\hat{q}_n = 0$, then $\hat{\lambda}_n = 1-\alpha$.
\end{itemize}

Combining these three cases with the previous characterization of $\hat{p}_n$ and $\hat{q}_n$ yields
$$
\hat{\lambda}_n \in
\begin{cases}
    -\alpha, & \text{if } s_n < \hat{t}_z(x_n), \\
    [-\alpha, 1-\alpha], & \text{if }  s_n = \hat{t}_z(x_n), \\
    1-\alpha, & \text{if } s_n > \hat{t}_z(x_n).
\end{cases}
$$
This completes the proof.
\qed

\subsubsection{Proof of Proposition~\ref{prop:dual_solution_monotone}}
\label{sec:proof_dual_solution_monotone}
Recall that $s_1\dots, s_{n-1}$ are fixed, while $s_n$ is treated as a parameter. Consider the dual problem \eqref{eq:dual_problem} in scalar form:
$$
\begin{aligned}
    &\maximize \quad F(\lambda; s_n) :=  \sum_{i=1}^{n} w_i\cdot\lambda_i s_i - \cR^*(w\odot\lambda), \\
    &\st \quad -\alpha \leq \lambda_i \leq 1-\alpha, \quad i=1,\dots,n.
\end{aligned}
$$
In the objective function $F$, the decision variable is $\lambda=[\lambda_i]_{1\leq i\leq n}\in\mathbb{R}^n$, and the parameter is $s_n\in\mathbb{R}$. Let $\hat{\lambda}(s_n)=[\hat{\lambda}_i(s_n)]_{1\le i\leq n}$ denote an optimal solution to this dual problem.

We prove the monotonicity of $\hat{\lambda}_n(\cdot)$ by contradiction. Suppose that $\hat{\lambda}_n(\cdot)$ is not monotonically increasing, i.e., there exists $s_n, s'_n \in \RR$ such that
$$
\rbr{s'_n - s_n}(\hat{\lambda}_n(s'_n) - \hat{\lambda}_n(s_n))< 0,
$$
or equivalently,
\begin{equation}
    \label{eq:contradiction_1}
    s'_n(\hat{\lambda}_n(s'_n) - \hat{\lambda}_n(s_n)) < s_n(\hat{\lambda}_n(s'_n) - \hat{\lambda}_n(s_n)).
\end{equation}
By the optimality of $\hat{\lambda}(s'_n)$ for the parameter value $s'_n$, we have $F(\hat{\lambda}(s'_n); s'_n)\geq F(\hat{\lambda}(s_n); s'_n)$, or equivalently,
$$
    \sum_{i=1}^{n-1} w_i\cdot\hat{\lambda}_i(s'_n) s_i + w_n\cdot\hat{\lambda}_n(s'_n) s'_n - \cR^*(w\odot\hat{\lambda}(s'_n)) \geq \sum_{i=1}^{n-1} w_i\cdot\hat{\lambda}_i(s_n) s_i + w_n\cdot\hat{\lambda}_n(s_n) s'_n - \cR^*(w\odot \hat{\lambda}(s_n)).
$$
Rearranging terms yields
\begin{equation}
    \label{eq:contradiction_2}
    w_n\cdot s'_n(\hat{\lambda}_n(s'_n) - \hat{\lambda}_n(s_n))\geq \sum_{i=1}^{n-1} w_i\cdot s_i(\hat{\lambda}_i(s_n) - \hat{\lambda}_i(s'_n)) - \rbr{\cR^*(w\odot\hat{\lambda}(s_n)) - \cR^*(w\odot\hat{\lambda}(s'_n))}.
\end{equation}
Combining \eqref{eq:contradiction_1} and \eqref{eq:contradiction_2}, we obtain
\begin{align*}
    w_n \cdot s_n(\hat{\lambda}_n(s'_n) - \hat{\lambda}_n(s_n)) &> w_n \cdot s'_n(\hat{\lambda}_n(s'_n) - \hat{\lambda}_n(s_n)) \\
    &\geq \sum_{i=1}^{n-1} w_i \cdot s_i(\hat{\lambda}_i(s_n) - \hat{\lambda}_i(s'_n)) - \rbr{\cR^*(w\odot\hat{\lambda}(s_n)) - \cR^*(w\odot\hat{\lambda}(s'_n))}.
\end{align*}
Rearranging further gives
$$
\sum_{i=1}^{n} w_i\cdot \hat{\lambda}_i(s'_n) s_i - \cR^*(w\odot\hat{\lambda}(s'_n)) > \sum_{i=1}^{n} w_i\cdot\hat{\lambda}_i(s_n) s_i - \cR^*(w\odot \hat{\lambda}(s_n)).
$$
This implies $F(\hat{\lambda}(s'_n); s_n) > F(\hat{\lambda}(s_n); s_n)$, which contradicts the optimality of $\hat{\lambda}(s_n)$ for the parameter value $s_n$. Hence $\hat{\lambda}_n(\cdot)$ must be monotonically increasing.
\qed

\subsection{Technical Lemmas}
\label{sec:technical_lemmas}

\begin{lemma}[Strong convexity]
    \label{lem:strong_convexity}
    Let $\Lambda:\cF_K\to\RR$ be convex, and suppose $t^\star$ minimizes $J(t):=\Lambda(t)+\lambda\|t\|_K^2$ over $\cF_K$. Then $J(t)-J(t^\star)\geq\lambda\|t-t^\star\|_K^2$ for all $t\in\cF_K$.
\end{lemma}
\begin{proof}
    Fix $t\in\cF_K$. For $\tau\in(0,1)$, let $t_\tau:=t^\star+\tau(t-t^\star)=(1-\tau)t^\star+\tau t$. It is easy to verify the identity
    $$
    \|t_\tau\|_K^2=(1-\tau)\|t^\star\|_K^2+\tau\|t\|_K^2-\tau(1-\tau)\|t-t^\star\|_K^2.
    $$
    Combining this identity with the convexity of $\Lambda$, we have
    \begin{align*}
        J(t^\star)\leq J(t_\tau)
        &\leq(1-\tau)\Lambda(t^\star)+\tau\Lambda(t)+\lambda\rbr{(1-\tau)\|t^\star\|_K^2+\tau\|t\|_K^2-\tau(1-\tau)\|t-t^\star\|_K^2}\\
        &=(1-\tau)J(t^\star)+\tau J(t)-\lambda\tau(1-\tau)\|t-t^\star\|_K^2,
    \end{align*}
    where the first inequality follows from the optimality of $t^\star$. Rearranging terms and dividing by $\tau$, we obtain
    $$
    J(t)-J(t^\star)\geq\lambda(1-\tau)\|t-t^\star\|_K^2.
    $$
    Letting $\tau\downarrow0$ completes the proof.
\end{proof}

\begin{lemma}[Leave-one-out stability]
    \label{lem:loo_stability}
    Fix points $o_1,\dots,o_m\in\cO$, scores $s_1,\dots,s_m\in\RR$, and nonnegative weights $w_1,\dots,w_m$. Let $\hat t$ minimize $J(t):=\sum_{i=1}^mw_i\ell_\alpha(t(o_i),s_i)+\lambda\|t\|_K^2$ over $\cF_K$, and let $\hat t^-$ minimize $J^-(t):=J(t)-w_1\ell_\alpha(t(o_1),s_1)$ over $\cF_K$, where the remaining weights are \emph{not} renormalized. Under Assumption~\ref{assumption:bounded_kernel},
    $$
    \|\hat t-\hat t^-\|_K\leq\frac{\kappa w_1}{2\lambda},\qquad\|\hat t-\hat t^-\|_\infty\leq\frac{\kappa^2w_1}{2\lambda}.
    $$
\end{lemma}
\begin{proof}
    Let $\Delta:=\hat t^--\hat t$. Both $J$ and $J^-$ are of the form in Lemma~\ref{lem:strong_convexity}, since the weighted sums of pinball losses are convex in $t$. Applying Lemma~\ref{lem:strong_convexity} to $J$ (with minimizer $\hat t$) and to $J^-$ (with minimizer $\hat t^-$), we have
    $$
    J(\hat t^-)-J(\hat t)\geq\lambda\|\Delta\|_K^2
    \qquad\text{and}\qquad
    J^-(\hat t)-J^-(\hat t^-)\geq\lambda\|\Delta\|_K^2.
    $$
    Adding the two inequalities and using $J(t)-J^-(t)=w_1\ell_\alpha(t(o_1),s_1)$, we obtain
    \begin{align*}
        2\lambda\|\Delta\|_K^2
        &\leq w_1\rbr{\ell_\alpha(\hat t^-(o_1),s_1)-\ell_\alpha(\hat t(o_1),s_1)}\\
        &\leq w_1|\Delta(o_1)|\\
        &\leq w_1\kappa\|\Delta\|_K,
    \end{align*}
    where the second inequality follows from the fact that $\ell_\alpha$ is $1$-Lipschitz in its first argument, and the last inequality follows from the reproducing property and the Cauchy--Schwarz inequality, $|\Delta(o)|=|\langle\Delta,K(\cdot,o)\rangle_K|\leq\kappa\|\Delta\|_K$. If $\Delta=0$, there is nothing to prove; otherwise, dividing both sides by $2\lambda\|\Delta\|_K$ gives the first bound. The second bound then follows from $\|\Delta\|_\infty\leq\kappa\|\Delta\|_K$. This completes the proof.
\end{proof}

\begin{lemma}[Well-posedness and measurability of the RKHS threshold]
    \label{lem:RKHS_technical}
    Suppose Assumption~\ref{assumption:bounded_kernel} holds, $\lambda>0$, and (as assumed throughout) the kernel $K$ and the maps $\obs$, $V$, $\psi$ and $z\mapsto g\cdot z$ are measurable.
    \begin{itemize}
        \item[(i)] For every $z$, the objectives in \eqref{eq:hat_t_RKHS} and \eqref{eq:hat_t_RKHS_loo} admit a unique minimizer over $\cF_K$.
        \item[(ii)] $\|\hat t^{\,K}_z\|_K\leq\kappa/(2\lambda)$ for every $z$.
        \item[(iii)] Every $f\in\cF_K$ is measurable, and the maps $z\mapsto\hat t^{\,K}_z(\obs(g\cdot z))$ ($g\in\cG$), $z\mapsto\langle\hat t^{\,K}_z,f\rangle_K$ and $z\mapsto\|\hat t^{\,K}_z\|_K$ are measurable; the same holds for $\hat t^{-e}_z$.
    \end{itemize}
\end{lemma}
\begin{proof}
    For notational convenience, fix $z$ and write $m:=[\cG:\cH]$, $o_i:=o_i(z)$ and $s_i:=s_i(z)$ for $1\leq i\leq m$. Both objectives in \eqref{eq:hat_t_RKHS} and \eqref{eq:hat_t_RKHS_loo} can be written as
    $$
    L(t):=\sum_{i=1}^mw_i\,\ell_\alpha\bigl(t(o_i),s_i\bigr)+\lambda\|t\|_K^2,
    $$
    where $w_i=1/[\cG:\cH]$ for all $i$ in \eqref{eq:hat_t_RKHS}, and $w_1=0$, $w_i=1/[\cG:\cH]$ for $i\geq2$ in \eqref{eq:hat_t_RKHS_loo}. In both cases, $w_i\geq0$ and $\sum_iw_i\leq1$. Let $\hat t$ denote a minimizer of $L$.

    \paragraph{Proof of (i).} We first show existence. By the reproducing property and the Cauchy--Schwarz inequality, $|t(o)-t'(o)|\leq\kappa\|t-t'\|_K$ for all $t,t'\in\cF_K$ and $o\in\cO$. Since $\ell_\alpha$ is convex and $1$-Lipschitz in its first argument, each loss term $t\mapsto\ell_\alpha(t(o_i),s_i)$ is convex and $\kappa$-Lipschitz with respect to $\|\cdot\|_K$. Hence $L$ is convex and continuous on $\cF_K$, and is therefore weakly lower semicontinuous. Moreover, since $\ell_\alpha\geq0$, we have
    $$
    L(t)\geq\lambda\|t\|_K^2\qquad\text{for all }t\in\cF_K.
    $$
    Therefore, any minimizing sequence of $L$ is bounded in $\cF_K$ and admits a weakly convergent subsequence. By weak lower semicontinuity, its weak limit is a minimizer of $L$. For uniqueness, suppose that $\hat t$ and $\hat t'$ are both minimizers. Applying Lemma~\ref{lem:strong_convexity} with $\Lambda(t)=\sum_iw_i\ell_\alpha(t(o_i),s_i)$ gives $0=L(\hat t')-L(\hat t)\geq\lambda\|\hat t'-\hat t\|_K^2$, and hence $\hat t'=\hat t$.

    \paragraph{Proof of (ii).} We apply the first-order argument in the proof of Theorem~\ref{thm:coverage} to $L$ in the direction $\hat t$. Specifically, the function $H(\varepsilon):=L(\hat t+\varepsilon\hat t)$ is a finite convex function on $\RR$ minimized at $\varepsilon=0$, so $H'_-(0)\leq0\leq H'_+(0)$. The penalty contributes $\frac{\diff}{\diff\varepsilon}\lambda(1+\varepsilon)^2\|\hat t\|_K^2\big|_{\varepsilon=0}=2\lambda\|\hat t\|_K^2$, and, exactly as in the derivation of \eqref{eq:optimality_condition}, there exist coefficients $c_i\in[\alpha-1,\alpha]$ such that
    $$
    \sum_{i=1}^mw_ic_i\,\hat t(o_i)+2\lambda\|\hat t\|_K^2=0.
    $$
    Since $|c_i|\leq1$, $|\hat t(o_i)|\leq\kappa\|\hat t\|_K$ and $\sum_iw_i\leq1$, we obtain
    $$
    2\lambda\|\hat t\|_K^2=-\sum_{i=1}^mw_ic_i\,\hat t(o_i)\leq\sum_{i=1}^mw_i|\hat t(o_i)|\leq\kappa\|\hat t\|_K.
    $$
    Dividing both sides by $2\lambda\|\hat t\|_K$ (the case $\hat t=0$ is trivial) gives $\|\hat t\|_K\leq\kappa/(2\lambda)$.

    \paragraph{Proof of (iii).} We first show that every $f\in\cF_K$ is measurable. For each $o\in\cO$, the function $K(\cdot,o)$ is measurable as a section of the jointly measurable function $K$, and hence so is every finite linear combination of such functions. Every $f\in\cF_K$ is a $\|\cdot\|_K$-limit of such combinations $f_k$. Since $|f_k(o)-f(o)|\leq\kappa\|f_k-f\|_K$, the convergence also holds pointwise, and thus $f$ is measurable.

    We next construct a measurable approximation of $\hat t$. Let $S_z:=\mathrm{span}\{K(o_i,\cdot)\}_{1\leq i\leq m}$. For any $t\in\cF_K$, let $\Pi t$ denote its orthogonal projection onto $S_z$. By the reproducing property,
    $$
    (\Pi t)(o_i)=\langle\Pi t,K(o_i,\cdot)\rangle_K=\langle t,K(o_i,\cdot)\rangle_K=t(o_i),\qquad\|\Pi t\|_K\leq\|t\|_K,
    $$
    so $L(\Pi t)\leq L(t)$. By uniqueness of the minimizer, $\hat t\in S_z$, and we can write $\hat t=\sum_ia_i^\star K(o_i,\cdot)$ for some $a^\star\in\RR^m$; no measurable choice of $a^\star$ is needed below. Let $\bm K_z:=(K(o_i,o_j))_{1\leq i,j\leq m}$ be the Gram matrix and $s:=(s_i)_{1\leq i\leq m}$. For $\epsilon>0$, a positive semidefinite matrix $\bm K\in\RR^{m\times m}$ and $s\in\RR^m$, define the perturbed finite-dimensional objective
    $$
    J_\epsilon(a;\bm K,s):=\sum_{i=1}^mw_i\,\ell_\alpha\bigl((\bm Ka)_i,s_i\bigr)+\lambda a^\top\bm Ka+\epsilon\|a\|_2^2,\qquad a\in\RR^m.
    $$
    The function $J_\epsilon$ is jointly continuous in $(a,\bm K,s)$ and $2\epsilon$-strongly convex in $a$, so it has a unique minimizer $a_\epsilon(\bm K,s)$. We claim that $(\bm K,s)\mapsto a_\epsilon(\bm K,s)$ is continuous. Indeed, comparing with $a=0$ and using $\ell_\alpha(0,s_i)\leq|s_i|$, we have
    $$
    \epsilon\|a_\epsilon(\bm K,s)\|_2^2\leq J_\epsilon(0;\bm K,s)\leq\sum_{i=1}^mw_i|s_i|,
    $$
    so the minimizers are locally bounded. Let $(\bm K_k,s_k)\to(\bm K,s)$, and let $\bar a$ be any limit point of $a_\epsilon(\bm K_k,s_k)$. For every $a\in\RR^m$, we have $J_\epsilon(a_\epsilon(\bm K_k,s_k);\bm K_k,s_k)\leq J_\epsilon(a;\bm K_k,s_k)$, and passing to the limit along the corresponding subsequence gives $J_\epsilon(\bar a;\bm K,s)\leq J_\epsilon(a;\bm K,s)$ by joint continuity. Hence $\bar a=a_\epsilon(\bm K,s)$ by uniqueness, which, together with local boundedness, proves the claim. Note that no inverse of $\bm K$ is used, so changes in the rank of $\bm K$ are allowed. Since $z\mapsto(\bm K_z,s)$ is measurable, so is $z\mapsto a_\epsilon:=a_\epsilon(\bm K_z,s)$. Define
    $$
    t_{z,\epsilon}:=\sum_{i=1}^m(a_\epsilon)_iK(o_i,\cdot)\in\cF_K.
    $$

    It remains to show that $t_{z,\epsilon}$ approximates $\hat t$. Since $t_{z,\epsilon}(o_i)=(\bm K_za_\epsilon)_i$ and $\|t_{z,\epsilon}\|_K^2=a_\epsilon^\top\bm K_za_\epsilon$, and similarly for $\hat t$ and $a^\star$, we have
    $$
    L(t_{z,\epsilon})\leq J_\epsilon(a_\epsilon;\bm K_z,s)\leq J_\epsilon(a^\star;\bm K_z,s)=L(\hat t)+\epsilon\|a^\star\|_2^2,
    $$
    where the second inequality follows from the optimality of $a_\epsilon$. By Lemma~\ref{lem:strong_convexity}, this implies
    $$
    \lambda\|t_{z,\epsilon}-\hat t\|_K^2\leq L(t_{z,\epsilon})-L(\hat t)\leq\epsilon\|a^\star\|_2^2\to0\qquad\text{as }\epsilon\downarrow0.
    $$
    Therefore, along $\epsilon=1/r$ with $r\to\infty$, the functions
    $$
    t_{z,\epsilon}(\obs(g\cdot z))=\sum_{i=1}^m(a_\epsilon)_iK\bigl(o_i,\obs(g\cdot z)\bigr),\qquad
    \langle t_{z,\epsilon},f\rangle_K=\sum_{i=1}^m(a_\epsilon)_if(o_i),\qquad
    \|t_{z,\epsilon}\|_K^2=a_\epsilon^\top\bm K_za_\epsilon
    $$
    converge pointwise in $z$ to $\hat t(\obs(g\cdot z))$, $\langle\hat t,f\rangle_K$ and $\|\hat t\|_K^2$, respectively. Each of these functions is measurable in $z$ by the measurability of $f$ and $z\mapsto a_\epsilon$ shown above, and pointwise limits of measurable functions are measurable. Since $L$ covers both objectives, the same conclusion holds for $\hat t^{-e}_z$. This completes the proof.
\end{proof}

\begin{lemma}[Invariance and exchangeability on compact groups]
    \label{lem:sampled_exchangeability}
    Let $\cG$ be a compact metrizable group with its Borel $\sigma$-field and uniform distribution $U$, whose action on $\cZ$ is jointly measurable, and suppose $Z\deq G\cdot Z$ for $G\sim U$. Then:
    \begin{itemize}
        \item[(i)] $g\cdot Z\deq Z$ for every $g\in\cG$. Consequently, for every measurable $h\geq0$, or measurable $h$ with $\EE|h(Z)|<\infty$, we have $\EE h(Z)=\EE\int_\cG h(g\cdot Z)\,\diff U(g)$.
        \item[(ii)] Let $G_1,\dots,G_N\simiid U$ be independent of $Z$, and let $G_0:=e$. Then the vector $(G_0\cdot Z,G_1\cdot Z,\dots,G_N\cdot Z)$ is exchangeable.
    \end{itemize}
\end{lemma}
\begin{proof}
    We prove the two parts separately.

    \paragraph{Proof of (i).} Let $G\sim U$ be independent of $Z$, and fix $g\in\cG$. Applying the measurable map $z\mapsto g\cdot z$ to both sides of $Z\deq G\cdot Z$, we have
    $$
    g\cdot Z\deq g\cdot(G\cdot Z)=(gG)\cdot Z\deq G\cdot Z\deq Z,
    $$
    where the first relation follows from the distributional invariance of $Z$, the equality follows from the definition of the action, the third relation follows from the fact that $gG\sim U$ is independent of $Z$ by the invariance of $U$, and the last relation follows from the distributional invariance of $Z$. Hence $\EE h(g\cdot Z)=\EE h(Z)$ for every $g\in\cG$. Since $(g,z)\mapsto h(g\cdot z)$ is jointly measurable, Fubini's theorem gives
    $$
    \EE\int_\cG h(g\cdot Z)\,\diff U(g)=\int_\cG\EE h(g\cdot Z)\,\diff U(g)=\int_\cG\EE h(Z)\,\diff U(g)=\EE h(Z).
    $$
    For integrable $h$, the use of Fubini's theorem is justified by applying the same identity to $|h|$.

    \paragraph{Proof of (ii).} Let $G\sim U$ be independent of $(Z,G_{1:N})$. Since $G\cdot Z\deq Z$ and $G_{1:N}$ is independent of $(Z,G)$, we have $(G\cdot Z,G_{1:N})\deq(Z,G_{1:N})$. Applying the measurable map $(z,g_{1:N})\mapsto(z,g_1\cdot z,\dots,g_N\cdot z)$ to both sides and using the definition of the action, we obtain
    $$
    (G_0\cdot Z,G_1\cdot Z,\dots,G_N\cdot Z)\deq(H_0\cdot Z,H_1\cdot Z,\dots,H_N\cdot Z),\qquad\text{where }H_0:=G,\ H_i:=G_iG\ (1\leq i\leq N).
    $$
    We next show that $Z,H_0,H_1,\dots,H_N$ are independent with $H_i\sim U$ for all $i$. Indeed, conditionally on $(Z,G)=(z,g)$, the elements $G_1g,\dots,G_Ng$ are i.i.d.\ with distribution $U$ by the right invariance of $U$. Since this conditional distribution does not depend on $(z,g)$, $(H_1,\dots,H_N)$ is independent of $(Z,H_0)$; moreover, $H_0=G\sim U$ is independent of $Z$. Consequently, for any permutation $\pi$ of $\{0,\dots,N\}$,
    $$
    (Z,H_{\pi(0)},\dots,H_{\pi(N)})\deq(Z,H_0,\dots,H_N).
    $$
    Applying the measurable map $(z,h_0,\dots,h_N)\mapsto(h_0\cdot z,\dots,h_N\cdot z)$ shows that $(H_i\cdot Z)_{0\leq i\leq N}$ is exchangeable, and hence so is $(G_i\cdot Z)_{0\leq i\leq N}$. This completes the proof.
\end{proof}

\begin{lemma}[Full RKHS threshold on a compact group]
    \label{lem:full_threshold}
    In the setting of Appendix~\ref{sec:sampled_adaptive_symmpi}, fix $z\in\cZ$ and let $P_z$ be the distribution of $(\obs(G\cdot z),\psi(G\cdot\tilde z))$ for $G\sim U$. Then:
    \begin{itemize}
        \item[(i)] The objective in \eqref{eq:hat_t_RKHS_full} admits a unique minimizer $\hat t_z:=\hat t^{\,K}_z$ over $\cF_K$.
        \item[(ii)] $\hat t_{g\cdot z}=\hat t_z$ for all $g\in\cG$.
        \item[(iii)] There exists a measurable function $c_z:\cO\times\RR\to[\alpha-1,\alpha]$ such that $c_z(o,s)=\alpha-\1\{s>\hat t_z(o)\}$ whenever $s\neq\hat t_z(o)$, and
        \begin{equation}
            \label{eq:full_first_order}
            \int c_z(o,s)\,h(o)\,\diff P_z(o,s)+2\lambda\langle\hat t_z,h\rangle_K=0\qquad\text{for all }h\in\cF_K.
        \end{equation}
        In particular, $c_z(o,s)$ is a subgradient of $\ell_\alpha(\cdot,s)$ at $\hat t_z(o)$ for every $(o,s)$.
        \item[(iv)] $\|\hat t_z\|_K\leq\kappa/(2\lambda)$.
    \end{itemize}
\end{lemma}
\begin{proof}
    Recall that the objective in \eqref{eq:hat_t_RKHS_full} is $\Lambda(t)+\lambda\|t\|_K^2$, where
    $$
    \Lambda(t):=\int\ell_\alpha\bigl(t(o),s\bigr)\,\diff P_z(o,s)=\EE_{G\sim U}\left[\ell_\alpha\bigl(t(\obs(G\cdot z)),\psi(G\cdot\tilde z)\bigr)\right].
    $$
    By Lemma~\ref{lem:RKHS_technical}(iii), every $t\in\cF_K$ is measurable. Moreover, $\ell_\alpha(t(o),s)\leq|t(o)|+|s|\leq\kappa\|t\|_K+|s|$, and $\int|s|\,\diff P_z=\int|\psi(g\cdot\tilde z)|\,\diff U(g)<\infty$ by assumption. Hence $\Lambda$ is finite on $\cF_K$.

    \paragraph{Proof of (i).} Since $|t(o)-t'(o)|\leq\kappa\|t-t'\|_K$ and $\ell_\alpha$ is convex and $1$-Lipschitz in its first argument, $\Lambda$ is convex and $\kappa$-Lipschitz with respect to $\|\cdot\|_K$. Hence the objective is convex and continuous, and is bounded below by $\lambda\|t\|_K^2$. The same argument as in the proof of Lemma~\ref{lem:RKHS_technical}(i) shows that a minimizer exists and, by Lemma~\ref{lem:strong_convexity}, is unique.

    \paragraph{Proof of (ii).} For any $g\in\cG$ and $t\in\cF_K$, the objective for $g\cdot z$ satisfies
    \begin{align*}
        \EE_{G\sim U}\left[\ell_\alpha\bigl(t(\obs(G\cdot(g\cdot z))),\psi(G\cdot V(g\cdot z))\bigr)\right]
        &=\EE_{G\sim U}\left[\ell_\alpha\bigl(t(\obs((Gg)\cdot z)),\psi((Gg)\cdot\tilde z)\bigr)\right]\\
        &=\EE_{G\sim U}\left[\ell_\alpha\bigl(t(\obs(G\cdot z)),\psi(G\cdot\tilde z)\bigr)\right],
    \end{align*}
    where the first equality follows from the definition of the action and the $\cG$-deterministic equivariance of $V$, and the second equality follows from the right invariance of $U$. Hence the objectives for $g\cdot z$ and $z$ coincide as functions on $\cF_K$, and the uniqueness in (i) gives $\hat t_{g\cdot z}=\hat t_z$.

    \paragraph{Proof of (iii).} For notational convenience, write $\hat t:=\hat t_z$, $P:=P_z$, and
    $$
    D:=\{(o,s):s=\hat t(o)\},\qquad a(o,s):=\alpha-\1\{s>\hat t(o)\}.
    $$
    Here $D$ is the set of interpolation points, and $a(o,s)$ is the derivative of $\ell_\alpha(\cdot,s)$ at $\hat t(o)$ for $(o,s)\notin D$. Our goal is to find a function $\eta$ with values in $[\alpha-1,\alpha]$ on $D$ such that $c_z:=a\1_{D^c}+\eta\1_D$ satisfies \eqref{eq:full_first_order}.

    We first compute the directional derivatives of the objective at $\hat t$. Fix $h\in\cF_K$. For each $(o,s)$, consider the difference quotient
    $$
    q_\varepsilon(o,s):=\frac{\ell_\alpha\bigl(\hat t(o)+\varepsilon h(o),s\bigr)-\ell_\alpha\bigl(\hat t(o),s\bigr)}{\varepsilon},\qquad\varepsilon>0.
    $$
    By the convexity of $\ell_\alpha$ in its first argument, $q_\varepsilon(o,s)$ is nonincreasing as $\varepsilon\downarrow0$. Since $\ell_\alpha$ is $1$-Lipschitz in its first argument, $|q_\varepsilon(o,s)|\leq|h(o)|\leq\kappa\|h\|_K$. Moreover, as $\varepsilon\downarrow0$,
    $$
    q_\varepsilon(o,s)\to
    \begin{cases}
        a(o,s)\,h(o), & \text{if }(o,s)\notin D,\\
        \max\{\alpha h(o),(\alpha-1)h(o)\}, & \text{if }(o,s)\in D,
    \end{cases}
    $$
    where the second case follows from $\ell_\alpha(s+\varepsilon h(o),s)=\varepsilon\bigl(\alpha[h(o)]_++(1-\alpha)[-h(o)]_+\bigr)$. In addition, $\varepsilon^{-1}\bigl(\lambda\|\hat t+\varepsilon h\|_K^2-\lambda\|\hat t\|_K^2\bigr)=2\lambda\langle\hat t,h\rangle_K+\lambda\varepsilon\|h\|_K^2\to2\lambda\langle\hat t,h\rangle_K$. Therefore, by the optimality of $\hat t$ and the dominated convergence theorem,
    \begin{equation}
        \label{eq:full_directional}
        \begin{split}
        0&\leq\lim_{\varepsilon\downarrow0}\frac{\Lambda(\hat t+\varepsilon h)+\lambda\|\hat t+\varepsilon h\|_K^2-\Lambda(\hat t)-\lambda\|\hat t\|_K^2}{\varepsilon}\\
        &=\int_{D^c}a\,h(o)\,\diff P+\int_D\max\{\alpha h(o),(\alpha-1)h(o)\}\,\diff P+2\lambda\langle\hat t,h\rangle_K.
        \end{split}
    \end{equation}

    We next reformulate \eqref{eq:full_first_order} as a statement about a convex set in $\cF_K$. The map $h\mapsto\int_{D^c}a\,h(o)\,\diff P+2\lambda\langle\hat t,h\rangle_K$ is a bounded linear functional on $\cF_K$; let $v_0\in\cF_K$ be its Riesz representer. Define the linear map $\mathsf A:L^\infty(P)\to\cF_K$ by
    $$
    \langle\mathsf A\eta,h\rangle_K:=\int\eta\,\1_D\,h(o)\,\diff P\qquad\text{for all }h\in\cF_K,
    $$
    which is well defined since $|\int\eta\1_Dh(o)\,\diff P|\leq\kappa\|\eta\|_\infty\|h\|_K$. Let
    $$
    B:=\{\eta\in L^\infty(P):\alpha-1\leq\eta\leq\alpha\ P\text{-a.s.}\},\qquad\cC:=v_0+\mathsf A(B)\subseteq\cF_K.
    $$
    By construction, for $\eta\in B$ and $c_z:=a\1_{D^c}+\eta\1_D$, \eqref{eq:full_first_order} is equivalent to $v_0+\mathsf A\eta=0$. Hence it suffices to show that $0\in\cC$. We first show that $\cC$ is closed and convex. Since $L^\infty(P)$ is the dual of $L^1(P)$, the set $B$ is convex, weak$^*$-closed and bounded, and is therefore weak$^*$-compact by the Banach--Alaoglu theorem. Moreover, $\mathsf A$ is weak$^*$-to-weak continuous, because for each $h\in\cF_K$, the map $\eta\mapsto\langle\mathsf A\eta,h\rangle_K$ is integration against $\1_D\,h(o)\in L^1(P)$. Hence $\cC$ is convex and weakly compact, and thus closed in $\cF_K$.

    We now prove $0\in\cC$ by contradiction. Suppose that $0\notin\cC$. Let $v^*$ be the projection of $0$ onto the closed convex set $\cC$, and let $h:=-v^*\neq0$. By the characterization of the projection,
    $$
    \langle v,h\rangle_K\leq-\|v^*\|_K^2<0\qquad\text{for all }v\in\cC.
    $$
    On the other hand, for every $\eta\in B$, we have $\eta\,\1_D\,h(o)\leq\1_D\max\{\alpha h(o),(\alpha-1)h(o)\}$, with equality for $\eta^*:=\alpha\1\{h(o)>0\}+(\alpha-1)\1\{h(o)\leq0\}\in B$. Therefore,
    $$
    \sup_{v\in\cC}\langle v,h\rangle_K=\int_{D^c}a\,h(o)\,\diff P+\int_D\max\{\alpha h(o),(\alpha-1)h(o)\}\,\diff P+2\lambda\langle\hat t,h\rangle_K\geq0
    $$
    by \eqref{eq:full_directional}, which is a contradiction. Hence $v_0+\mathsf A\eta=0$ for some $\eta\in B$. Choosing a version of $\eta$ with values in $[\alpha-1,\alpha]$ everywhere and setting $c_z:=a\1_{D^c}+\eta\1_D$, we obtain \eqref{eq:full_first_order}. Finally, $c_z(o,s)=a(o,s)$ is the derivative of $\ell_\alpha(\cdot,s)$ at $\hat t(o)$ for $(o,s)\notin D$, and $c_z(o,s)\in[\alpha-1,\alpha]$, the subdifferential of $\ell_\alpha(\cdot,s)$ at $s$, for $(o,s)\in D$. Hence $c_z(o,s)$ is a subgradient of $\ell_\alpha(\cdot,s)$ at $\hat t(o)$ for every $(o,s)$.

    \paragraph{Proof of (iv).} Taking $h=\hat t$ in \eqref{eq:full_first_order} and using $|c_z|\leq1$ and $|\hat t(o)|\leq\kappa\|\hat t\|_K$, we have
    $$
    2\lambda\|\hat t\|_K^2=-\int c_z(o,s)\,\hat t(o)\,\diff P(o,s)\leq\kappa\|\hat t\|_K,
    $$
    which gives $\|\hat t\|_K\leq\kappa/(2\lambda)$. This completes the proof.
\end{proof}

\begin{lemma}[Sampled versus full threshold]
    \label{lem:sampled_stability}
    In the setting of Appendix~\ref{sec:sampled_adaptive_symmpi}, fix $z\in\cZ$ and let $c_z$ be as in Lemma~\ref{lem:full_threshold}(iii). For $g\in\cG$ and $g_{1:N}\in\cG^N$ with $g_0:=e$, define $\xi_z(g):=c_z\bigl(\obs(g\cdot z),\psi(g\cdot\tilde z)\bigr)K(\obs(g\cdot z),\cdot)\in\cF_K$ and
    $$
    \Xi_N(z,g_{1:N}):=\frac{1}{N+1}\sum_{i=0}^N\xi_z(g_i)+2\lambda\hat t^{\,K}_z.
    $$
    \begin{itemize}
        \item[(a)] For every $g_{1:N}\in\cG^N$, $\|\hat t^{\,K}_{z,\samp}-\hat t^{\,K}_z\|_K\leq\|\Xi_N(z,g_{1:N})\|_K/(2\lambda)$, where $\hat t^{\,K}_{z,\samp}$ is computed from $G_{1:N}=g_{1:N}$.
        \item[(b)] If $N\geq1$ and $G_1,\dots,G_N\simiid U$, then $\EE\|\Xi_N(z,G_{1:N})\|_K\leq\kappa(\sqrt N+2)/(N+1)$, and for every $\delta\in(0,1]$, with probability at least $1-\delta$, $\|\hat t^{\,K}_{z,\samp}-\hat t^{\,K}_z\|_\infty\leq u_N(\delta)$.
    \end{itemize}
\end{lemma}
\begin{proof}
    For notational convenience, write $\hat t:=\hat t^{\,K}_z$, $\hat t_\samp:=\hat t^{\,K}_{z,\samp}$, $\vartheta:=\hat t_\samp-\hat t$, $o_i:=\obs(g_i\cdot z)$ and $s_i:=\psi(g_i\cdot\tilde z)$ for $0\leq i\leq N$, and let
    $$
    \hat L(t):=\frac{1}{N+1}\sum_{i=0}^N\ell_\alpha\bigl(t(o_i),s_i\bigr)+\lambda\|t\|_K^2
    $$
    denote the objective in \eqref{eq:hat_t_sampled} with the RKHS penalty, so that $\hat t_\samp$ minimizes $\hat L$.

    \paragraph{Proof of (a).} We bound $\hat L(\hat t_\samp)-\hat L(\hat t)$ from both sides. For the lower bound, since $c_z(o_i,s_i)$ is a subgradient of $\ell_\alpha(\cdot,s_i)$ at $\hat t(o_i)$ by Lemma~\ref{lem:full_threshold}(iii), and $\vartheta(o_i)=\langle\vartheta,K(o_i,\cdot)\rangle_K$, we have
    $$
    \ell_\alpha\bigl(\hat t_\samp(o_i),s_i\bigr)\geq\ell_\alpha\bigl(\hat t(o_i),s_i\bigr)+\bigl\langle c_z(o_i,s_i)K(o_i,\cdot),\vartheta\bigr\rangle_K=\ell_\alpha\bigl(\hat t(o_i),s_i\bigr)+\langle\xi_z(g_i),\vartheta\rangle_K.
    $$
    Averaging over $0\leq i\leq N$ and using $\lambda\|\hat t_\samp\|_K^2=\lambda\|\hat t\|_K^2+2\lambda\langle\hat t,\vartheta\rangle_K+\lambda\|\vartheta\|_K^2$, we obtain
    $$
    \hat L(\hat t_\samp)-\hat L(\hat t)\geq\Bigl\langle\frac{1}{N+1}\sum_{i=0}^N\xi_z(g_i)+2\lambda\hat t,\ \vartheta\Bigr\rangle_K+\lambda\|\vartheta\|_K^2=\langle\Xi_N,\vartheta\rangle_K+\lambda\|\vartheta\|_K^2.
    $$
    For the upper bound, since $\hat t_\samp$ minimizes $\hat L$, Lemma~\ref{lem:strong_convexity} gives $\hat L(\hat t)-\hat L(\hat t_\samp)\geq\lambda\|\vartheta\|_K^2$. Adding the two inequalities and applying the Cauchy--Schwarz inequality, we obtain
    $$
    2\lambda\|\vartheta\|_K^2\leq-\langle\Xi_N,\vartheta\rangle_K\leq\|\Xi_N\|_K\|\vartheta\|_K.
    $$
    Dividing both sides by $2\lambda\|\vartheta\|_K$ (the case $\vartheta=0$ is trivial) proves (a).

    \paragraph{Proof of (b).} We first compute the mean of $\xi_z(G)$ for $G\sim U$. Let $\bar\xi:=-2\lambda\hat t$. For every $h\in\cF_K$, by the definition of $P_z$ and \eqref{eq:full_first_order},
    $$
    \EE\langle\xi_z(G),h\rangle_K=\EE\left[c_z\bigl(\obs(G\cdot z),\psi(G\cdot\tilde z)\bigr)h(\obs(G\cdot z))\right]=\int c_z(o,s)h(o)\,\diff P_z(o,s)=\langle\bar\xi,h\rangle_K.
    $$
    Moreover, $\|\xi_z(g)\|_K\leq\kappa$ since $|c_z|\leq1$, and $\|\bar\xi\|_K=2\lambda\|\hat t\|_K\leq\kappa$ by Lemma~\ref{lem:full_threshold}(iv). Writing $\xi_i:=\xi_z(G_i)$ for $0\leq i\leq N$, we decompose
    $$
    \Xi_N=\frac{1}{N+1}\sum_{i=0}^N(\xi_i-\bar\xi)=\frac{1}{N+1}\sum_{i=1}^N(\xi_i-\bar\xi)+\frac{1}{N+1}(\xi_0-\bar\xi),
    $$
    where the last term is deterministic since $G_0=e$.

    We next bound $\EE\|\Xi_N\|_K$. The inner products $\langle\xi_i,\xi_j\rangle_K=c_z(o_i,s_i)c_z(o_j,s_j)K(o_i,o_j)$ are measurable. For $1\leq i\neq j\leq N$, conditioning on $G_j$ and applying the above mean identity with $h=\xi_j-\bar\xi$ gives $\EE\langle\xi_i-\bar\xi,\xi_j-\bar\xi\rangle_K=0$. Hence
    $$
    \EE\Bigl\|\sum_{i=1}^N(\xi_i-\bar\xi)\Bigr\|_K^2=\sum_{i=1}^N\EE\|\xi_i-\bar\xi\|_K^2=N\bigl(\EE\|\xi_1\|_K^2-\|\bar\xi\|_K^2\bigr)\leq N\kappa^2.
    $$
    Combining this bound with Jensen's inequality, the triangle inequality and $\|\xi_0-\bar\xi\|_K\leq2\kappa$, we obtain
    $$
    \EE\|\Xi_N\|_K\leq\frac{1}{N+1}\rbr{\sqrt{N\kappa^2}+2\kappa}=\frac{\kappa(\sqrt N+2)}{N+1}.
    $$

    Finally, we establish the high-probability bound. Changing a single $G_i$, $1\leq i\leq N$, changes $\|\Xi_N\|_K$ by at most $2\kappa/(N+1)$. By McDiarmid's inequality,
    $$
    \PP\bigl(\|\Xi_N\|_K\geq\EE\|\Xi_N\|_K+t\bigr)\leq\exp\cbr{-\frac{(N+1)^2t^2}{2N\kappa^2}},\qquad t>0.
    $$
    Taking $t=\kappa\sqrt{2N\log(1/\delta)}/(N+1)$, we have, with probability at least $1-\delta$,
    $$
    \|\Xi_N\|_K\leq\frac{\kappa}{N+1}\Bigl(\sqrt N+2+\sqrt{2N\log(1/\delta)}\Bigr).
    $$
    Combining this bound with (a) and $\|\cdot\|_\infty\leq\kappa\|\cdot\|_K$ gives $\|\hat t_\samp-\hat t\|_\infty\leq\kappa\|\Xi_N\|_K/(2\lambda)\leq u_N(\delta)$. This completes the proof.
\end{proof}

\begin{lemma}[Measurability of the full RKHS threshold]
    \label{lem:full_measurability}
    In the setting of Appendix~\ref{sec:sampled_adaptive_symmpi}, the map $(z,g)\mapsto\hat t^{\,K}_z(\obs(g\cdot z))$ is measurable.
\end{lemma}
\begin{proof}
    We approximate the full threshold by the average of sampled thresholds and pass to the limit.

    We first show that the sampled threshold is jointly measurable. The proof of Lemma~\ref{lem:RKHS_technical}(iii) applies verbatim to the sampled threshold, with the $N+1$ points $\obs(g_i\cdot z)$, the scores $\psi(g_i\cdot\tilde z)$ and the weights $1/(N+1)$. Since the coefficients $a_\epsilon$ in that proof are continuous in the Gram matrix and the scores, which are measurable in $(z,g_{1:N})$, and $K(\obs(g_i\cdot z),\obs(g\cdot z))$ is measurable in $(z,g_{1:N},g)$, the map
    $$
    (z,g_{1:N},g)\mapsto\hat t^{\,K}_{z,\samp}(\obs(g\cdot z))
    $$
    is jointly measurable. Moreover, by the argument in the proof of Lemma~\ref{lem:RKHS_technical}(ii), it is bounded by $\kappa^2/(2\lambda)$. Hence, by Fubini's theorem, the function
    $$
    \tau_N(z,g):=\int\hat t^{\,K}_{z,\samp}(\obs(g\cdot z))\,\diff U^{\otimes N}(g_{1:N})
    $$
    is measurable in $(z,g)$.

    We next show that $\tau_N$ converges to the full threshold. By Lemma~\ref{lem:sampled_stability}(a) and (b), for every $(z,g)$,
    \begin{align*}
        \bigl|\tau_N(z,g)-\hat t^{\,K}_z(\obs(g\cdot z))\bigr|
        &\leq\int\bigl|\hat t^{\,K}_{z,\samp}(\obs(g\cdot z))-\hat t^{\,K}_z(\obs(g\cdot z))\bigr|\,\diff U^{\otimes N}(g_{1:N})\\
        &\leq\frac{\kappa\,\EE\|\Xi_N(z,G_{1:N})\|_K}{2\lambda}\leq\frac{\kappa^2(\sqrt N+2)}{2\lambda(N+1)}\to0\qquad\text{as }N\to\infty.
    \end{align*}
    Therefore, $\hat t^{\,K}_z(\obs(g\cdot z))=\lim_{N\to\infty}\tau_N(z,g)$ is a pointwise limit of measurable functions, and is thus measurable. This completes the proof.
\end{proof}

\end{document}